\documentclass[10pt]{article}
\usepackage{epsf}
\usepackage{axodraw}
\usepackage{epsfig}                            
\usepackage{graphicx}
\usepackage{rotate}
\usepackage{latexsym}
\renewcommand{\cal}{\mathcal}
\begin{document}
%
%
%-- ROSETTA.TEX
%
%================
% General utility
%================
%
\newcommand{\nl}{\nonumber\\}
\newcommand{\nn}{\nonumber}
\newcommand{\ds}{\displaystyle}
\newcommand{\mpar}[1]{{\marginpar{\hbadness10000%
                      \sloppy\hfuzz10pt\boldmath\bf#1}}%
                      \typeout{marginpar: #1}\ignorespaces}
\def\mnew{\mpar{\hfil NEW \hfil}\ignorespaces}
\newcommand{\lpar}{\left(}                            % bracketing
\newcommand{\rpar}{\right)} 
\newcommand{\lrbr}{\left[}
\newcommand{\rrbr}{\right]}
\newcommand{\lcbr}{\left\{}
\newcommand{\rcbr}{\right\}} 
\newcommand{\rbrak}[1]{\lrbr#1\rrbr}
\newcommand{\bq}{\begin{equation}}                    % equationing
\newcommand{\eq}{\end{equation}}
\newcommand{\bqa}{\begin{eqnarray}}
\newcommand{\eqa}{\end{eqnarray}}
\newcommand{\ba}[1]{\begin{array}{#1}}
\newcommand{\ea}{\end{array}}
\newcommand{\ben}{\begin{enumerate}}
\newcommand{\een}{\end{enumerate}}
\newcommand{\bei}{\begin{itemize}}
\newcommand{\eei}{\end{itemize}}
\newcommand{\bec}{\begin{center}}
\newcommand{\eec}{\end{center}}
\newcommand{\eqn}[1]{Eq.(\ref{#1})}
\newcommand{\eqns}[2]{Eqs.(\ref{#1}--\ref{#2})}
\newcommand{\eqnss}[1]{Eqs.(\ref{#1})}
\newcommand{\eqnsc}[2]{Eqs.(\ref{#1},~\ref{#2})}
\newcommand{\tbn}[1]{Tab.~\ref{#1}}
\newcommand{\tbns}[2]{Tabs.~\ref{#1}--\ref{#2}}
\newcommand{\tbnsc}[2]{Tabs.~\ref{#1},~\ref{#2}}
\newcommand{\fig}[1]{Fig.~\ref{#1}}
\newcommand{\figs}[2]{Figs.~\ref{#1}--\ref{#2}}
\newcommand{\sect}[1]{Sect.~\ref{#1}}
\newcommand{\subsect}[1]{Sub-Sect.~\ref{#1}}
%
% Miscellanea of symbols:
%========================
%
\newcommand{\TeV}{\;\mathrm{TeV}}                     
\newcommand{\GeV}{\;\mathrm{GeV}}
\newcommand{\MeV}{\;\mathrm{MeV}}
\newcommand{\nb}{\;\mathrm{nb}}
\newcommand{\pb}{\;\mathrm{pb}}
\newcommand{\fb}{\;\mathrm{fb}}
\def\Re{\mathop{\operator@font Re}\nolimits}
\def\Im{\mathop{\operator@font Im}\nolimits}
\newcommand{\ord}[1]{{\cal O}\lpar#1\rpar}
\newcommand{\group}{SU(2)\otimes U(1)}
\newcommand{\ib}{i}
\newcommand{\asums}[1]{\sum_{#1}}
\newcommand{\asumt}[2]{\sum_{#1}^{#2}}
\newcommand{\asum}[3]{\sum_{#1=#2}^{#3}}
%
% Powers of 10:
%==============
%
\newcommand{\tmi}{\times 10^{-1}}
\newcommand{\tmii}{\times 10^{-2}}
\newcommand{\tmiii}{\times 10^{-3}}
\newcommand{\tmiv}{\times 10^{-4}}
\newcommand{\tmfv}{\times 10^{-5}}
\newcommand{\tmfvi}{\times 10^{-6}}
\newcommand{\tmfvii}{\times 10^{-7}}
\newcommand{\tmfviii}{\times 10^{-8}}
\newcommand{\tmfix}{\times 10^{-9}}
\newcommand{\tmfx}{\times 10^{-10}}
%
% Fields:
%========
%
\newcommand{\fer}{{\rm{fer}}}
\newcommand{\bos}{{\rm{bos}}}
\newcommand{\lep}{{l}}
\newcommand{\had}{{h}}
\newcommand{\gen}{\rm{g}}
\newcommand{\dbl}{\rm{d}}
\newcommand{\philone}{\phi}
\newcommand{\philoneb}{\phi_{0}}
\newcommand{\phiind}[1]{\phi_{#1}}
\newcommand{\gBi}[2]{B_{#1}^{#2}}
\newcommand{\gBn}[1]{B_{#1}}
%
% vector-bosons
%--------------
%
\newcommand{\ph}{\gamma}
\newcommand{\ab}{A}
\newcommand{\abr}{A^r}
\newcommand{\abb}{A^{0}}
\newcommand{\abi}[1]{A_{#1}}
\newcommand{\abri}[1]{A^r_{#1}}
\newcommand{\abbi}[1]{A^{0}_{#1}}
\newcommand{\wb}{W}            
\newcommand{\wbi}[1]{W_{#1}}           
\newcommand{\wbp}{W^{+}}
\newcommand{\wbm}{W^{-}}
\newcommand{\wbpm}{W^{\pm}}
\newcommand{\wbpi}[1]{W^{+}_{#1}}
\newcommand{\wbmi}[1]{W^{-}_{#1}}
\newcommand{\wbpmi}[1]{W^{\pm}_{#1}}
\newcommand{\wbli}[1]{W^{[+}_{#1}}
\newcommand{\wbri}[1]{W^{-]}_{#1}}
\newcommand{\zb}{Z}
\newcommand{\zbi}[1]{Z_{#1}}
\newcommand{\vb}{V}
\newcommand{\vbi}[1]{V_{#1}}      
\newcommand{\vbiv}[1]{V^{*}_{#1}}      
\newcommand{\Pb}{P}
\newcommand{\Sb}{S}
\newcommand{\Bb}{B}
%
% Higgs-Kibble ghosts
%--------------------
%
\newcommand{\hk}{K}
\newcommand{\hKi}[1]{K_{#1}}
\newcommand{\hkg}{\phi}
\newcommand{\hkn}{\phi^{0}}                 
\newcommand{\hkp}{\phi^{+}}
\newcommand{\hkm}{\phi^{-}}
\newcommand{\hkpm}{\phi^{\pm}}
\newcommand{\hkmp}{\phi^{\mp}}
\newcommand{\hki}[1]{\phi^{#1}}
\newcommand{\hb}{H}
\newcommand{\hbi}[1]{H_{#1}}
\newcommand{\hkl}{\phi^{[+\cgfi\cgfi}}
\newcommand{\hkr}{\phi^{-]}}
%
% FP-ghosts
%----------
%
\newcommand{\fpx}{X}
\newcommand{\fpy}{Y}
\newcommand{\fpxp}{X^+}
\newcommand{\fpxm}{X^-}
\newcommand{\fpxpm}{X^{\pm}}
\newcommand{\fpxi}[1]{X^{#1}}
\newcommand{\fpyZ}{Y^{\ssZ}}
\newcommand{\fpyA}{Y^{\ssA}}
\newcommand{\fpyZA}{Y_{\ssZ,\ssA}}
\newcommand{\fpbxi}[1]{{\overline{X}}^{#1}}
\newcommand{\fpbyZ}{{\overline{Y}}^{\ssZ}}
\newcommand{\fpbyA}{{\overline{Y}}^{\ssA}}
\newcommand{\fpbyZA}{{\overline{Y}}^{\ssZ,\ssA}}
%
% Fermionic fields
%-----------------
%
\newcommand{\Flone}{F}
\newcommand{\fpsi}{\psi}
\newcommand{\fpsii}[1]{\psi^{#1}}
\newcommand{\fpsib}{\psi^{0}}
\newcommand{\fpsir}{\psi^r}
\newcommand{\fpsiL}{\psi_{_L}}
\newcommand{\fpsiR}{\psi_{_R}}
\newcommand{\fpsiLi}[1]{\psi_{_L}^{#1}}
\newcommand{\fpsiRi}[1]{\psi_{_R}^{#1}}
\newcommand{\fpsiLbi}[1]{\psi_{_{0L}}^{#1}}
\newcommand{\fpsiRbi}[1]{\psi_{_{0R}}^{#1}}
\newcommand{\fpsiLR}{\psi_{_{L,R}}}
\newcommand{\fbpsi}{{\overline{\psi}}}
\newcommand{\fbpsii}[1]{{\overline{\psi}}^{#1}}
\newcommand{\fbpsir}{{\overline{\psi}}^r}
\newcommand{\fbpsiL}{{\overline{\psi}}_{_L}}
\newcommand{\fbpsiR}{{\overline{\psi}}_{_R}}
\newcommand{\fbpsiLi}[1]{\overline{\psi_{_L}}^{#1}}
\newcommand{\fbpsiRi}[1]{\overline{\psi_{_R}}^{#1}}
\newcommand{\fe}{e}
\newcommand{\ff}{f}
\newcommand{\fep}{e^{+}}
\newcommand{\fem}{e^{-}}
\newcommand{\fepm}{e^{\pm}}
\newcommand{\fp}{f^{+}}
\newcommand{\fm}{f^{-}}
\newcommand{\fhp}{h^{+}}
\newcommand{\fhm}{h^{-}}
\newcommand{\fh}{h}
\newcommand{\flm}{\mu}
\newcommand{\flmp}{\mu^{+}}
\newcommand{\flmm}{\mu^{-}}
\newcommand{\fll}{l}
\newcommand{\fllp}{l^{+}}
\newcommand{\fllm}{l^{-}}
\newcommand{\flt}{\tau}
\newcommand{\fltp}{\tau^{+}}
\newcommand{\fltm}{\tau^{-}}
\newcommand{\fq}{q}
\newcommand{\fqi}[1]{\fq_{#1}}
\newcommand{\bfqi}[1]{\barq_{#1}}
\newcommand{\ffQ}{Q}
\newcommand{\fu}{u}
\newcommand{\fd}{d}
\newcommand{\fc}{c}
\newcommand{\fs}{s}
\newcommand{\fqp}{q'}
\newcommand{\fup}{u'}
\newcommand{\fdp}{d'}
\newcommand{\fcp}{c'}
\newcommand{\fsp}{s'}
\newcommand{\fdpp}{d''}
\newcommand{\ffi}[1]{f_{#1}}
\newcommand{\bffi}[1]{{\overline{f}}_{#1}}
\newcommand{\ffpi}[1]{f'_{#1}}
\newcommand{\bffpi}[1]{{\overline{f}}'_{#1}}
\newcommand{\ft}{t}
\newcommand{\ffb}{b}
\newcommand{\ffp}{f'}
\newcommand{\fft}{{\tilde{f}}}
\newcommand{\fl}{l}
\newcommand{\fli}[1]{\fl_{#1}}
\newcommand{\fnu}{\nu}
\newcommand{\fU}{U}
\newcommand{\fD}{D}
\newcommand{\fUc}{\overline{U}}
\newcommand{\fDc}{\overline{D}}
\newcommand{\fnul}{\nu_l}
\newcommand{\fnue}{\nu_e}
\newcommand{\fnum}{\nu_{\mu}}
\newcommand{\fnut}{\nu_{\tau}}
\newcommand{\fbe}{{\overline{e}}}
\newcommand{\fbu}{{\overline{u}}}
\newcommand{\fbd}{{\overline{d}}}
\newcommand{\fbf}{{\overline{f}}}
\newcommand{\fbfp}{{\overline{f}}'}
\newcommand{\fbl}{{\overline{l}}}
\newcommand{\fbnu}{{\overline{\nu}}}
\newcommand{\fbnul}{{\overline{\nu}}_{\fl}}
\newcommand{\fbnue}{{\overline{\nu}}_{\fe}}
\newcommand{\fbnum}{{\overline{\nu}}_{\flm}}
\newcommand{\fbnut}{{\overline{\nu}}_{\flt}}
\newcommand{\fuL}{u_{_L}}
\newcommand{\fdL}{d_{_L}}
\newcommand{\ffL}{f_{_L}}
\newcommand{\fbuL}{{\overline{u}}_{_L}}
\newcommand{\fbdL}{{\overline{d}}_{_L}}
\newcommand{\fbfL}{{\overline{f}}_{_L}}
\newcommand{\fuR}{u_{_R}}
\newcommand{\fdR}{d_{_R}}
\newcommand{\ffR}{f_{_R}}
\newcommand{\fbuR}{{\overline{u}}_{_R}}
\newcommand{\fbdR}{{\overline{d}}_{_R}}
\newcommand{\fbfR}{{\overline{f}}_{_R}}
%
% anti-fermions, GP's realization
%--------------------------------
%
\newcommand{\barf}{\overline f}                
\newcommand{\barl}{\overline l}
\newcommand{\barq}{\overline q}
\newcommand{\barqp}{\overline{q}'}
\newcommand{\barb}{\overline b}
\newcommand{\bart}{\overline t}
\newcommand{\barc}{\overline c}
\newcommand{\baru}{\overline u}
\newcommand{\bard}{\overline d}
\newcommand{\bars}{\overline s}
\newcommand{\barv}{\overline v}
\newcommand{\barnu}{\overline{\nu}}
\newcommand{\barne}{\overline{\nu}_{\fe}}
\newcommand{\barnm}{\overline{\nu}_{\flm}}
\newcommand{\barnt}{\overline{\nu}_{\flt}}
%
% gluon
%------
%
\newcommand{\glu}{g}
%
% (anti)proton
%-------------
%
\newcommand{\prot}{p}
\newcommand{\aprot}{{\bar{p}}}
\newcommand{\Nucln}{N}
%
% Vector resonances
%------------------
%
\newcommand{\tM}{{\tilde M}}
\newcommand{\tMs}{{\tilde M}^2}
\newcommand{\tW}{{\tilde \Gamma}}
\newcommand{\tWs}{{\tilde\Gamma}^2}
\newcommand{\fphi}{\phi}
\newcommand{\fJpsi}{J/\psi}
\newcommand{\fgpsi}{\psi}
\newcommand{\Glone}{\Gamma}
\newcommand{\Gloni}[1]{\Gamma_{#1}}
\newcommand{\Glones}{\Gamma^2}
\newcommand{\Glonec}{\Gamma^3}
\newcommand{\glone}{\gamma}
\newcommand{\glones}{\gamma^2}
\newcommand{\gloneq}{\gamma^4}
\newcommand{\gloni}[1]{\gamma_{#1}}
\newcommand{\glonis}[1]{\gamma^2_{#1}}
\newcommand{\Grest}[2]{\Gamma_{#1}^{#2}}
\newcommand{\grest}[2]{\gamma_{#1}^{#2}}
\newcommand{\resampl}{A_{_R}}
\newcommand{\resasyi}[1]{{\cal{A}}_{#1}}
\newcommand{\sSrest}[1]{\sigma_{#1}}
\newcommand{\Srest}[2]{\sigma_{#1}\lpar{#2}\rpar}
\newcommand{\Gdist}[1]{{\cal{G}}\lpar{#1}\rpar}
\newcommand{\sGdist}{{\cal{G}}}
\newcommand{\Aarea}{A_{0}}
\newcommand{\Aareai}[1]{{\cal{A}}\lpar{#1}\rpar}
\newcommand{\sAarea}{{\cal{A}}}
\newcommand{\resolw}{\sigma_{\Energ}}
\newcommand{\chizer}{\chi_{_0}}
\newcommand{\ini}{\rm{in}}
\newcommand{\fin}{\rm{fin}}
\newcommand{\ifi}{\rm{if}}
\newcommand{\ipf}{\rm{i+f}}
\newcommand{\tot}{\rm{tot}}
\newcommand{\Bac}{Q}
\newcommand{\Res}{R}
\newcommand{\Int}{I}
\newcommand{\NRe}{NR}
\newcommand{\ratoe}{\delta}
\newcommand{\ratoes}{\delta^2}
%
% QED-boxes
%----------
%
\newcommand{\Fbox}[2]{f^{\rm{box}}_{#1}\lpar{#2}\rpar}
\newcommand{\Dbox}[2]{\delta^{\rm{box}}_{#1}\lpar{#2}\rpar}
\newcommand{\Bbox}[3]{{\cal{B}}_{#1}^{#2}\lpar{#3}\rpar}
%
% Masses:
%========
%
\newcommand{\phm}{\lambda}
\newcommand{\phms}{\lambda^2}
\newcommand{\mV}{M_{_V}}
\newcommand{\mw}{M_{_W}}
\newcommand{\mX}{M_{_X}}
\newcommand{\mY}{M_{_Y}}
\newcommand{\LM}{M}
\newcommand{\mz}{M_{_Z}}
\newcommand{\bzm}{M_{_0}}
\newcommand{\mh}{M_{_H}}
\newcommand{\bhm}{M_{_{0H}}}
\newcommand{\mf}{m_f}
\newcommand{\mfp}{m_{f'}}
\newcommand{\mfh}{m_{h}}
\newcommand{\mt}{m_t}
\newcommand{\me}{m_e}
\newcommand{\mm}{m_{\mu}}
\newcommand{\mtau}{m_{\tau}}
\newcommand{\muq}{m_u}
\newcommand{\md}{m_d}
\newcommand{\muqp}{m'_u}
\newcommand{\mdqp}{m'_d}
\newcommand{\mc}{m_c}
\newcommand{\ms}{m_s}
\newcommand{\mb}{m_b}
\newcommand{\mup}{M_u}                              % pole masses
\newcommand{\mdp}{M_d}
\newcommand{\mcp}{M_c}
\newcommand{\msp}{M_s}
\newcommand{\mbp}{M_b}
%
% Masses squared, some cubed
%---------------------------
%
\newcommand{\mls}{m^2_l}
\newcommand{\mVs}{M^2_{_V}}
\newcommand{\mws}{M^2_{_W}}
\newcommand{\mwc}{M^3_{_W}}
\newcommand{\LMs}{M^2}
\newcommand{\LMc}{M^3}
\newcommand{\mzs}{M^2_{_Z}}
\newcommand{\mzc}{M^3_{_Z}}
\newcommand{\bzms}{M^2_{_0}}
\newcommand{\bzmc}{M^3_{_0}}
\newcommand{\bhms}{M^2_{_{0H}}}
\newcommand{\mhs}{M^2_{_H}}
\newcommand{\mfs}{m^2_f}
\newcommand{\mfc}{m^3_f}
\newcommand{\mfps}{m^2_{f'}}
\newcommand{\mfhs}{m^2_{h}}
\newcommand{\mfpc}{m^3_{f'}}
\newcommand{\mts}{m^2_t}
\newcommand{\mes}{m^2_e}
\newcommand{\mms}{m^2_{\mu}}
\newcommand{\mmc}{m^3_{\mu}}
\newcommand{\mmfour}{m^4_{\mu}}
\newcommand{\mmf}{m^5_{\mu}}
\newcommand{\mmfive}{m^5_{\mu}}
\newcommand{\mmsix}{m^6_{\mu}}
\newcommand{\mminv}{\frac{1}{m_{\mu}}}
\newcommand{\mtaus}{m^2_{\tau}}
\newcommand{\mus}{m^2_u}
\newcommand{\mds}{m^2_d}
\newcommand{\muqps}{m'^2_u}
\newcommand{\mdqps}{m'^2_d}
\newcommand{\mcs}{m^2_c}
\newcommand{\mss}{m^2_s}
\newcommand{\mbs}{m^2_b}
\newcommand{\mups}{M^2_u}
\newcommand{\mdps}{M^2_d}
\newcommand{\mcps}{M^2_c}
\newcommand{\msps}{M^2_s}
\newcommand{\mbps}{M^2_b}
%
% Some ratios
%------------
%
\newcommand{\muf}{\mu_{\ff}}
\newcommand{\mufs}{\mu^2_{\ff}}
\newcommand{\mufq}{\mu^4_{\ff}}
\newcommand{\mufx}{\mu^6_{\ff}}
\newcommand{\muz}{\mu_{_{\zb}}}
\newcommand{\muw}{\mu_{_{\wb}}}
\newcommand{\mut}{\mu_{\ft}}
\newcommand{\muzs}{\mu^2_{_{\zb}}}
\newcommand{\muws}{\mu^2_{_{\wb}}}
\newcommand{\muts}{\mu^2_{\ft}}
\newcommand{\muSW}{\mu^2_{_{\wb}}}
\newcommand{\muwq}{\mu^4_{_{\wb}}}
\newcommand{\muwsx}{\mu^6_{_{\wb}}}
\newcommand{\muwms}{\mu^{-2}_{_{\wb}}}
\newcommand{\muhs}{\mu^2_{_{\hb}}}
\newcommand{\muhq}{\mu^4_{_{\hb}}}
\newcommand{\muhsx}{\mu^6_{_{\hb}}}
\newcommand{\mutq}{\mu^4_{_{\hb}}}   % bardinworry to be checked by grep 
\newcommand{\mutsx}{\mu^6_{_{\hb}}}  %     -/-
\newcommand{\muL}{\mu}
\newcommand{\muS}{\mu^2}
\newcommand{\muQ}{\mu^4}
\newcommand{\muizs}{\mu^2_{0}}
\newcommand{\muizq}{\mu^4_{0}}
\newcommand{\muis}{\mu^2_{1}}
\newcommand{\muiis}{\mu^2_{2}}
\newcommand{\muiiis}{\mu^2_{3}}
\newcommand{\muii}[1]{\mu_{#1}}
\newcommand{\muisi}[1]{\mu^2_{#1}}
\newcommand{\muiqi}[1]{\mu^4_{#1}}
\newcommand{\muixi}[1]{\mu^6_{#1}}
\newcommand{\zm}{z_m}
\newcommand{\ri}[1]{r_{#1}}
\newcommand{\xw}{x_w}
\newcommand{\xws}{x^2_w}
\newcommand{\xwc}{x^3_w}
\newcommand{\xth}{x_t}
\newcommand{\xths}{x^2_t}
\newcommand{\xthc}{x^3_t}
\newcommand{\xthf}{x^4_t}
\newcommand{\xthv}{x^5_t}
\newcommand{\xthx}{x^6_t}
\newcommand{\xh}{x_h}
\newcommand{\xhs}{x^2_h}
\newcommand{\xhc}{x^3_h}
\newcommand{\Rl}{R_{\fl}}
\newcommand{\Rb}{R_{\ffb}}
\newcommand{\Rc}{R_{\fc}}
%
% Masses quartic
%---------------
%
\newcommand{\mwq}{M^4_{_\wb}}
\newcommand{\mwf}{M^4_{_\wb}}
\newcommand{\LMq}{M^4}
\newcommand{\mzq}{M^4_{_Z}}
\newcommand{\bzmq}{M^4_{_0}}
\newcommand{\mhq}{M^4_{_H}}
\newcommand{\mfq}{m^4_f}
\newcommand{\mfpq}{m^4_{f'}}
\newcommand{\mtq}{m^4_t}
\newcommand{\meq}{m^4_e}
\newcommand{\mmq}{m^4_{\mu}}
\newcommand{\mtauq}{m^4_{\tau}}
\newcommand{\muqq}{m^4_u}
\newcommand{\mdq}{m^4_d}
\newcommand{\mcq}{m^4_c}
\newcommand{\msq}{m^4_s}
\newcommand{\mbq}{m^4_b}
\newcommand{\mupq}{M^4_u}
\newcommand{\mdpq}{M^4_d}
\newcommand{\mcpq}{M^4_c}
\newcommand{\mspq}{M^4_s}
\newcommand{\mbpq}{M^4_b}
%
% Masses sixtupled
%-----------------
%
\newcommand{\mwx}{M^6_{_W}}
\newcommand{\mzx}{M^6_{_Z}}
\newcommand{\mfx}{m^6_f}
\newcommand{\mfpx}{m^6_{f'}}
\newcommand{\LMx}{M^6}
%
% More masses
%------------
%
\newcommand{\mer}{m_{er}}
\newcommand{\mlep}{m_l}
\newcommand{\mleps}{m^2_l}
\newcommand{\mone}{m_1}
\newcommand{\mtwo}{m_2}
\newcommand{\mtre}{m_3}
\newcommand{\mfor}{m_4}
\newcommand{\mlone}{m}
\newcommand{\mloneb}{\bar{m}}
\newcommand{\mind}[1]{m_{#1}}
\newcommand{\mones}{m^2_1}
\newcommand{\mtwos}{m^2_2}
\newcommand{\mtres}{m^2_3}
\newcommand{\mfors}{m^2_4}
\newcommand{\mlones}{m^2}
\newcommand{\minds}[1]{m^2_{#1}}
\newcommand{\moneq}{m^4_1}
\newcommand{\mtwoq}{m^4_2}
\newcommand{\mtreq}{m^4_3}
\newcommand{\mforq}{m^4_4}
\newcommand{\mloneq}{m^4}
\newcommand{\mindq}[1]{m^4_{#1}}
\newcommand{\mlonev}{m^5}
\newcommand{\mindv}[1]{m^5_{#1}}
\newcommand{\monex}{m^6_1}
\newcommand{\mtwox}{m^6_2}
\newcommand{\mtrex}{m^6_3}
\newcommand{\mforx}{m^6_4}
\newcommand{\mlonex}{m^6}
\newcommand{\mindx}[1]{m^6_{#1}}
\newcommand{\Mone}{M_1}
\newcommand{\Mtwo}{M_2}
\newcommand{\Mtre}{M_3}
\newcommand{\Mfor}{M_4}
\newcommand{\Mlone}{M}
\newcommand{\Mlonep}{M'}
\newcommand{\Miind}{M_i}
\newcommand{\Mind}[1]{M_{#1}}
\newcommand{\Minds}[1]{M^2_{#1}}
\newcommand{\Mindc}[1]{M^3_{#1}}
\newcommand{\Mindf}[1]{M^4_{#1}}
\newcommand{\Mones}{M^2_1}
\newcommand{\Mtwos}{M^2_2}
\newcommand{\Mtres}{M^2_3}
\newcommand{\Mfors}{M^2_4}
\newcommand{\Mlones}{M^2}
\newcommand{\Mloneps}{M'^2}
\newcommand{\Miinds}{M^2_i}
\newcommand{\Mlonec}{M^3}
\newcommand{\Monec}{M^3_1}
\newcommand{\Mtwoc}{M^3_2}
\newcommand{\Moneq}{M^4_1}
\newcommand{\Mtwoq}{M^4_2}
\newcommand{\Mtreq}{M^4_3}
\newcommand{\Mforq}{M^4_4}
\newcommand{\Mloneq}{M^4}
\newcommand{\Miindq}{M^4_i}
\newcommand{\Monex}{M^6_1}
\newcommand{\Mtwox}{M^6_2}
\newcommand{\Mtrex}{M^6_3}
\newcommand{\Mforx}{M^6_4}
\newcommand{\Mlonex}{M^6}
\newcommand{\Miindx}{M^6_i}
\newcommand{\meb}{m_0}
\newcommand{\mebs}{m^2_0}
%
% Pole masses again
%------------------
%
\newcommand{\Mq }{M_q  }
\newcommand{\MqS}{M^2_q}
\newcommand{\Ms }{M_s  }
\newcommand{\MsS}{M^2_s}
\newcommand{\Mc }{M_c  }
\newcommand{\McS}{M^2_c}
\newcommand{\Mb }{M_b  }
\newcommand{\MbS}{M^2_b}
\newcommand{\Mt }{M_t  }
\newcommand{\MtS}{M^2_t}
%
% Some quark masses
%------------------
%
\newcommand{\mq}{m_q}
\newcommand{\mqs}{m^2_q}
\newcommand{\mqS}{m^2_q}
\newcommand{\mqQ}{m^4_q}
\newcommand{\mqX}{m^6_q}
\newcommand{\mqp}{m'_q }
\newcommand{\mqpS}{m'^2_q}
\newcommand{\mqpQ}{m'^4_q}
%
% Some logs of mass ratios
%=========================
%
\newcommand{\lL}{l}
\newcommand{\ls}{l^2}
\newcommand{\LL}{L}
\newcommand{\LcalL}{\cal{L}}
\newcommand{\LS}{L^2}
\newcommand{\LC}{L^3}
\newcommand{\LQ}{L^4}
\newcommand{\lw}{l_w}
\newcommand{\Lw}{L_w}
\newcommand{\Lws}{L^2_w}
\newcommand{\Lz}{L_z}
\newcommand{\Lzs}{L^2_z}
\newcommand{\Li}[1]{L_{#1}}
\newcommand{\Lis}[1]{L^2_{#1}}
\newcommand{\Lic}[1]{L^3_{#1}}
%
% Mandelstam variables
%=====================
%
\newcommand{\sman}{s}
\newcommand{\tman}{t}
\newcommand{\uman}{u}
\newcommand{\smani}[1]{s_{#1}}
\newcommand{\bsmani}[1]{{\bar{s}}_{#1}}
\newcommand{\smans}{s^2}
\newcommand{\tmans}{t^2}
\newcommand{\umans}{u^2}
\newcommand{\shat}{{\hat s}}
\newcommand{\that}{{\hat t}}
\newcommand{\uhat}{{\hat u}}
\newcommand{\hq}{{\hat Q}}
%
% More invariant variables
%-------------------------
%
\newcommand{\smanp}{s'}
\newcommand{\smanpi}[1]{s'_{#1}}
\newcommand{\tmanp}{t'}
\newcommand{\umanp}{u'}
\newcommand{\kappi}[1]{\kappa_{#1}}
\newcommand{\zetai}[1]{\zeta_{#1}}
%
% QED
%====
%
% Phase space and QED-varia
%--------------------------
%
\newcommand{\Phaspi}[1]{\Gamma_{#1}}
\newcommand{\rbetai}[1]{\beta_{#1}}
\newcommand{\ralphai}[1]{\alpha_{#1}}
\newcommand{\rbetais}[1]{\beta^2_{#1}}
\newcommand{\Lambdi}[1]{\Lambda_{#1}}
\newcommand{\Nomini}[1]{N_{#1}}
\newcommand{\smlone}{\frac{-\sman-\ib\ep}{\mlones}}
%
% Angles in bremsstrahlung
%-------------------------
%
\newcommand{\theti}[1]{\theta_{#1}}
\newcommand{\delti}[1]{\delta_{#1}}
\newcommand{\phigi}[1]{\phi_{#1}}
\newcommand{\acoli}[1]{\xi_{#1}}
\newcommand{\scats}{s}
\newcommand{\scatss}{s^2}
\newcommand{\scatsi}[1]{s_{#1}}
\newcommand{\scatsis}[1]{s^2_{#1}}
\newcommand{\scatst}[2]{s_{#1}^{#2}}
\newcommand{\scatc}{c}
\newcommand{\scatcs}{c^2}
\newcommand{\scatci}[1]{c_{#1}}
\newcommand{\scatcis}[1]{c^2_{#1}}
\newcommand{\scatct}[2]{c_{#1}^{#2}}
\newcommand{\angamt}[2]{\gamma_{#1}^{#2}}
%
% More about bremsstrahlung
%--------------------------
%
\newcommand{\Regia}{{\cal{R}}}
\newcommand{\Iconi}[2]{{\cal{I}}_{#1}\lpar{#2}\rpar}
\newcommand{\sIcon}[1]{{\cal{I}}_{#1}}
\newcommand{\betaf}{\beta_{\ff}}
\newcommand{\betafs}{\beta^2_{\ff}}
\newcommand{\Kfact}[2]{{\cal{K}}_{#1}\lpar{#2}\rpar}
%
% Structure and flux functions
%-----------------------------
%
\newcommand{\Struf}[4]{{\cal D}^{#1}_{#2}\lpar{#3;#4}\rpar}
\newcommand{\sStruf}[2]{{\cal D}^{#1}_{#2}}
\newcommand{\Fluxf}[2]{H\lpar{#1;#2}\rpar}
\newcommand{\Fluxfi}[4]{H_{#1}^{#2}\lpar{#3;#4}\rpar}
\newcommand{\sFluxf}{H}
\newcommand{\Bflux}[2]{{\cal{B}}_{#1}\lpar{#2}\rpar}
\newcommand{\bflux}[2]{{\cal{B}}_{#1}\lpar{#2}\rpar}
\newcommand{\Fluxd}[2]{D_{#1}\lpar{#2}\rpar}
\newcommand{\fluxd}[2]{C_{#1}\lpar{#2}\rpar}
\newcommand{\Fluxh}[4]{{\cal{H}}_{#1}^{#2}\lpar{#3;#4}\rpar}
\newcommand{\Sluxh}[4]{{\cal{S}}_{#1}^{#2}\lpar{#3;#4}\rpar}
\newcommand{\Fluxhb}[4]{{\overline{{\cal{H}}}}_{#1}^{#2}\lpar{#3;#4}\rpar}
\newcommand{\sFluxhb}{{\overline{{\cal{H}}}}}
\newcommand{\Sluxhb}[4]{{\overline{{\cal{S}}}}_{#1}^{#2}\lpar{#3;#4}\rpar}
\newcommand{\sSluxhb}[2]{{\overline{{\cal{S}}}}_{#1}^{#2}}
\newcommand{\fluxh}[4]{h_{#1}^{#2}\lpar{#3;#4}\rpar}
\newcommand{\fluxhs}[3]{h_{#1}^{#2}\lpar{#3}\rpar}
\newcommand{\sfluxhs}[2]{h_{#1}^{#2}}
\newcommand{\fluxhb}[4]{{\overline{h}}_{#1}^{#2}\lpar{#3;#4}\rpar}
\newcommand{\Strufd}[2]{D\lpar{#1;#2}\rpar}
%
% Mass and momenta squared ratios
%================================
%
\newcommand{\rMQ}[1]{r^2_{#1}}
\newcommand{\rMQs}[1]{r^4_{#1}}
\newcommand{\rf}{w_{\ff}}
\newcommand{\zf}{z_{\ff}}
\newcommand{\rfs}{w^2_{\ff}}
\newcommand{\zfs}{z^2_{\ff}}
\newcommand{\rfc}{w^3_{\ff}}
\newcommand{\zfc}{z^3_{\ff}}
\newcommand{\df}{d_{\ff}}
\newcommand{\rfp}{w_{\ffp}}
\newcommand{\rfps}{w^2_{\ffp}}
\newcommand{\rfpc}{w^3_{\ffp}}
\newcommand{\rt}{w_{\ft}}
\newcommand{\rts}{w^2_{\ft}}
\newcommand{\dt}{d_{\ft}}
\newcommand{\dts}{d^2_{\ft}}
\newcommand{\rh}{r_{h}}
\newcommand{\Lnrt}{\ln{\rt}}
\newcommand{\Rw}{R_{_{\wb}}}
\newcommand{\Rws}{R^2_{_{\wb}}}
\newcommand{\Rz}{R_{_{\zb}}}
\newcommand{\Rzp}{R^{+}_{_{\zb}}}
\newcommand{\Rzm}{R^{-}_{_{\zb}}}
\newcommand{\Rzs}{R^2_{_{\zb}}}
\newcommand{\Rzc}{R^3_{_{\zb}}}
\newcommand{\Rv}{R_{_{\vb}}}
\newcommand{\rhw}{r_{_{\wb}}}
\newcommand{\rhz}{r_{_{\zb}}}
\newcommand{\rhws}{r^2_{_{\wb}}}
\newcommand{\rhzs}{r^2_{_{\zb}}}
%
% More ratios
%------------
%
\newcommand{\vqrato}{z}
\newcommand{\vqrats}{w}
\newcommand{\vqratq}{w^2}
\newcommand{\seyrat}{z}
\newcommand{\sexrat}{w}
\newcommand{\sehrat}{h}
\newcommand{\sewrat}{w}
\newcommand{\sezrat}{z}
\newcommand{\zetav}{\zeta}
\newcommand{\zetavi}[1]{\zeta_{#1}}
\newcommand{\bpo}{\beta^2}
\newcommand{\bpos}{\beta^4}
\newcommand{\bpt}{{\tilde\beta}^2}
\newcommand{\lap}{\kappa}
\newcommand{\hw}{h_{_{\wb}}}
\newcommand{\hz}{h_{_{\zb}}}
%
% Couplings
%==========
%
\newcommand{\ec}{e}
\newcommand{\ecs}{e^2}
\newcommand{\ect}{e^3}
\newcommand{\ecq}{e^4}
\newcommand{\ecb}{e_{_0}}
\newcommand{\ecbs}{e^2_{_0}}
\newcommand{\ecbq}{e^4_{_0}}
\newcommand{\eci}[1]{e_{#1}}
\newcommand{\ecis}[1]{e^2_{#1}}
\newcommand{\hate}{{\hat e}}
\newcommand{\gss}{g_{_S}}
\newcommand{\gsss}{g^2_{_S}}
\newcommand{\gssb}{g^2_{_{S_0}}}
\newcommand{\als}{\alpha_{_S}}
\newcommand{\as}{a_{_S}}
\newcommand{\ass}{a^2_{_S}}
\newcommand{\gf}{G_{\ssF}}
\newcommand{\gfs}{G^2_{\ssF}}
\newcommand{\gb}{g} 
\newcommand{\gbi}[1]{g_{#1}}
\newcommand{\gbb}{g_{0}}
\newcommand{\gbs}{g^2}
\newcommand{\gbc}{g^3}
\newcommand{\gbf}{g^4}
\newcommand{\gpb}{g'}
\newcommand{\gpbs}{g'^2}
\newcommand{\vc}[1]{v_{#1}}
\newcommand{\ac}[1]{a_{#1}}
\newcommand{\vcc}[1]{v^*_{#1}}
\newcommand{\acc}[1]{a^*_{#1}}
\newcommand{\hatv}[1]{{\hat v}_{#1}}
\newcommand{\vcs}[1]{v^2_{#1}}
\newcommand{\acs}[1]{a^2_{#1}}
\newcommand{\gcv}[1]{g^{#1}_{\ssV}}
\newcommand{\gca}[1]{g^{#1}_{\ssA}}
\newcommand{\gcp}[1]{g^{+}_{#1}}
\newcommand{\gcm}[1]{g^{-}_{#1}}
\newcommand{\gcpm}[1]{g^{\pm}_{#1}}
\newcommand{\vci}[2]{v^{#2}_{#1}}
\newcommand{\aci}[2]{a^{#2}_{#1}}
\newcommand{\vceff}[1]{v^{#1}_{\rm{eff}}}
\newcommand{\hvc}[1]{\hat{v}_{#1}}
\newcommand{\hvcs}[1]{\hat{v}^2_{#1}}
\newcommand{\Vc}[1]{V_{#1}}
\newcommand{\Ac}[1]{A_{#1}}
\newcommand{\Vcs}[1]{V^2_{#1}}
\newcommand{\Acs}[1]{A^2_{#1}}
\newcommand{\vpa}[2]{\sigma_{#1}^{#2}}
\newcommand{\vma}[2]{\delta_{#1}^{#2}}
\newcommand{\vfw}{\sigma^{a}_{\ff}}
\newcommand{\vfpw}{\sigma^{a}_{\ffp}}
\newcommand{\vfwi}[1]{\sigma^{a}_{#1}}
\newcommand{\vfwsi}[1]{\lpar\sigma^{a}_{#1}\rpar^2}
\newcommand{\vvfw}{\sigma^{a}_{\ff}}
\newcommand{\vvew}{\sigma^{a}_{\fe}}
%-------------------------------------------------> g,G - couplings 
\newcommand{\gv}{g_{_V}}
\newcommand{\ga}{g_{_A}}
\newcommand{\gve}{g^{\fe}_{_{V}}}
\newcommand{\gae}{g^{\fe}_{_{A}}}
\newcommand{\gvf}{g^{\ff}_{_{V}}}
\newcommand{\gaf}{g^{\ff}_{_{A}}}
\newcommand{\gva}{g_{_{V,A}}}
\newcommand{\gvae}{g^{\fe}_{_{V,A}}}
\newcommand{\gvaf}{g^{\ff}_{_{V,A}}}
\newcommand{\sGv}{{\cal{G}}_{_V}}
\newcommand{\cGa}{{\cal{G}}^{*}_{_A}}
\newcommand{\cGv}{{\cal{G}}^{*}_{_V}}
\newcommand{\sGa}{{\cal{G}}_{_A}}
\newcommand{\Gvf}{{\cal{G}}^{\ff}_{_{V}}}
\newcommand{\Gaf}{{\cal{G}}^{\ff}_{_{A}}}
\newcommand{\Gvaf}{{\cal{G}}^{\ff}_{_{V,A}}}
\newcommand{\Gve}{{\cal{G}}^{\fe}_{_{V}}}
\newcommand{\Gae}{{\cal{G}}^{\fe}_{_{A}}}
\newcommand{\Gvae}{{\cal{G}}^{\fe}_{_{V,A}}}
\newcommand{\gvl}{g^{\fl}_{_{V}}}
\newcommand{\gal}{g^{\fl}_{_{A}}}
\newcommand{\gval}{g^{\fl}_{_{V,A}}}
\newcommand{\gvb}{g^{\ffb}_{_{V}}}
\newcommand{\gab}{g^{\ffb}_{_{A}}}
\newcommand{\fvf}{F_{_V}^{\ff}}
\newcommand{\faf}{F_{_A}^{\ff}}
\newcommand{\fvl}{F_{_V}^{\fl}}
\newcommand{\fal}{F_{_A}^{\fl}}
\newcommand{\corat}{\kappa}
\newcommand{\corats}{\kappa^2}
%
% Deltology-rhoology-kappaology
%==============================
%                              
\newcommand{\dr}{\Delta r}
\newcommand{\drl}{\Delta r_{_L}}
\newcommand{\drh}{\Delta{\hat r}}
\newcommand{\drhw}{\Delta{\hat r}_{_W}}
\newcommand{\rhou}{\rho_{_U}}
\newcommand{\rhoz}{\rho_{_\zb}}
\newcommand{\rZ}{\rho_{_\zb}}
\newcommand{\rhob}{\rho_{_0}}
\newcommand{\rZf}{\rho^{\ff}_{_\zb}}
\newcommand{\rhoe}{\rho_{\fe}}
\newcommand{\rhof}{\rho_{\ff}}
\newcommand{\rhoi}[1]{\rho_{#1}}
\newcommand{\kZf}{\kappa^{\ff}_{_\zb}}
\newcommand{\rWf}{\rho^{\ff}_{_\wb}}
\newcommand{\brWf}{{\bar{\rho}}^{\ff}_{_\wb}}
\newcommand{\rHf}{\rho^{\ff}_{_\hb}}
\newcommand{\brHf}{{\bar{\rho}}^{\ff}_{_\hb}}
\newcommand{\rhoR}{\rho^R_{_{\zb}}}
\newcommand{\hatrh}{{\hat\rho}}
\newcommand{\ku}{\kappa_{_U}}
\newcommand{\rZdf}[1]{\rho^{#1}_{_\zb}}
\newcommand{\kZdf}[1]{\kappa^{#1}_{_\zb}}
\newcommand{\rdfL}[1]{\rho^{#1}_{_L}}
\newcommand{\kdfL}[1]{\kappa^{#1}_{_L}}
\newcommand{\rdfR}[1]{\rho^{#1}_{\rm{rem}}}
\newcommand{\kdfR}[1]{\kappa^{#1}_{\rm{rem}}}
\newcommand{\bark}{\overline\kappa}
%
% Weak mixing angles
%===================
%
\newcommand{\stw}{s_{\theta}}             % bare, Lagrangian parameters
\newcommand{\ctw}{c_{\theta}}
\newcommand{\stws}{s_{\theta}^2}
\newcommand{\stwc}{s_{\theta}^3}
\newcommand{\stwf}{s_{\theta}^4}
\newcommand{\stwx}{s_{\theta}^6}
\newcommand{\ctws}{c_{\theta}^2}
\newcommand{\ctwc}{c_{\theta}^3}
\newcommand{\ctwf}{c_{\theta}^4}
\newcommand{\ctwx}{c_{\theta}^6}
\newcommand{\stwfiv}{s_{\theta}^5}
\newcommand{\ctwfiv}{c_{\theta}^5}
\newcommand{\stwsix}{s_{\theta}^6}
\newcommand{\ctwsix}{c_{\theta}^6}
%
% on-shell sines
%---------------
%
\newcommand{\siw}{s_{_W}}           
\newcommand{\cow}{c_{_W}}
\newcommand{\siws}{s^2_{_W}}
\newcommand{\cows}{c^2_{_W}}
\newcommand{\siwc}{s^3_{_W}}
\newcommand{\cowc}{c^3_{_W}}
\newcommand{\siwf}{s^4_{_W}}
\newcommand{\cowf}{c^4_{_W}}
\newcommand{\siwx}{s^6_{_W}}
\newcommand{\cowx}{c^6_{_W}}
\newcommand{\sons}{s_{_W}}
\newcommand{\sonss}{s^2_{_W}}
\newcommand{\cons}{c_{_W}}
\newcommand{\cooss}{c^2_{_W}}
%
% effective and other weak mixing angles
%---------------------------------------
%
\newcommand{\szs}{{\overline s}^2}
\newcommand{\szq}{{\overline s}^4}
\newcommand{\czs}{{\overline c}^2}
\newcommand{\sbs}{s_{_0}^2}
\newcommand{\cbs}{c_{_0}^2}
\newcommand{\dss}{\Delta s^2}
\newcommand{\snes}{s_{\nu e}^2}
\newcommand{\cnes}{c_{\nu e}^2}
\newcommand{\shs}{{\hat s}^2}
\newcommand{\chs}{{\hat c}^2}
\newcommand{\chl}{{\hat c}}
\newcommand{\seffs}{s^2_{\rm{eff}}}
\newcommand{\seffsf}[1]{\sin^2\theta^{#1}_{\rm{eff}}}
\newcommand{\sress}{s^2_{\rm res}}                
\newcommand{\sR}{s_{_R}}
\newcommand{\sRs}{s^2_{_R}}
\newcommand{\ctwe}{c_{\theta}^6}
\newcommand{\sany}{s}
\newcommand{\cany}{c}
\newcommand{\sanys}{s^2}
\newcommand{\canys}{c^2}
%
% Spinology etc.
%===============
%
\newcommand{\sip}{u}                             % incoming particle
\newcommand{\siap}{{\bar{v}}}                    %    "     anti-p
\newcommand{\sop}{{\bar{u}}}                     % outgoing p
\newcommand{\soap}{v}                            %    "     anti-p
\newcommand{\ip}[1]{u\lpar{#1}\rpar}             % incoming particle
\newcommand{\iap}[1]{{\bar{v}}\lpar{#1}\rpar}    %    "     anti-p
\newcommand{\op}[1]{{\bar{u}}\lpar{#1}\rpar}     % outgoing p
\newcommand{\oap}[1]{v\lpar{#1}\rpar}            %    "     anti-p
%
% With polarization
%------------------
%
\newcommand{\ipp}[2]{u\lpar{#1,#2}\rpar}         % incoming particle
\newcommand{\ipap}[2]{{\bar v}\lpar{#1,#2}\rpar} %    "     anti-p
\newcommand{\opp}[2]{{\bar u}\lpar{#1,#2}\rpar}  % outgoing p
\newcommand{\opap}[2]{v\lpar{#1,#2}\rpar}        %    "     anti-p
\newcommand{\upspi}[1]{u\lpar{#1}\rpar}
\newcommand{\vpspi}[1]{v\lpar{#1}\rpar}
\newcommand{\wpspi}[1]{w\lpar{#1}\rpar}
\newcommand{\ubpspi}[1]{{\bar{u}}\lpar{#1}\rpar}
\newcommand{\vbpspi}[1]{{\bar{v}}\lpar{#1}\rpar}
\newcommand{\wbpspi}[1]{{\bar{w}}\lpar{#1}\rpar}
\newcommand{\udpspi}[1]{u^{\dagger}\lpar{#1}\rpar}
\newcommand{\vdpspi}[1]{v^{\dagger}\lpar{#1}\rpar}
\newcommand{\wdpspi}[1]{w^{\dagger}\lpar{#1}\rpar}
\newcommand{\Ubilin}[1]{U\lpar{#1}\rpar}
\newcommand{\Vbilin}[1]{V\lpar{#1}\rpar}
\newcommand{\Xbilin}[1]{X\lpar{#1}\rpar}
\newcommand{\Ybilin}[1]{Y\lpar{#1}\rpar}
\newcommand{\up}[2]{u_{#1}\lpar #2\rpar}
\newcommand{\vp}[2]{v_{#1}\lpar #2\rpar}
\newcommand{\ubp}[2]{{\overline u}_{#1}\lpar #2\rpar}
\newcommand{\vbp}[2]{{\overline v}_{#1}\lpar #2\rpar}
\newcommand{\Pje}[1]{\frac{1}{2}\lpar 1 + #1\,\gfd\rpar}
\newcommand{\Pj}[1]{\Pi_{#1}}
\newcommand{\trace}{\mbox{Tr}}
%
% polarization operators and related things
%==========================================
%
\newcommand{\Poper}[2]{P_{#1}\lpar{#2}\rpar}
\newcommand{\Loper}[2]{\Lambda_{#1}\lpar{#2}\rpar}
\newcommand{\proj}[3]{P_{#1}\lpar{#2,#3}\rpar}
\newcommand{\sproj}[1]{P_{#1}}
\newcommand{\Nden}[3]{N_{#1}^{#2}\lpar{#3}\rpar}
\newcommand{\sNden}[1]{N_{#1}}
\newcommand{\nden}[2]{n_{#1}^{#2}}
%
% Wave functions
%===============
%
\newcommand{\vwf}[2]{e_{#1}\lpar#2\rpar}             % vector wave funct.
\newcommand{\vwfb}[2]{{\overline e}_{#1}\lpar#2\rpar}
\newcommand{\pwf}[2]{\epsilon_{#1}\lpar#2\rpar}      % photon wave funct.
\newcommand{\sla}[1]{/\!\!\!#1}
\newcommand{\slac}[1]{/\!\!\!\!#1}
%
% Momenta
%========
%
\newcommand{\iemom}{p_{_-}}                    % 2f incoming momenta
\newcommand{\ipmom}{p_{_+}}
\newcommand{\oemom}{q_{_-}}                    % 2f outgoing momenta
\newcommand{\opmom}{q_{_+}}
%
% Scalar product of two momenta
%==============================
%
\newcommand{\spro}[2]{{#1}\cdot{#2}}
%
% gammas
%=======
%
\newcommand{\gfour}{\gamma_4}                    
\newcommand{\gfd}{\gamma_5}                    
\newcommand{\gap}{\lpar 1+\gamma_5\rpar}
\newcommand{\gam}{\lpar 1-\gamma_5\rpar}
\newcommand{\gdp}{\gamma_+}
\newcommand{\gdm}{\gamma_-}
\newcommand{\gdpm}{\gamma_{\pm}}
\newcommand{\gad}{\gamma}
\newcommand{\gapm}{\lpar 1\pm\gamma_5\rpar}
\newcommand{\gadi}[1]{\gamma_{#1}}
\newcommand{\gadu}[1]{\gamma_{#1}}
\newcommand{\gapu}[1]{\gamma^{#1}}
\newcommand{\sigd}[2]{\sigma_{#1#2}}
\newcommand{\sumsp}{\overline{\sum_{\mbox{spins}}}}
%
% Special functions & integrals
%==============================
%
\newcommand{\li}[2]{\mathrm{Li}_{#1}\lpar\displaystyle{#2}\rpar} % polylog
\newcommand{\etaf}[2]{\eta\lpar#1,#2\rpar}
\newcommand{\lkall}[3]{\lambda\lpar#1,#2,#3\rpar}       % Kallen's lambda
\newcommand{\slkall}[3]{\lambda^{1/2}\lpar#1,#2,#3\rpar}%\sqrt
\newcommand{\segam}{\Gamma}                             % Euler's Gamma
\newcommand{\egam}[1]{\Gamma\lpar#1\rpar}               % Euler's Gamma
\newcommand{\ebe}[2]{B\lpar#1,#2\rpar}                  % Euler's beta
\newcommand{\ddel}[1]{\delta\lpar#1\rpar}               % Dirac's delta
\newcommand{\drii}[2]{\delta_{#1#2}}                    % Kronecker's delta
\newcommand{\driv}[4]{\delta_{#1#2#3#4}}                % gen.   "
\newcommand{\intmomi}[2]{\int\,d^{#1}#2}
\newcommand{\intmomii}[3]{\int\,d^{#1}#2\,\int\,d^{#1}#3}
\newcommand{\intfx}[1]{\int_{\scriptstyle 0}^{\scriptstyle 1}\,d#1}
\newcommand{\intfxy}[2]{\int_{\scriptstyle 0}^{\scriptstyle 1}\,d#1\,
                        \int_{\scriptstyle 0}^{\scriptstyle #1}\,d#2}
\newcommand{\intfxyz}[3]{\int_{\scriptstyle 0}^{\scriptstyle 1}\,d#1\,
                         \int_{\scriptstyle 0}^{\scriptstyle #1}\,d#2\,
                         \int_{\scriptstyle 0}^{\scriptstyle #2}\,d#3}
\newcommand{\Beta}[2]{{\rm{B}}\lpar #1,#2\rpar}
\newcommand{\sBeta}{\rm{B}}
\newcommand{\sign}[1]{{\rm{sign}}\lpar{#1}\rpar}
%
% Z widths
%=========
%
\newcommand{\gn}{\Gamma_{\nu}}
\newcommand{\gel}{\Gamma_{\fe}}
\newcommand{\gmu}{\Gamma_{\mu}}
\newcommand{\gff}{\Gamma_{\ff}}
\newcommand{\gt}{\Gamma_{\tau}}
\newcommand{\gl}{\Gamma_{\fl}}
\newcommand{\gq}{\Gamma_{\fq}}
\newcommand{\gu}{\Gamma_{\fu}}
\newcommand{\gd}{\Gamma_{\fd}}
\newcommand{\gc}{\Gamma_{\fc}}
\newcommand{\gs}{\Gamma_{\fs}}
\newcommand{\gbq}{\Gamma_{\ffb}}
\newcommand{\gz}{\Gamma_{_{\zb}}}
\newcommand{\gw}{\Gamma_{_{\wb}}}
\newcommand{\gh}{\Gamma_{_{h}}}
\newcommand{\ghb}{\Gamma_{_{\hb}}}
\newcommand{\gi}{\Gamma_{\rm{inv}}}
\newcommand{\gzs}{\Gamma^2_{_{\zb}}}
%
% Quantum numbers
%================
%
\newcommand{\tcie}{I^{(3)}_{\fe}}
\newcommand{\tcim}{I^{(3)}_{\flm}}
\newcommand{\tcif}{I^{(3)}_{\ff}}
\newcommand{\tciq}{I^{(3)}_{\fq}}
\newcommand{\tcib}{I^{(3)}_{\ffb}}
\newcommand{\tcih}{I^{(3)}_h}
\newcommand{\tcii}{I^{(3)}_i}
\newcommand{\tcift}{I^{(3)}_{\tilde f}}
\newcommand{\tcifp}{I^{(3)}_{f'}}
\newcommand{\wispt}[1]{I^{(3)}_{#1}}
\newcommand{\ql}{Q_l}
\newcommand{\qe}{Q_e}
\newcommand{\qu}{Q_u}
\newcommand{\qd}{Q_d}
\newcommand{\qb}{Q_b}
\newcommand{\qt}{Q_t}
\newcommand{\qup}{Q'_u}
\newcommand{\qdp}{Q'_d}
\newcommand{\qmu}{Q_{\mu}}
\newcommand{\qes}{Q^2_e}
\newcommand{\qec}{Q^3_e}
\newcommand{\qus}{Q^2_u}
\newcommand{\qds}{Q^2_d}
\newcommand{\qbs}{Q^2_b}
\newcommand{\qts}{Q^2_t}
\newcommand{\qbc}{Q^3_b}
\newcommand{\qf}{Q_f}
\newcommand{\qfs}{Q^2_f}
\newcommand{\qfc}{Q^3_f}
\newcommand{\qff}{Q^4_f}
\newcommand{\qep}{Q_{e'}}
\newcommand{\qfp}{Q_{f'}}
\newcommand{\qfps}{Q^2_{f'}}
\newcommand{\qfpc}{Q^3_{f'}}
\newcommand{\qq}{Q_q}
\newcommand{\qqs}{Q^2_q}
\newcommand{\qi}{Q_i}
\newcommand{\qis}{Q^2_i}
\newcommand{\qj}{Q_j}
\newcommand{\qjs}{Q^2_j}
\newcommand{\QW}{Q_{_\wb}}
\newcommand{\QWs}{Q^2_{_\wb}}
\newcommand{\Qd}{Q_d}
\newcommand{\Qds}{Q^2_d}
\newcommand{\Qu}{Q_u}
\newcommand{\Qus}{Q^2_u}
\newcommand{\vi}{v_i}
\newcommand{\vis}{v^2_i}
\newcommand{\ai}{a_i}
\newcommand{\ais}{a^2_i}
%
% Self-energies
%==============
%
\newcommand{\piv}{\Pi_{_V}}
\newcommand{\pia}{\Pi_{_A}}
\newcommand{\piva}{\Pi_{_{V,A}}}
\newcommand{\pivi}[1]{\Pi^{({#1})}_{_V}}
\newcommand{\piai}[1]{\Pi^{({#1})}_{_A}}
\newcommand{\pivai}[1]{\Pi^{({#1})}_{_{V,A}}}
\newcommand{\pih}{{\hat\Pi}}
\newcommand{\sgh}{{\hat\Sigma}}
\newcommand{\Pgg}{\Pi_{\ph\ph}}
\newcommand{\Ptg}{\Pi_{_{3Q}}}
\newcommand{\Ptt}{\Pi_{_{33}}}
\newcommand{\Pzg}{\Pi_{_{\zb\ab}}}
\newcommand{\Pzga}[2]{\Pi^{#1}_{_{\zb\ab}}\lpar#2\rpar}
\newcommand{\Pf}{\Pi_{_F}}
\newcommand{\Sgg}{\Sigma_{_{\ab\ab}}}
\newcommand{\Szg}{\Sigma_{_{\zb\ab}}}
\newcommand{\SVV}{\Sigma_{_{\vb\vb}}}
\newcommand{\USvv}{{\hat\Sigma}_{_{\vb\vb}}}
\newcommand{\Sww}{\Sigma_{_{\wb\wb}}}
\newcommand{\Swwg}{\Sigma^{_G}_{_{\wb\wb}}}
\newcommand{\Szz}{\Sigma_{_{\zb\zb}}}
\newcommand{\Shh}{\Sigma_{_{\hb\hb}}}
\newcommand{\Spzz}{\Sigma'_{_{\zb\zb}}}
\newcommand{\Stg}{\Sigma_{_{3Q}}}
\newcommand{\Stt}{\Sigma_{_{33}}}
\newcommand{\bSww}{{\overline\Sigma}_{_{WW}}}
\newcommand{\bStg}{{\overline\Sigma}_{_{3Q}}}
\newcommand{\bStt}{{\overline\Sigma}_{_{33}}}
\newcommand{\Sssn}{\Sigma_{_{\hkn\hkn}}}
\newcommand{\Sssc}{\Sigma_{_{\phi\phi}}}
\newcommand{\Szn}{\Sigma_{_{\zb\hkn}}}
\newcommand{\Swc}{\Sigma_{_{\wb\hkg}}}
\newcommand{\mix}[2]{{\cal{M}}^{#1}\lpar{#2}\rpar}
\newcommand{\bmix}[2]{\Pi^{{#1},F}_{_{\zb\ab}}\lpar{#2}\rpar}
\newcommand{\hPgg}[2]{{\hat{\Pi}^{{#1},F}}_{_{\ph\ph}}\lpar{#2}\rpar}
\newcommand{\hmix}[2]{{\hat{\Pi}^{{#1},F}}_{_{\zb\ab}}\lpar{#2}\rpar}
\newcommand{\Dz}[2]{{\cal{D}}_{_{\zb}}^{#1}\lpar{#2}\rpar}
\newcommand{\bDz}[2]{{\cal{D}}^{{#1},F}_{_{\zb}}\lpar{#2}\rpar}
\newcommand{\hDz}[2]{{\hat{\cal{D}}}^{{#1},F}_{_{\zb}}\lpar{#2}\rpar}
\newcommand{\Szzd}[2]{\Sigma'^{#1}_{_{\zb\zb}}\lpar{#2}\rpar}
\newcommand{\Swwd}[2]{\Sigma'^{#1}_{_{\wb\wb}}\lpar{#2}\rpar}
\newcommand{\Shhd}[2]{\Sigma'^{#1}_{_{\hb\hb}}\lpar{#2}\rpar}
\newcommand{\ZFren}[2]{{\cal{Z}}^{#1}\lpar{#2}\rpar}
\newcommand{\WFren}[2]{{\cal{W}}^{#1}\lpar{#2}\rpar}
\newcommand{\HFren}[2]{{\cal{H}}^{#1}\lpar{#2}\rpar}
\newcommand{\WI}{\cal{W}}
%
% QCD varia
%==========
%
\newcommand{\cf}{c_f}
\newcommand{\Cf}{C_{_F}}
\newcommand{\Nf}{N_f}
\newcommand{\Nc}{N_c}
\newcommand{\Ncs}{N^2_c}
\newcommand{\nf }{n_f}
\newcommand{\nfs}{n^2_f}
\newcommand{\nfc}{n^3_f}
\newcommand{\MSB}{\overline{MS}}
\newcommand{\LMSB}{\Lambda_{\overline{\mathrm{MS}}}}
\newcommand{\LMSBp}{\Lambda'_{\overline{\mathrm{MS}}}}
\newcommand{\LMSBS}{\Lambda^2_{\overline{\mathrm{MS}}}}
\newcommand{\LMSBv }{\mbox{$\Lambda^{(5)}_{\overline{\mathrm{MS}}}$}}
\newcommand{\LMSBvS}{\mbox{$\left(\Lambda^{(5)}_{\overline{\mathrm{MS}}}\right)^2$}}
\newcommand{\LMSBt }{\mbox{$\Lambda^{(3)}_{\overline{\mathrm{MS}}}$}}
\newcommand{\LMSBtS}{\mbox{$\left(\Lambda^{(3)}_{\overline{\mathrm{MS}}}\right)^2$}}
\newcommand{\LMSBf }{\mbox{$\Lambda^{(4)}_{\overline{\mathrm{MS}}}$}}
\newcommand{\LMSBfS}{\mbox{$\left(\Lambda^{(4)}_{\overline{\mathrm{MS}}}\right)^2$}}
\newcommand{\LMSBn }{\mbox{$\Lambda^{(\nf)}_{\overline{\mathrm{MS}}}$}}
\newcommand{\LMSBnS}{\mbox{$\left(\Lambda^{(\nf)}_{\overline{\mathrm{MS}}}\right)^2$}}
\newcommand{\LMSBnml }{\mbox{$\Lambda^{(\nf-1)}_{\overline{\mathrm{MS}}}$}}
\newcommand{\LMSBnmlS}{\mbox{$\left(\Lambda^{(\nf-1)}_{\overline{\mathrm{MS}}}\right)^2$}}
\newcommand{\Bnf}{\lpar\nf \rpar}
\newcommand{\Bnfm}{\lpar\nf-1 \rpar}
\newcommand{\LuM}{L_{_M}}
\newcommand{\bef}{\beta_{\ff}}
\newcommand{\befs}{\beta^2_{\ff}}
\newcommand{\befc}{\beta^3_{f}}
\newcommand{\alsp}{\alpha'_{_S}}
\newcommand{\api}{\displaystyle \frac{\als(s)}{\pi}}
\newcommand{\alss}{\alpha^2_{_S}}
\newcommand{\ztwo}{\zeta(2)}
\newcommand{\ztri}{\zeta(3)}
\newcommand{\zfor}{\zeta(4)}
\newcommand{\zfiv}{\zeta(5)}
\newcommand{\bi}[1]{b_{#1}}
\newcommand{\ci}[1]{c_{#1}}
\newcommand{\Ci}[1]{C_{#1}}
\newcommand{\bip}[1]{b'_{#1}}
\newcommand{\cip}[1]{c'_{#1}}
%
% Numerical factors
%==================
%
\newcommand{\osps}{16\,\pi^2}
\newcommand{\srt}{\sqrt{2}}
\newcommand{\ospsi}{\displaystyle{\frac{i}{16\,\pi^2}}}
%
% 2f processes
%=============
%
\newcommand{\tfpromu}{\mbox{$e^+e^-\to \mu^+\mu^-$}}
\newcommand{\tfprotau}{\mbox{$e^+e^-\to \tau^+\tau^-$}}
\newcommand{\tfproe}{\mbox{$e^+e^-\to e^+e^-$}}
\newcommand{\tfpronu}{\mbox{$e^+e^-\to \barnu\nu$}}
\newcommand{\tfproqq}{\mbox{$e^+e^-\to \barq q$}}
\newcommand{\tfprohad}{\mbox{$e^+e^-\to\,$} hadrons}
%
% brems. processes
%-----------------
%
\newcommand{\bpromu}{\mbox{$e^+e^-\to \mu^+\mu^-\ph$}}
\newcommand{\bprotau}{\mbox{$e^+e^-\to \tau^+\tau^-\ph$}}
\newcommand{\bproe}{\mbox{$e^+e^-\to e^+e^-\ph$}}
\newcommand{\bpronu}{\mbox{$e^+e^-\to \barnu\nu\ph$}}
\newcommand{\bproqq}{\mbox{$e^+e^-\to \barq q \ph$}}
%
% 2b processes
%-------------
%
\newcommand{\tbprow} {\mbox{$e^+e^-\to \wbp \wbm $}}
\newcommand{\tbproz} {\mbox{$e^+e^-\to \zb  \zb  $}}
\newcommand{\tbproh} {\mbox{$e^+e^-\to \zb  \hb  $}}
\newcommand{\tbprozg}{\mbox{$e^+e^-\to \zb  \ph  $}}
\newcommand{\tbprog} {\mbox{$e^+e^-\to \ph  \ph  $}}
%
% Line-style for propagators
%===========================
%
\newcommand{\Fermionline}[1]{
\vcenter{\hbox{
  \begin{picture}(60,20)(0,{#1})
  \SetScale{2.}
    \ArrowLine(0,5)(30,5)
  \end{picture}}}
}
\newcommand{\AntiFermionline}[1]{
\vcenter{\hbox{
  \begin{picture}(60,20)(0,{#1})
  \SetScale{2.}
    \ArrowLine(30,5)(0,5)
  \end{picture}}}
}
\newcommand{\Photonline}[1]{
\vcenter{\hbox{
  \begin{picture}(60,20)(0,{#1})
  \SetScale{2.}
    \Photon(0,5)(30,5){2}{6.5}
  \end{picture}}}
}
\newcommand{\Gluonline}[1]{
\vcenter{\hbox{
  \begin{picture}(60,20)(0,{#1})
  \SetScale{2.}
    \Gluon(0,5)(30,5){2}{6.5}
  \end{picture}}}
}
\newcommand{\Wbosline}[1]{
\vcenter{\hbox{
  \begin{picture}(60,20)(0,{#1})
  \SetScale{2.}
    \Photon(0,5)(30,5){2}{4}
    \ArrowLine(13.3,3.1)(16.9,7.2)
  \end{picture}}}
}
\newcommand{\Zbosline}[1]{
\vcenter{\hbox{
  \begin{picture}(60,20)(0,{#1})
  \SetScale{2.}
    \Photon(0,5)(30,5){2}{4}
  \end{picture}}}
}
\newcommand{\Philine}[1]{
\vcenter{\hbox{
  \begin{picture}(60,20)(0,{#1})
  \SetScale{2.}
    \DashLine(0,5)(30,5){2}
  \end{picture}}}
}
\newcommand{\Phicline}[1]{
\vcenter{\hbox{
  \begin{picture}(60,20)(0,{#1})
  \SetScale{2.}
    \DashLine(0,5)(30,5){2}
    \ArrowLine(14,5)(16,5)
  \end{picture}}}
}
\newcommand{\Ghostline}[1]{
\vcenter{\hbox{
  \begin{picture}(60,20)(0,{#1})
  \SetScale{2.}
    \DashLine(0,5)(30,5){.5}
    \ArrowLine(14,5)(16,5)
  \end{picture}}}
}
%
% SM Lagrangian in \Rxi
%======================
%
\newcommand{\gauge}{g}
\newcommand{\gpar}{\xi}
\newcommand{\gparA}{\xi_{_A}}
\newcommand{\gparZ}{\xi_{_Z}}
\newcommand{\gpari}[1]{\gpar_{#1}}
\newcommand{\gparis}[1]{\gpar^2_{#1}}
\newcommand{\gpariq}[1]{\gpar^4_{#1}}
\newcommand{\gpars}{\xi^2}
\newcommand{\dgpar}{\Delta\gpar}
\newcommand{\dgparA}{\Delta\gparA}
\newcommand{\dgparZ}{\Delta\gparZ}
\newcommand{\gparq}{\xi^4}
\newcommand{\gparAs}{\xi^2_{_A}}
\newcommand{\gparAq}{\xi^4_{_A}}
\newcommand{\gparZs}{\xi^2_{_Z}}
\newcommand{\gparZq}{\xi^4_{_Z}}
\newcommand{\Rxi}{R_{\gpar}}
\newcommand{\hxi}{\chi}
%
% Lagrangiains
%-------------
%
\newcommand{\LSM}{{\cal{L}}_{_{\rm{SM}}}}
\newcommand{\LSMr}{{\cal{L}}^{\rm{ren}}_{_{\rm{SM}}}}
\newcommand{\LYM}{{\cal{L}}_{_{YM}}}
\newcommand{\Lzer}{{\cal{L}}_{_{0}}}
\newcommand{\Lone}{{\cal{L}}^{{\bos},I}}
\newcommand{\Lpro}{{\cal{L}}_{\rm{prop}}}
\newcommand{\Ls  }{{\cal{L}}_{_{S}}}
\newcommand{\Lsi }{{\cal{L}}^{I}_{_{S}}}
\newcommand{\Lgf }{{\cal{L}}_{gf  }}
\newcommand{\Lgfi}{{\cal{L}}^{I}_{gf}}
\newcommand{\Lf  }{{\cal{L}}^{{\fer},I}_{\ssV}}
\newcommand{\LHf }{{\cal{L}}^{\fer}_{\ssS}}
\newcommand{\LHfi}{{\cal{L}}^{{\fer},I}_{\ssS}}
\newcommand{\Lren}{{\cal{L}}_{\rm{ren}}}
\newcommand{\Lct}{{\cal{L}}_{\rm{ct}}}
\newcommand{\Lcti}[1]{{\cal{L}}^{#1}_{\rm{ct}}}
\newcommand{\LctI}{{\cal{L}}^{(2)}_{\rm{ct}}}
\newcommand{\Llone}{{\cal{L}}}
\newcommand{\LQED}{{\cal{L}}_{_{\rm{QED}}}}
\newcommand{\LQEDr}{{\cal{L}}^{\rm{ren}}_{_{\rm{QED}}}}
\newcommand{\FST}[3]{F_{#1#2}^{#3}}
\newcommand{\cD}[1]{D_{#1}}
\newcommand{\pd}[1]{\partial_{#1}}
\newcommand{\tgen}[1]{\tau^{#1}}
\newcommand{\gbl}{g_1}
\newcommand{\lctt}[3]{\varepsilon_{#1#2#3}}
\newcommand{\lctf}[4]{\varepsilon_{#1#2#3#4}}
\newcommand{\lctfb}[4]{\varepsilon\lpar{#1#2#3#4}\rpar}
\newcommand{\slct}{\varepsilon}
%gauge fixing
\newcommand{\cgfi}[1]{{\cal{C}}^{#1}}
\newcommand{\cgfZ}{{\cal{C}}_{_Z}}
\newcommand{\cgfA}{{\cal{C}}_{_A}}
\newcommand{\cgfZs}{{\cal{C}}^2_{_Z}}
\newcommand{\cgfAs}{{\cal{C}}^2_{_A}}
%parameters of scalar potential
\newcommand{\hpms}{\mu^2}
\newcommand{\hpal}{\alpha_{_H}}
\newcommand{\hpals}{\alpha^2_{_H}}
\newcommand{\hpbe}{\beta_{_H}}
\newcommand{\hpbep}{\beta^{'}_{_H}}
\newcommand{\hpla}{\lambda}
\newcommand{\hpalf}{\alpha_{f}}
\newcommand{\hpbef}{\beta_{f}}
%transformation parameters
\newcommand{\tpar}[1]{\Lambda^{#1}}
%M,L-operators
\newcommand{\Mop}[2]{{\rm{M}}^{#1#2}}
\newcommand{\Lop}[2]{{\rm{L}}^{#1#2}}
\newcommand{\Lgen}[1]{T^{#1}}
\newcommand{\Rgen}[1]{t^{#1}}
\newcommand{\fpari}[1]{\lambda_{#1}}
\newcommand{\fQ}[1]{Q_{#1}}
\newcommand{\unm}{I}
\newcommand{\cDsla}{/\!\!\!\!D}
%
% A-B-C-D functions
%==================
%
\newcommand{\saff}[1]{A_{#1}}                    % A form-factors
\newcommand{\aff}[2]{A_{#1}\lpar #2\rpar}                   
\newcommand{\sbff}[1]{B_{#1}}                    % B form-factors
\newcommand{\sfbff}[1]{B^{F}_{#1}}
\newcommand{\bff}[4]{B_{#1}\lpar #2;#3,#4\rpar}             
\newcommand{\bfft}[3]{B_{#1}\lpar #2,#3\rpar}             
\newcommand{\fbff}[4]{B^{F}_{#1}\lpar #2;#3,#4\rpar}        
\newcommand{\cdbff}[4]{\Delta B_{#1}\lpar #2;#3,#4\rpar}             
\newcommand{\sdbff}[4]{\delta B_{#1}\lpar #2;#3,#4\rpar}             
\newcommand{\cdbfft}[3]{\Delta B_{#1}\lpar #2,#3\rpar}             
\newcommand{\sdbfft}[3]{\delta B_{#1}\lpar #2,#3\rpar}             
\newcommand{\scff}[1]{C_{#1}}                    % C form-factors
\newcommand{\scffo}[2]{C_{#1}\lpar{#2}\rpar}                
\newcommand{\cff}[7]{C_{#1}\lpar #2,#3,#4;#5,#6,#7\rpar}    
\newcommand{\sccff}[5]{c_{#1}\lpar #2;#3,#4,#5\rpar} 
\newcommand{\sdff}[1]{D_{#1}}                    % D form-factors
\newcommand{\dffp}[7]{D_{#1}\lpar #2,#3,#4,#5,#6,#7;}       
\newcommand{\dffm}[4]{#1,#2,#3,#4\rpar}                     
\newcommand{\bzfa}[2]{B^{F}_{_{#2}}\lpar{#1}\rpar}
\newcommand{\bzfaa}[3]{B^{F}_{_{#2#3}}\lpar{#1}\rpar}
\newcommand{\shcff}[4]{C_{_{#2#3#4}}\lpar{#1}\rpar}
\newcommand{\shdff}[6]{D_{_{#3#4#5#6}}\lpar{#1,#2}\rpar}
\newcommand{\scdff}[3]{d_{#1}\lpar #2,#3\rpar} 
\newcommand{\scaldff}[1]{{\cal{D}}^{#1}}
\newcommand{\caldff}[2]{{\cal{D}}^{#1}\lpar{#2}\rpar}
\newcommand{\caldfft}[3]{{\cal{D}}_{#1}^{#2}\lpar{#3}\rpar}
%
% a-b-c-d functions
%------------------
%
\newcommand{\slaff}[1]{a_{#1}}                        
\newcommand{\slbff}[1]{b_{#1}}                        
\newcommand{\slbffh}[1]{{\hat{b}}_{#1}}    
\newcommand{\ssldff}[1]{d_{#1}}                        
\newcommand{\sslcff}[1]{c_{#1}}                        
\newcommand{\slcff}[2]{c_{#1}^{(#2)}}                        
\newcommand{\sldff}[2]{d_{#1}^{(#2)}}                        
\newcommand{\lbff}[3]{b_{#1}\lpar #2;#3\rpar}         
\newcommand{\lbffh}[2]{{\hat{b}}_{#1}\lpar #2\rpar}   
\newcommand{\lcff}[8]{c_{#1}^{(#2)}\lpar  #3,#4,#5;#6,#7,#8\rpar}         
\newcommand{\ldffp}[8]{d_{#1}^{(#2)}\lpar #3,#4,#5,#6,#7,#8;}
\newcommand{\ldffm}[4]{#1,#2,#3,#4\rpar}                   
%
% I-J functions
%--------------
%
\newcommand{\Iff}[4]{I_{#1}\lpar #2;#3,#4 \rpar}
\newcommand{\Jff}[4]{J_{#1}\lpar #2;#3,#4 \rpar}
\newcommand{\Jds}[5]{{\bar{J}}_{#1}\lpar #2,#3;#4,#5 \rpar}
%--
% n-dimension and epsilons
%=========================
%
\newcommand{\nhmt}{\frac{n}{2}-2}
\newcommand{\nhmts}{{n}/{2}-2}
\newcommand{\omnh}{1-\frac{n}{2}}
\newcommand{\nhmo}{\frac{n}{2}-1}
\newcommand{\fmon}{4-n}
\newcommand{\lpi}{\ln\pi}
\newcommand{\lmass}[1]{\ln #1}
\newcommand{\egnh}{\egam{\frac{n}{2}}}
\newcommand{\egomnh}{\egam{1-\frac{n}{2}}}
\newcommand{\egtmnh}{\egam{2-\frac{n}{2}}}
\newcommand{\Ddr}{{\ds\frac{1}{{\bar{\varepsilon}}}}}
\newcommand{\Ddrs}{{\ds\frac{1}{{\bar{\varepsilon}^2}}}}
\newcommand{\Ddrd}{{\bar{\varepsilon}}}
\newcommand{\ept}{\hat\varepsilon}
\newcommand{\Ddrh}{{\ds\frac{1}{\hat{\varepsilon}}}}
\newcommand{\Ddrp}{{\ds\frac{1}{\varepsilon'}}}
\newcommand{\Ddrps}{\lpar{\ds{\frac{1}{\varepsilon'}}}\rpar^2}
\newcommand{\dre}{\varepsilon}
\newcommand{\drei}[1]{\varepsilon_{#1}}
\newcommand{\epp}{\varepsilon'}
\newcommand{\eps}{\varepsilon^*}
%--
%-- Im for masses and propagators
%================================
\newcommand{\ep}{\epsilon}
%--
\newcommand{\propbt}[6]{{{#1_{#2}#1_{#3}}\over{\lpar #1^2 + #4 
-\ib\ep\rpar\lpar\lpar #5\rpar^2 + #6 -\ib\ep\rpar}}}
\newcommand{\propbo}[5]{{{#1_{#2}}\over{\lpar #1^2 + #3 - \ib\ep\rpar
\lpar\lpar #4\rpar^2 + #5 -\ib\ep\rpar}}}
\newcommand{\propc}[6]{{1\over{\lpar #1^2 + #2 - \ib\ep\rpar
\lpar\lpar #3\rpar^2 + #4 -\ib\ep\rpar
\lpar\lpar #5\rpar^2 + #6 -\ib\ep\rpar}}}
%--
\newcommand{\propa}[2]{{1\over {#1^2 + #2^2 - \ib\ep}}}
\newcommand{\propb}[4]{{1\over {\lpar #1^2 + #2 - \ib\ep\rpar
\lpar\lpar #3\rpar^2 + #4 -\ib\ep\rpar}}}
\newcommand{\propbs}[4]{{1\over {\lpar\lpar #1\rpar^2 + #2 - \ib\ep\rpar
\lpar\lpar #3\rpar^2 + #4 -\ib\ep\rpar}}}
\newcommand{\propat}[4]{{#3_{#1}#3_{#2}\over {#3^2 + #4^2 - \ib\ep}}}
\newcommand{\propaf}[6]{{#5_{#1}#5_{#2}#5_{#3}#5_{#4}\over 
{#5^2 + #6^2 -\ib\ep}}}
\newcommand{\momeps}[1]{#1^2 - \ib\ep}
\newcommand{\mopeps}[1]{#1^2 + \ib\ep}
%--
\newcommand{\propz}[1]{{1\over{#1^2 + \mzs - \ib\ep}}}
\newcommand{\propw}[1]{{1\over{#1^2 + \mws - \ib\ep}}}
\newcommand{\proph}[1]{{1\over{#1^2 + \mhs - \ib\ep}}}
\newcommand{\propf}[2]{{1\over{#1^2 + #2}}}
\newcommand{\propzrg}[3]{{{\delta_{#1#2}}\over{{#3}^2 + \mzs - \ib\ep}}}
\newcommand{\propwrg}[3]{{{\delta_{#1#2}}\over{{#3}^2 + \mws - \ib\ep}}}
\newcommand{\propzug}[3]{{
      {\delta_{#1#2} + \displaystyle{{{#3}^{#1}{#3}^{#2}}\over{\mzs}}}
                         \over{{#3}^2 + \mzs - \ib\ep}}}
\newcommand{\propwug}[3]{{
      {\delta_{#1#2} + \displaystyle{{{#3}^{#1}{#3}^{#2}}\over{\mws}}}
                        \over{{#3}^2 + \mws - \ib\ep}}}
%----------------------------------------
\newcommand{\thf}[1]{\theta\lpar #1\rpar}
\newcommand{\epf}[1]{\varepsilon\lpar #1\rpar}
\newcommand{\singp}{\stackrel{sing}{\rightarrow}}
\newcommand{\aint}[3]{\int_{#1}^{#2}\,d #3}
\newcommand{\aroot}[1]{\sqrt{#1}}
\newcommand{\gramc}{\Delta_3}
\newcommand{\gramd}{\Delta_4}
\newcommand{\pinch}[2]{P^{(#1)}\lpar #2\rpar}
\newcommand{\pinchc}[2]{C^{(#1)}_{#2}}
\newcommand{\pinchd}[2]{D^{(#1)}_{#2}}
\newcommand{\loarg}[1]{\ln\lpar #1\rpar}
\newcommand{\loargr}[1]{\ln\lrbr #1\rrbr}
\newcommand{\lsoarg}[1]{\ln^2\lpar #1\rpar}
\newcommand{\ltarg}[2]{\ln\lpar #1\rpar\lpar #2\rpar}
\newcommand{\rfun}[2]{R\lpar #1,#2\rpar}
\newcommand{\pinchb}[3]{B_{#1}\lpar #2,#3\rpar}
\newcommand{\lga}{\ph}
\newcommand{\lzga}{\ssZ\ph}
%
% Auxiliary functions
%
\newcommand{\afa}[5]{A_{#1}^{#2}\lpar #3;#4,#5\rpar}
\newcommand{\bfa}[5]{B_{#1}^{#2}\lpar #3;#4,#5\rpar} 
\newcommand{\hfa}[5]{H_{#1}^{#2}\lpar #3;#4,#5\rpar}
\newcommand{\rfa}[5]{R_{#1}^{#2}\lpar #3;#4,#5\rpar}
\newcommand{\afao}[3]{A_{#1}^{#2}\lpar #3\rpar}
\newcommand{\bfao}[3]{B_{#1}^{#2}\lpar #3\rpar}
\newcommand{\hfao}[3]{H_{#1}^{#2}\lpar #3\rpar}
\newcommand{\rfao}[3]{R_{#1}^{#2}\lpar #3\rpar}
\newcommand{\afas}[2]{A_{#1}^{#2}}
\newcommand{\bfas}[2]{B_{#1}^{#2}}
\newcommand{\hfas}[2]{H_{#1}^{#2}}
\newcommand{\rfas}[2]{R_{#1}^{#2}}
\newcommand{\tfas}[2]{T_{#1}^{#2}}
\newcommand{\afaR}[6]{A_{#1}^{\gpar}\lpar #2;#3,#4,#5,#6 \rpar}
\newcommand{\bfaR}[6]{B_{#1}^{\gpar}\lpar #2;#3,#4,#5,#6 \rpar}
\newcommand{\hfaR}[6]{H_{#1}^{\gpar}\lpar #2;#3,#4,#5,#6 \rpar}
\newcommand{\shfaR}[1]{H_{#1}^{\gpar}}
\newcommand{\rfaR}[6]{R_{#1}^{\gpar}\lpar #2;#3,#4,#5,#6 \rpar}
\newcommand{\srfaR}[1]{R_{#1}^{\gpar}}
\newcommand{\afaRg}[5]{A_{#1 \lga}^{\gpar}\lpar #2;#3,#4,#5 \rpar}
\newcommand{\bfaRg}[5]{B_{#1 \lga}^{\gpar}\lpar #2;#3,#4,#5 \rpar}
\newcommand{\hfaRg}[5]{H_{#1 \lga}^{\gpar}\lpar #2;#3,#4,#5 \rpar}
\newcommand{\shfaRg}[1]{H_{#1\lga}^{\gpar}}
\newcommand{\rfaRg}[5]{R_{#1 \lga}^{\gpar}\lpar #2;#3,#4,#5 \rpar}
\newcommand{\srfaRg}[1]{R_{#1\lga}^{\gpar}}
\newcommand{\afaRt}[3]{A_{#1}^{\gpar}\lpar #2,#3 \rpar}
\newcommand{\hfaRt}[3]{H_{#1}^{\gpar}\lpar #2,#3 \rpar}
\newcommand{\hfaRf}[4]{H_{#1}^{\gpar}\lpar #2,#3,#4 \rpar}
\newcommand{\afasm}[4]{A_{#1}^{\lpar #2,#3,#4 \rpar}}
\newcommand{\bfasm}[4]{B_{#1}^{\lpar #2,#3,#4 \rpar}}
\newcommand{\color}[1]{c_{#1}}
\newcommand{\htf}[2]{H_2\lpar #1,#2\rpar}
\newcommand{\rof}[2]{R_1\lpar #1,#2\rpar}
\newcommand{\rtf}[2]{R_3\lpar #1,#2\rpar}
\newcommand{\rtrans}[2]{R_{#1}^{#2}}
\newcommand{\momf}[2]{#1^2_{#2}}
\newcommand{\Scalvert}[8][70]{
  \vcenter{\hbox{
  \SetScale{0.8}
  \begin{picture}(#1,50)(15,15)
    \Line(25,25)(50,50)      \Text(34,20)[lc]{#6} \Text(11,20)[lc]{#3}
    \Line(50,50)(25,75)      \Text(34,60)[lc]{#7} \Text(11,60)[lc]{#4}
    \Line(50,50)(90,50)      \Text(11,40)[lc]{#2} \Text(55,33)[lc]{#8}
    \GCirc(50,50){10}{1}          \Text(60,48)[lc]{#5} 
  \end{picture}}}
  }
%
% db-s additions, beware, I've modified above also
%
\newcommand{\tHs}{\mu}
\newcommand{\tHsz}{\mu_{_0}}
\newcommand{\tHss}{\mu^2}
\newcommand{\Reb}{{\rm{Re}}}
\newcommand{\Imb}{{\rm{Im}}}
%
% gp's additions 
%
\newcommand{\spd}{\partial}
\newcommand{\fun}[1]{f\lpar{#1}\rpar}
\newcommand{\ffun}[2]{F_{#1}\lpar #2\rpar}
\newcommand{\gfun}[2]{G_{#1}\lpar #2\rpar}
\newcommand{\sffun}[1]{F_{#1}}
\newcommand{\csffun}[1]{{\cal{F}}_{#1}}
\newcommand{\sgfun}[1]{G_{#1}}
\newcommand{\tpfi}{\lpar 2\pi\rpar^4\ib}
\newcommand{\ffv}{F_{_V}}
\newcommand{\fga}{G_{_A}}
\newcommand{\ffm}{F_{_M}}
\newcommand{\ffs}{F_{_S}}
\newcommand{\fgp}{G_{_P}}
\newcommand{\fge}{G_{_E}}
\newcommand{\ffa}{F_{_A}}
\newcommand{\ffps}{F_{_P}}
\newcommand{\ffe}{F_{_E}}
\newcommand{\gacom}[2]{\lpar #1 + #2\gfd\rpar}
\newcommand{\mft}{m_{\tilde f}}
\newcommand{\qft}{Q_{f'}}
\newcommand{\vft}{v_{\tilde f}}
\newcommand{\subb}[2]{b_{#1}\lpar #2 \rpar}
\newcommand{\fwfr}[5]{\Sigma\lpar #1,#2,#3;#4,#5 \rpar}
\newcommand{\slim}[2]{\lim_{#1 \to #2}}
\newcommand{\sprop}[3]{
{#1\over {\lpar q^2\rpar^2\lpar \lpar q+ #2\rpar^2+#3^2\rpar }}}
%
% roots, variables, coefficients
%
\newcommand{\xroot}[1]{x_{#1}}
\newcommand{\yroot}[1]{y_{#1}}
\newcommand{\zroot}[1]{z_{#1}}
\newcommand{\lvar}{l}
\newcommand{\rvar}{r}
\newcommand{\tvar}{t}
\newcommand{\uvar}{u}
\newcommand{\vvar}{v}
\newcommand{\xvar}{x}
\newcommand{\yvar}{y}
\newcommand{\zvar}{z}
\newcommand{\yvarp}{y'}
\newcommand{\rvars}{r^2}
\newcommand{\vvars}{v^2}
\newcommand{\xvars}{x^2}
\newcommand{\yvars}{y^2}
\newcommand{\zvars}{z^2}
\newcommand{\rvarc}{r^3}
\newcommand{\xvarc}{x^3}
\newcommand{\yvarc}{y^3}
\newcommand{\zvarc}{z^3}
\newcommand{\rvarq}{r^4}
\newcommand{\xvarq}{x^4}
\newcommand{\yvarq}{y^4}
\newcommand{\zvarq}{z^4}
\newcommand{\avar}{a}
\newcommand{\avars}{a^2}
\newcommand{\avarc}{a^3}
\newcommand{\avari}[1]{a_{#1}}
\newcommand{\avart}[2]{a_{#1}^{#2}}
\newcommand{\delvari}[1]{\delta_{#1}}
\newcommand{\rvari}[1]{r_{#1}}
\newcommand{\xvari}[1]{x_{#1}}
\newcommand{\yvari}[1]{y_{#1}}
\newcommand{\zvari}[1]{z_{#1}}
\newcommand{\rvart}[2]{r_{#1}^{#2}}
\newcommand{\xvart}[2]{x_{#1}^{#2}}
\newcommand{\yvart}[2]{y_{#1}^{#2}}
\newcommand{\zvart}[2]{z_{#1}^{#2}}
\newcommand{\rvaris}[1]{r^2_{#1}}
\newcommand{\xvaris}[1]{x^2_{#1}}
\newcommand{\yvaris}[1]{y^2_{#1}}
\newcommand{\zvaris}[1]{z^2_{#1}}
\newcommand{\Xvar}{X}
\newcommand{\Xvars}{X^2}
\newcommand{\Xvari}[1]{X_{#1}}
\newcommand{\Xvaris}[1]{X^2_{#1}}
\newcommand{\Yvar}{Y}
\newcommand{\Yvars}{Y^2}
\newcommand{\Yvari}[1]{Y_{#1}}
\newcommand{\Yvaris}[1]{Y^2_{#1}}
%---
\newcommand{\lnx}{\ln\xvar}
\newcommand{\lnz}{\ln\zvar}
\newcommand{\lnsx}{\ln^2\xvar}
\newcommand{\lnsz}{\ln^2\zvar}
\newcommand{\lncz}{\ln^3\zvar}
\newcommand{\lnomz}{\ln\lpar 1-\zvar\rpar}
\newcommand{\lnsomz}{\ln^2\lpar 1-\zvar\rpar}
\newcommand{\ccoefi}[1]{c_{#1}}
\newcommand{\ccoeft}[2]{c^{#1}_{#2}}
%
% Matrices
%
\newcommand{\Smat}{{\cal{S}}}
\newcommand{\Mmat}{{\cal{M}}}
\newcommand{\Xmat}[1]{X_{#1}}
\newcommand{\XmatI}[1]{X^{-1}_{#1}}
\newcommand{\unitmat}{I}
\newcommand{\Kmat}{{C}}
\newcommand{\Kmatc}{{C}^{\dagger}}
\newcommand{\Kmati}[1]{{C}_{#1}}
\newcommand{\Kmatci}[1]{{C}^{\dagger}_{#1}}
\newcommand{\ffac}[2]{f_{#1}^{#2}}
\newcommand{\Ffac}[1]{F_{#1}}
\newcommand{\Rvec}[2]{R^{(#1)}_{#2}}
\newcommand{\momfl}[2]{#1_{#2}}
\newcommand{\momfs}[2]{#1^2_{#2}}
\newcommand{\fpseZ}{A^{^{FP,Z}}}
\newcommand{\fpseA}{A^{^{FP,A}}}
\newcommand{\fptZ}{T^{^{FP,Z}}}
\newcommand{\fptA}{T^{^{FP,A}}}
\newcommand{\dprop}{\overline\Delta}
\newcommand{\dpropi}[1]{d_{#1}}
\newcommand{\dpropic}[1]{d^{c}_{#1}}
\newcommand{\dpropii}[2]{d_{#1}\lpar #2\rpar}
\newcommand{\dpropis}[1]{d^2_{#1}}
\newcommand{\dproppi}[1]{d'_{#1}}
\newcommand{\psf}[4]{P\lpar #1;#2,#3,#4\rpar}
\newcommand{\ssf}[5]{S^{(#1)}\lpar #2;#3,#4,#5\rpar}
\newcommand{\csf}[5]{C_{_S}^{(#1)}\lpar #2;#3,#4,#5\rpar}
%
% polarization vectors
%=====================
%
\newcommand{\lvec}{l}
\newcommand{\lvecs}{l^2}
\newcommand{\lveci}[1]{l_{#1}}
\newcommand{\mvec}{m}
\newcommand{\mvecs}{m^2}
\newcommand{\mveci}[1]{m_{#1}}
\newcommand{\nvec}{n}
\newcommand{\nvecs}{n^2}
\newcommand{\nveci}[1]{n_{#1}}
\newcommand{\epi}[1]{\epsilon_{#1}}
\newcommand{\phep}[1]{\ep_{#1}}
\newcommand{\sphep}{\ep}
\newcommand{\vbep}[1]{e_{#1}}
\newcommand{\vbepp}[1]{e^{+}_{#1}}
\newcommand{\vbepm}[1]{e^{-}_{#1}}
\newcommand{\svbep}{e}
%
% longitudinal polarizations
%===========================
%
\newcommand{\lpol}{\lambda}
\newcommand{\spol}{\sigma}
\newcommand{\rpol}{\rho  }
\newcommand{\kpol}{\kappa}
\newcommand{\lpols}{\lambda^2}
\newcommand{\spols}{\sigma^2}
\newcommand{\rpols}{\rho^2}
\newcommand{\kpols}{\kappa^2}
\newcommand{\lpoli}[1]{\lambda_{#1}}
\newcommand{\spoli}[1]{\sigma_{#1}}
\newcommand{\rpoli}[1]{\rho_{#1}}
\newcommand{\kpoli}[1]{\kappa_{#1}}
%
% some vectors
%=============
%
\newcommand{\uvec}{u}
\newcommand{\uveci}[1]{u_{#1}}
%
% Momenta:
%=========
%
\newcommand{\imom}{q}
\newcommand{\imomi}[1]{q_{#1}}
\newcommand{\imoms}{q^2}
\newcommand{\pmom}{p}
\newcommand{\pmomp}{p'}
\newcommand{\pmoms}{p^2}
\newcommand{\pmomq}{p^4}
\newcommand{\pmomx}{p^6}
\newcommand{\pmomi}[1]{p_{#1}}
\newcommand{\pmomis}[1]{p^2_{#1}}
%--
\newcommand{\Pmom}{P}
\newcommand{\Pmoms}{P^2}
\newcommand{\Pmomi}[1]{P_{#1}}
\newcommand{\Pmomis}[1]{P^2_{#1}}
\newcommand{\Kmom}{K}
\newcommand{\Kmoms}{K^2}
\newcommand{\Kmomi}[1]{K_{#1}}
\newcommand{\Kmomis}[1]{K^2_{#1}}
%--
\newcommand{\kmom}{k}
\newcommand{\kmoms}{k^2}
\newcommand{\kmomi}[1]{k_{#1}}
\newcommand{\lmom}{l}
\newcommand{\lmoms}{l^2}
\newcommand{\lmomi}[1]{l_{#1}}
\newcommand{\qmom}{q}
\newcommand{\qmoms}{q^2}
\newcommand{\qmomi}[1]{q_{#1}}
\newcommand{\qmomis}[1]{q^2_{#1}}
\newcommand{\smom}{s}
\newcommand{\smoms}{s^2}
\newcommand{\smomi}[1]{s_{#1}}
\newcommand{\tmom}{t}
\newcommand{\tmoms}{t^2}
\newcommand{\tmomi}[1]{t_{#1}}
\newcommand{\Trmom}{Q}
\newcommand{\Prmom}{P}
\newcommand{\gmv}{Q^2}
\newcommand{\Trmoms}{Q^2}
\newcommand{\Prmoms}{P^2}
\newcommand{\Ptmoms}{T^2}
\newcommand{\Pumoms}{U^2}
\newcommand{\Trmomq}{Q^4}
\newcommand{\Prmomq}{P^4}
\newcommand{\Ptmomq}{T^4}
\newcommand{\Pumomq}{U^4}
\newcommand{\Trmomx}{Q^6}
\newcommand{\Trmomi}[1]{Q_{#1}}
\newcommand{\Trmomis}[1]{Q^2_{#1}}
\newcommand{\Prmomi}[1]{P_{#1}}
\newcommand{\pone}{p_1}
\newcommand{\ptwo}{p_2}
\newcommand{\ptre}{p_3}
\newcommand{\pfor}{p_4}
\newcommand{\pones}{p_1^2}
\newcommand{\ptwos}{p_2^2}
\newcommand{\ptres}{p_3^2}
\newcommand{\pfors}{p_4^2}
\newcommand{\poneq}{p_1^4}
\newcommand{\ptwoq}{p_2^4}
\newcommand{\ptreq}{p_3^4}
\newcommand{\pforq}{p_4^4}
\newcommand{\modmom}[1]{\mid{\vec{#1}}\mid}
\newcommand{\modmomi}[2]{\mid{\vec{#1}}_{#2}\mid}
\newcommand{\vect}[1]{{\vec{#1}}}
\newcommand{\Energ}{E}
\newcommand{\Energp}{E'}
\newcommand{\Energpp}{E''}
\newcommand{\Energs}{E^2}
\newcommand{\Energc}{E^3}
\newcommand{\Energf}{E^4}
\newcommand{\Energv}{E^5}
\newcommand{\Energx}{E^6}
\newcommand{\Energi}[1]{E_{#1}}
\newcommand{\Energt}[2]{E_{#1}^{#2}}
\newcommand{\Energis}[1]{E^2_{#1}}
\newcommand{\energ}{e}
\newcommand{\energp}{e'}
\newcommand{\energpp}{e''}
\newcommand{\energs}{e^2}
\newcommand{\energi}[1]{e_{#1}}
\newcommand{\energt}[2]{e_{#1}^{#2}}
\newcommand{\energis}[1]{e^2_{#1}}
\newcommand{\wenerg}{w}
\newcommand{\wenergs}{w^2}
\newcommand{\wenergi}[1]{w_{#1}}
\newcommand{\wenergp}{w'}
\newcommand{\wenergpp}{w''}
%
% kinematical cuts
%=================
%
\newcommand{\ecut}{e}
\newcommand{\ecuts}{e^2}
\newcommand{\ecuti}[1]{e^{#1}}
\newcommand{\ccut}{c_m}
\newcommand{\ccuti}[1]{c_{#1}}
\newcommand{\ccuts}{c^2_m}
\newcommand{\scuts}{s^2_m}
\newcommand{\ccutis}[1]{c^2_{#1}}
\newcommand{\ccutic}[1]{c^3_{#1}}
\newcommand{\ccutc}{c^3_m}
\newcommand{\rcut}{\varrho}
\newcommand{\rcuts}{\varrho^2}
\newcommand{\rcuti}[1]{\varrho_{#1}}
\newcommand{\rcutu}[1]{\varrho^{#1}}
\newcommand{\Dcut}{\Delta}
%
%-----LIB_VERT_XI1.TEX--------------------------
\newcommand{\dwf}{\delta_{_{WF}}}
\newcommand{\gbar}{\overline g}
\newcommand{\PP}{\mbox{PP}}
\newcommand{\mv}{m_{_V}}
\newcommand{\bGv}{{\overline\Gamma}_{_V}}
\newcommand{\Umuv}{\hat{\mu}_\ssV}
\newcommand{\Svv}{{\Sigma}_\ssV}
\newcommand{\muv}{p_\ssV}
\newcommand{\muvb}{\mu_{\ssV_{0}}}
\newcommand{\URPvv}{{P}_\ssV}
\newcommand{\RPvv}{{P}_\ssV}
\newcommand{\Svvrem}{{\Sigma}_\ssV^{\mathrm{rem}}}
\newcommand{\USvvrem}{\hat{\Sigma}_\ssV^{\mathrm{rem}}}
\newcommand{\Gv}{\Gamma_{_V}}
%
% Renormalization 
%
\newcommand{\param}{p}
\newcommand{\parami}[1]{p^{#1}}
\newcommand{\paramb}{p_{0}}
\newcommand{\Zcon}{Z}
\newcommand{\Zconi}[1]{Z_{#1}}
\newcommand{\zconi}[1]{z_{#1}}
\newcommand{\Zconim}[1]{{Z^-_{#1}}}
\newcommand{\zconim}[1]{{z^-_{#1}}}
\newcommand{\Zcont}[2]{Z_{#1}^{#2}}
\newcommand{\zcont}[2]{z_{#1}^{#2}}
\newcommand{\zcontm}[2]{z_{#1}^{{#2}-}}
\newcommand{\sZconi}[2]{\sqrt{Z_{#1}}^{\;#2}}
\newcommand{\php}[3]{e^{#1}_{#2}\lpar #3 \rpar}
\newcommand{\gacome}[1]{\lpar #1 - \gfd\rpar}
\newcommand{\sPj}[2]{\Lambda^{#1}_{#2}}
\newcommand{\sPjs}[2]{\Lambda_{#1,#2}}
\newcommand{\amos}{\mbox{$M^2_{_1}$}}
\newcommand{\amts}{\mbox{$M^2_{_2}$}}
\newcommand{\er}{e_{_{R}}}
\newcommand{\epr}{e'_{_{R}}}
\newcommand{\ers}{e^2_{_{R}}}
\newcommand{\erc}{e^3_{_{R}}}
\newcommand{\erq}{e^4_{_{R}}}
\newcommand{\erf}{e^5_{_{R}}}
\newcommand{\sour}{J}
\newcommand{\sourb}{\overline J}
\newcommand{\lrm}{M_{_R}}
%
% db: fermionic self-energies and vertex libraries
%
\newcommand{\vlami}[1]{\lambda_{#1}}
\newcommand{\vlamis}[1]{\lambda^2_{#1}}
\newcommand{\Vvert}{V}
\newcommand{\Avert}{A}
\newcommand{\Svert}{S}
\newcommand{\Pvert}{P}
\newcommand{\vvert}{F}
\newcommand{\Cvert}{\cal{V}}
\newcommand{\Bvert}{\cal{B}}
\newcommand{\Vveri}[2]{V_{#1}^{#2}}
\newcommand{\Fveri}[1]{{\cal{F}}^{#1}}
\newcommand{\Cveri}[1]{{\cal{V}}\lpar{#1}\rpar}
\newcommand{\Bveri}[1]{{\cal{B}}\lpar{#1}\rpar}
\newcommand{\Vverti}[3]{V_{#1}^{#2}\lpar{#3}\rpar}
\newcommand{\Averti}[3]{A_{#1}^{#2}\lpar{#3}\rpar}
\newcommand{\Gverti}[3]{G_{#1}^{#2}\lpar{#3}\rpar}
\newcommand{\Zverti}[3]{Z_{#1}^{#2}\lpar{#3}\rpar}
\newcommand{\Hverti}[2]{H^{#1}\lpar{#2}\rpar}
\newcommand{\Wverti}[3]{W_{#1}^{#2}\lpar{#3}\rpar}
\newcommand{\Cverti}[2]{{\cal{V}}_{#1}^{#2}}
\newcommand{\vverti}[3]{F^{#1}_{#2}\lpar{#3}\rpar}
\newcommand{\averti}[3]{{\overline{F}}^{#1}_{#2}\lpar{#3}\rpar}
\newcommand{\fveone}[1]{f_{#1}}
\newcommand{\fvetri}[3]{f^{#1}_{#2}\lpar{#3}\rpar}
\newcommand{\gvetri}[3]{g^{#1}_{#2}\lpar{#3}\rpar}
\newcommand{\cvetri}[3]{{\cal{F}}^{#1}_{#2}\lpar{#3}\rpar}
\newcommand{\hvetri}[3]{{\hat{\cal{F}}}^{#1}_{#2}\lpar{#3}\rpar}
\newcommand{\avetri}[3]{{\overline{\cal{F}}}^{#1}_{#2}\lpar{#3}\rpar}
\newcommand{\fverti}[2]{F^{#1}_{#2}}
\newcommand{\cverti}[2]{{\cal{F}}_{#1}^{#2}}
\newcommand{\fV}{f_{_{\Vvert}}}
\newcommand{\gA}{g_{_{\Avert}}}
\newcommand{\fVi}[1]{f^{#1}_{_{\Vvert}}}
\newcommand{\seai}[1]{a_{#1}}
\newcommand{\seapi}[1]{a'_{#1}}
\newcommand{\seAi}[2]{A_{#1}^{#2}}
\newcommand{\sewi}[1]{w_{#1}}
\newcommand{\seWi}[1]{W_{#1}}
\newcommand{\seWsi}[1]{W^{*}_{#1}}
\newcommand{\seWti}[2]{W_{#1}^{#2}}
\newcommand{\sewti}[2]{w_{#1}^{#2}}
\newcommand{\seSig}[1]{\Sigma_{#1}\lpar\sla{\pmom}\rpar}
\newcommand{\ww}{w}
%
% gp: sm_renorm_oneloop
%
\newcommand{\bbff}[1]{{\overline B}_{#1}}
\newcommand{\sW}{p_{_W}}
\newcommand{\sZ}{p_{_Z}}
\newcommand{\ssp}{s_p}
\newcommand{\fW}{f_{_W}}
\newcommand{\fZ}{f_{_Z}}
\newcommand{\tabn}[1]{Tab.(\ref{#1})}
\newcommand{\subMSB}[1]{{#1}_{\mbox{$\overline{\scriptscriptstyle MS}$}}}
\newcommand{\supMSB}[1]{{#1}^{\mbox{$\overline{\scriptscriptstyle MS}$}}}
\newcommand{\redMSB}{{\mbox{$\overline{\scriptscriptstyle MS}$}}}
\newcommand{\gpbb}{g'_{0}}
\newcommand{\Zconip}[1]{Z'_{#1}}
\newcommand{\bpff}[4]{B'_{#1}\lpar #2;#3,#4\rpar}             % B' form-factor
\newcommand{\xidf}{\xi^2-1}
\newcommand{\tDdr}{1/{\bar{\varepsilon}}}
\newcommand{\cRz}{{\cal R}_{_Z}}
\newcommand{\cRg}{{\cal R}_{\gamma}}
\newcommand{\Sz}{\Sigma_{_Z}}
\newcommand{\alh}{{\hat\alpha}}
\newcommand{\alhz}{\alpha_{_Z}}
\newcommand{\Phzg}{{\hat\Pi}_{_{\zb\ab}}}
\newcommand{\fvvert}{F^{\rm vert}_{_V}}
\newcommand{\gavert}{G^{\rm vert}_{_A}}
\newcommand{\bmv}{{\overline m}_{_V}}
\newcommand{\Sgn}{\Sigma_{\gamma\hkn}}
\newcommand{\tabns}[2]{Tabs.(\ref{#1}--\ref{#2})}
\newcommand{\rmboxd}{{\rm Box}_d\lpar s,t,u;M_1,M_2,M_3,M_4\rpar}
\newcommand{\rmboxc}{{\rm Box}_c\lpar s,t,u;M_1,M_2,M_3,M_4\rpar}
%
% D-functions
%
\newcommand{\Afaci}[1]{A_{#1}}
\newcommand{\Afacis}[1]{A^2_{#1}}
\newcommand{\upar}[1]{u}
\newcommand{\upari}[1]{u_{#1}}
\newcommand{\vpari}[1]{v_{#1}}
\newcommand{\lpari}[1]{l_{#1}}
\newcommand{\Lpari}[1]{l_{#1}}
\newcommand{\Nff}[2]{N^{(#1)}_{#2}}
\newcommand{\Sff}[2]{S^{(#1)}_{#2}}
\newcommand{\sSff}{S}
\newcommand{\FQED}[2]{F_{#1#2}}
\newcommand{\fbpsif}{{\overline{\psi}_f}}
\newcommand{\fpsif}{\psi_f}
\newcommand{\etafd}[2]{\eta_d\lpar#1,#2\rpar}
\newcommand{\sigdu}[2]{\sigma_{#1#2}}
\newcommand{\scalc}[4]{c_{_0}\lpar #1;#2,#3,#4\rpar}
\newcommand{\scald}[2]{d_{_0}\lpar #1,#2\rpar}
\newcommand{\pir}[1]{\Pi^{\rm ren}\lpar #1\rpar}
\newcommand{\sigh}{\sigma_{\rm had}}
\newcommand{\dah}{\Delta\alpha^{(5)}_{\rm had}}
\newcommand{\dat}{\Delta\alpha_{\rm top}}
\newcommand{\Vqed}[3]{V_1^{\rm sub}\lpar#1;#2,#3\rpar}
\newcommand{\thetah}{{\hat\theta}}
\newcommand{\mtsix}{m^6_t}
\newcommand{\smlon}{\frac{\mlones}{s}}
\newcommand{\lntwo}{\ln 2}
\newcommand{\wmin}{w_{\rm min}}
\newcommand{\kmin}{k_{\rm min}}
\newcommand{\scaldi}[3]{d_{_0}^{#1}\lpar #2,#3\rpar}
\newcommand{\mdls}{\Big|}
\newcommand{\smf}{\frac{\mfs}{s}}
\newcommand{\bint}{\beta_{\rm int}}
\newcommand{\IRv}{V_{_{\rm IR}}}
\newcommand{\IRr}{R_{_{\rm IR}}}
\newcommand{\fssts}{\frac{s^2}{t^2}}
\newcommand{\fssus}{\frac{s^2}{u^2}}
\newcommand{\optM}{1+\frac{t}{M^2}}
\newcommand{\opuM}{1+\frac{u}{M^2}}
\newcommand{\ftM}{\lpar -\frac{t}{M^2}\rpar}
\newcommand{\fuM}{\lpar -\frac{u}{M^2}\rpar}
\newcommand{\omsM}{1-\frac{s}{M^2}}
\newcommand{\xsf}{\sigma_{_{\rm F}}}
\newcommand{\xsb}{\sigma_{_{\rm B}}}
\newcommand{\afb}{A_{_{\rm FB}}}
\newcommand{\rsoft}{\rm soft}
\newcommand{\rms}{\rm s}
\newcommand{\rsmx}{\sqrt{s_{\rm max}}}
\newcommand{\rspm}{\sqrt{s_{\pm}}}
\newcommand{\rsp}{\sqrt{s_{+}}}
\newcommand{\rsm}{\sqrt{s_{-}}}
\newcommand{\sigmx}{\sigma_{\rm max}}
\newcommand{\gG}[2]{G_{#1}^{#2}}
\newcommand{\gacomm}[2]{\lpar #1 - #2\gfd\rpar}
\newcommand{\fcsx}{\frac{1}{\ctwsix}}
\newcommand{\fcq}{\frac{1}{\ctwf}}
\newcommand{\fcs}{\frac{1}{\ctws}}
\newcommand{\affs}[2]{{\cal A}_{#1}\lpar #2\rpar}                   % A
\newcommand{\stwei}{s_{\theta}^8}
\def\mdan{\vspace{1mm}\mpar{\hfil$\downarrow$new\hfil}\vspace{-1mm}
          \ignorespaces}
\def\muan{\vspace{-1mm}\mpar{\hfil$\uparrow$new\hfil}\vspace{1mm}\ignorespaces}
\def\mlan{\vspace{-1mm}\mpar{\hfil$\rightarrow$new\hfil}\vspace{1mm}\ignorespaces}
\def\mnnew{\mpar{\hfil NEWNEW \hfil}\ignorespaces}
%
% db's of libraries
%                  
\newcommand{\boxc}[2]{{\cal{B}}_{#1}^{#2}}
\newcommand{\boxct}[3]{{\cal{B}}_{#1}^{#2}\lpar{#3}\rpar}
\newcommand{\hboxc}[3]{\hat{{\cal{B}}}_{#1}^{#2}\lpar{#3}\rpar}
\newcommand{\vev}{\langle v \rangle}
\newcommand{\vevi}[1]{\langle v_{#1}\rangle}
\newcommand{\vevs}{\langle v^2   \rangle}
\newcommand{\fwfrV}[5]{\Sigma_{_V}\lpar #1,#2,#3;#4,#5 \rpar}
\newcommand{\fwfrS}[7]{\Sigma_{_S}\lpar #1,#2,#3;#4,#5;#6,#7 \rpar}
\newcommand{\fSi}[1]{f^{#1}_{_{\Svert}}}
\newcommand{\fPi}[1]{f^{#1}_{_{\Pvert}}}
\newcommand{\mXs}{m_{_X}}
\newcommand{\mXss}{m^2_{_X}}
\newcommand{\mYs}{M^2_{_Y}}
\newcommand{\xik}{\xi_k}
\newcommand{\xiks}{\xi^2_k}
\newcommand{\mpls}{m^2_+}
\newcommand{\mmis}{m^2_-}
%
%--
\newcommand{\SN}{\Sigma_{_N}}
\newcommand{\SC}{\Sigma_{_C}}
\newcommand{\SPN}{\Sigma'_{_N}}
\newcommand{\PFf}{\Pi^{\fer}_{_F}}
\newcommand{\PFb}{\Pi^{\bos}_{_F}}
\newcommand{\dPZ}{\Delta{\hat\Pi}_{_Z}}
\newcommand{\Sfin}{\Sigma_{_F}}
\newcommand{\Sfir}{\Sigma_{_R}}
\newcommand{\Sfinh}{{\hat\Sigma}_{_F}}
\newcommand{\Sfinf}{\Sigma^{\fer}_{_F}}
\newcommand{\Sfinbh}{\Sigma^{\bos}_{_F}}
\newcommand{\alf}{\alpha^{\fer}}
\newcommand{\alhfz}{\alpha^{\fer}\lpar{\ssZ}\rpar}
\newcommand{\alhfs}{\alpha^{\fer}\lpar{\sman}\rpar}
\newcommand{\gfQ}{g^f_{_{Q}}}
\newcommand{\gfL}{g^f_{_{L}}}
\newcommand{\ccf}{\frac{\gbs}{16\,\pi^2}}
\newcommand{\chq}{{\hat c}^4}
\newcommand{\muuq}{m_{u'}}
\newcommand{\muus}{m^2_{u'}}
\newcommand{\mdd}{m_{d'}}
%--
\newcommand{\clf}[2]{\mathrm{Cli}_{_#1}\lpar\displaystyle{#2}\rpar}
\def\stes{\sin^2\theta}
\def\acal{\cal A}
\def\alr{A_{_{\rm{LR}}}}
\newcommand{\barQ}{\overline Q}
\newcommand{\Sptg}{\Sigma'_{_{3Q}}}
\newcommand{\Sptt}{\Sigma'_{_{33}}}
\newcommand{\Ppgg}{\Pi'_{\ph\ph}}
\newcommand{\Pww}{\Pi_{_{\wb\wb}}}
\newcommand{\capV}[2]{{\cal F}^{#2}_{_{#1}}}
%--
\newcommand{\bt}{\beta_t}
\newcommand{\mhsix}{M^6_{_H}}
\newcommand{\topt}{{\cal T}_{33}}
\newcommand{\topq}{{\cal T}_4}
\newcommand{\Phzgf}{{\hat\Pi}^{\fer}_{_{\zb\ab}}}
\newcommand{\Phzgb}{{\hat\Pi}^{\bos}_{_{\zb\ab}}}
\newcommand{\Sfirh}{{\hat\Sigma}_{_R}}
\newcommand{\Szgh}{{\hat\Sigma}_{_{\zb\ab}}}
\newcommand{\Szghb}{{\hat\Sigma}^{\bos}_{_{\zb\ab}}}
\newcommand{\Szghf}{{\hat\Sigma}^{\fer}_{_{\zb\ab}}}
\newcommand{\Szgb}{\Sigma^{\bos}_{_{\zb\ab}}}
\newcommand{\Szgf}{\Sigma^{\fer}_{_{\zb\ab}}}
%--
\newcommand{\chig}{\chi_{_{\ph}}}
\newcommand{\chiz}{\chi_{_{\zb}}}
\newcommand{\Sfih}{{\hat\Sigma}}
\newcommand{\Szzh}{\hat{\Sigma}_{_{\zb\zb}}}
\newcommand{\dPZf}{\Delta{\hat\Pi}^f_{_{\zb}}}
\newcommand{\khZdf}[1]{{\hat\kappa}^{#1}_{_{\zb}}}
\newcommand{\chf}{{\hat c}^4}
%---
%for process sm_ola
\newcommand{\amp}[2]{{\cal{A}}_{_{#1}}^{\rm{#2}}}
%--
\newcommand{\hatvm}[1]{{\hat v}^-_{#1}}
\newcommand{\hatvp}[1]{{\hat v}^+_{#1}}
\newcommand{\hatvpm}[1]{{\hat v}^{\pm}_{#1}}
\newcommand{\kvz}[1]{\kappa^{\zb #1}_{_V}}
%--
\newcommand{\barp}{\overline p}                
\newcommand{\delw}{\Delta_{_{\wb}}}
\newcommand{\bdelw}{{\bar{\Delta}}_{_{\wb}}}
\newcommand{\bdelf}{{\bar{\Delta}}_{\ff}}
\newcommand{\delz}{\Delta_{_\zb}}
\newcommand{\deli}[1]{\Delta\lpar{#1}\rpar}
\newcommand{\chizb}{\chi_{_\zb}}
\newcommand{\Swwp}{\Sigma'_{_{\wb\wb}}}
\newcommand{\epph}{\varepsilon'/2}
\newcommand{\sbffp}[1]{B'_{#1}}                    
\newcommand{\epss}{\varepsilon^*}
%--
\newcommand{\Ddrhs}{{\ds\frac{1}{\hat{\varepsilon}^2}}}
\newcommand{\lnmsb}{L_{_\wb}}
\newcommand{\lnsmsb}{L^2_{_\wb}}
\newcommand{\tpni}{\lpar 2\pi\rpar^n\ib}
\newcommand{\tpn}{2^n\,\pi^{n-2}}
%--
\newcommand{\cmf}{M_f}
\newcommand{\cmfs}{M^2_f}
\newcommand{\toDdr}{{\ds\frac{2}{{\bar{\varepsilon}}}}}
\newcommand{\troDdr}{{\ds\frac{3}{{\bar{\varepsilon}}}}}
\newcommand{\totDdr}{{\ds\frac{3}{{2\,\bar{\varepsilon}}}}}
\newcommand{\foDdr}{{\ds\frac{4}{{\bar{\varepsilon}}}}}
\newcommand{\smh}{m_{_H}}
\newcommand{\smhs}{m^2_{_H}}
\newcommand{\Ph}{\Pi_{_\hb}}
\newcommand{\Sphh}{\Sigma'_{_{\hb\hb}}}
\newcommand{\bh}{\beta}
\newcommand{\alsn}{\alpha^{(n_f)}_{_S}}
\newcommand{\smq}{m_q}
\newcommand{\smqp}{m_{q'}}
\newcommand{\shb}{h}
\newcommand{\hab}{A}
\newcommand{\hbpm}{H^{\pm}}
\newcommand{\hbp}{H^{+}}
\newcommand{\hbm}{H^{-}}
\newcommand{\msh}{M_h}
\newcommand{\mha}{M_{_A}}
\newcommand{\mhc}{M_{_{H^{\pm}}}}
\newcommand{\mshs}{M^2_h}
\newcommand{\mhas}{M^2_{_A}}
\newcommand{\barfp}{\overline{f'}}                
\newcommand{\chiii}{{\hat c}^3}
\newcommand{\chiv}{{\hat c}^4}
\newcommand{\chv}{{\hat c}^5}
\newcommand{\chvi}{{\hat c}^6}
\newcommand{\alsvi}{\alpha^{6}_{_S}}
\newcommand{\tww}{t_{_W}}
\newcommand{\ti}{t_{_1}}
\newcommand{\tii}{t_{_2}}
\newcommand{\tiii}{t_{_3}}
\newcommand{\tiv}{t_{_4}}
\newcommand{\psla}{\hbox{\rlap/p}}
\newcommand{\qsla}{\hbox{\rlap/q}}
\newcommand{\nsla}{\hbox{\rlap/n}}
\newcommand{\lsla}{\hbox{\rlap/l}}
\newcommand{\msla}{\hbox{\rlap/m}}
\newcommand{\cnsla}{\hbox{\rlap/N}}
\newcommand{\clsla}{\hbox{\rlap/L}}
\newcommand{\cmsla}{\hbox{\rlap/M}}
\newcommand{\blmt}{\lrbr - 3\rrbr}
\newcommand{\blfo}{\lrbr 4 1\rrbr}
\newcommand{\bltp}{\lrbr 2 +\rrbr}
%---------------------------------
% mixed QCD
\newcommand{\clitwo}[1]{{\rm{Li}}_{2}\lpar{#1}\rpar}
\newcommand{\clitri}[1]{{\rm{Li}}_{3}\lpar{#1}\rpar}
% subleadings
\newcommand{\xt}{x_{\ft}}
\newcommand{\zt}{z_{\ft}}
\newcommand{\Ht}{h_{\ft}}
\newcommand{\xts}{x^2_{\ft}}
\newcommand{\zts}{z^2_{\ft}}
\newcommand{\Hts}{h^2_{\ft}}
\newcommand{\ztc}{z^3_{\ft}}
\newcommand{\Htc}{h^3_{\ft}}
\newcommand{\ztq}{z^4_{\ft}}
\newcommand{\Htq}{h^4_{\ft}}
\newcommand{\ztv}{z^5_{\ft}}
\newcommand{\Htv}{h^5_{\ft}}
\newcommand{\ztx}{z^6_{\ft}}
\newcommand{\Htx}{h^6_{\ft}}
\newcommand{\ztz}{z^7_{\ft}}
\newcommand{\Htz}{h^7_{\ft}}
\newcommand{\sht}{\sqrt{\Ht}}
\newcommand{\atan}[1]{{\rm{arctan}}\lpar{#1}\rpar}
\newcommand{\dbff}[3]{{\hat{B}}_{_{{#2}{#3}}}\lpar{#1}\rpar}
\newcommand{\ztbs}{{\bar{z}}^{2}_{\ft}}
\newcommand{\ztb}{{\bar{z}}_{\ft}}
\newcommand{\Htbs}{{\bar{h}}^{2}_{\ft}}
\newcommand{\Htb}{{\bar{h}}_{\ft}}
\newcommand{\Hztb}{{\bar{hz}}_{\ft}}
\newcommand{\Ln}[1]{{\rm{Ln}}\lpar{#1}\rpar}
\newcommand{\Lns}[1]{{\rm{Ln}}^2\lpar{#1}\rpar}
\newcommand{\wt}{w_{\ft}}
\newcommand{\wts}{w^2_{\ft}}
\newcommand{\wtb}{\overline{w}}
%--
\newcommand{\fra}{\frac{1}{2}}
\newcommand{\frb}{\frac{1}{4}}
\newcommand{\frc}{\frac{3}{2}}
\newcommand{\frd}{\frac{3}{4}}
\newcommand{\fre}{\frac{9}{2}}
\newcommand{\frf}{\frac{9}{4}}
\newcommand{\frg}{\frac{5}{4}}
\newcommand{\frh}{\frac{5}{2}}
\newcommand{\fri}{\frac{1}{8}}
\newcommand{\frj}{\frac{7}{4}}
\newcommand{\frl}{\frac{7}{8}}
\newcommand{\Spzzh}{\hat{\Sigma}'_{_{\zb\zb}}}
\newcommand{\sss}{s\sqrt{s}}
\newcommand{\sqs}{\sqrt{s}}
\newcommand{\Rtg}{R_{_{3Q}}}
\newcommand{\Rtt}{R_{_{33}}}
\newcommand{\Rww}{R_{_{\wb\wb}}}
\newcommand{\ssZ}{{\scriptscriptstyle{\zb}}}
\newcommand{\ssW}{{\scriptscriptstyle{\wb}}}
\newcommand{\ssH}{{\scriptscriptstyle{\hb}}}
\newcommand{\ssV}{{\scriptscriptstyle{\vb}}}
\newcommand{\ssA}{{\scriptscriptstyle{A}}}
\newcommand{\ssB}{{\scriptscriptstyle{B}}}
\newcommand{\ssC}{{\scriptscriptstyle{C}}}
\newcommand{\ssD}{{\scriptscriptstyle{D}}}
\newcommand{\ssF}{{\scriptscriptstyle{F}}}
\newcommand{\ssG}{{\scriptscriptstyle{G}}}
\newcommand{\ssL}{{\scriptscriptstyle{L}}}
\newcommand{\ssM}{{\scriptscriptstyle{M}}}
\newcommand{\ssN}{{\scriptscriptstyle{N}}}
\newcommand{\ssP}{{\scriptscriptstyle{P}}}
\newcommand{\ssQ}{{\scriptscriptstyle{Q}}}
\newcommand{\ssR}{{\scriptscriptstyle{R}}}
\newcommand{\ssS}{{\scriptscriptstyle{S}}}
\newcommand{\ssT}{{\scriptscriptstyle{T}}}
\newcommand{\ssU}{{\scriptscriptstyle{U}}}
\newcommand{\ssX}{{\scriptscriptstyle{X}}}
\newcommand{\ssY}{{\scriptscriptstyle{Y}}}
\newcommand{\ssWF}{{\scriptscriptstyle{WF}}}
%--
\newcommand{\DiagramFermionToBosonFullWithMomenta}[8][70]{
  \vcenter{\hbox{
  \SetScale{0.8}
  \begin{picture}(#1,50)(15,15)
    \put(27,22){$\nearrow$}      
    \put(27,54){$\searrow$}    
    \put(59,29){$\to$}    
    \ArrowLine(25,25)(50,50)      \Text(34,20)[lc]{#6} \Text(11,20)[lc]{#3}
    \ArrowLine(50,50)(25,75)      \Text(34,60)[lc]{#7} \Text(11,60)[lc]{#4}
    \Photon(50,50)(90,50){2}{8}   \Text(80,40)[lc]{#2} \Text(55,33)[ct]{#8}
    \Vertex(50,50){2,5}          \Text(60,48)[cb]{#5} 
    \Vertex(90,50){2}
  \end{picture}}}
  }
\newcommand{\DiagramFermionToBosonPropagator}[4][85]{
  \vcenter{\hbox{
  \SetScale{0.8}
  \begin{picture}(#1,50)(15,15)
    \ArrowLine(25,25)(50,50)
    \ArrowLine(50,50)(25,75)
    \Photon(50,50)(105,50){2}{8}   \Text(90,40)[lc]{#2}
%    \GCirc(50,50){10}{0.5}         \Text(80,48)[cb]{#3}
    \Vertex(50,50){0.5}         \Text(80,48)[cb]{#3}
    \GCirc(82,50){8}{1}            \Text(55,48)[cb]{#4}
    \Vertex(105,50){2}
  \end{picture}}}
  }
\newcommand{\DiagramFermionToBosonEffective}[3][70]{
  \vcenter{\hbox{
  \SetScale{0.8}
  \begin{picture}(#1,50)(15,15)
    \ArrowLine(25,25)(50,50)
    \ArrowLine(50,50)(25,75)
    \Photon(50,50)(90,50){2}{8}   \Text(80,40)[lc]{#2}
    \BBoxc(50,50)(5,5)            \Text(55,48)[cb]{#3}
    \Vertex(90,50){2}
  \end{picture}}}
  }
\newcommand{\DiagramFermionToBosonFull}[3][70]{
  \vcenter{\hbox{
  \SetScale{0.8}
  \begin{picture}(#1,50)(15,15)
    \ArrowLine(25,25)(50,50)
    \ArrowLine(50,50)(25,75)
    \Photon(50,50)(90,50){2}{8}   \Text(80,40)[lc]{#2}
    \Vertex(50,50){2.5}          \Text(60,48)[cb]{#3}
    \Vertex(90,50){2}
  \end{picture}}}
  }
%--
\newcommand{\expgw}{\frac{\gf\mws}{2\srt\,\pi^2}}
\newcommand{\expgz}{\frac{\gf\mzs}{2\srt\,\pi^2}}
\newcommand{\Spww}{\Sigma'_{_{\wb\wb}}}
\newcommand{\shf}{{\hat s}^4}
%--
\newcommand{\acz}{\scff{0}}
\newcommand{\acoo}{\scff{11}}
\newcommand{\acod}{\scff{12}}
\newcommand{\acdo}{\scff{21}}
\newcommand{\acdd}{\scff{22}}
\newcommand{\acdt}{\scff{23}}
\newcommand{\acdq}{\scff{24}}
\newcommand{\acto}{\scff{31}}
\newcommand{\actd}{\scff{32}}
\newcommand{\actt}{\scff{33}}
\newcommand{\actq}{\scff{34}}
\newcommand{\actc}{\scff{35}}
\newcommand{\acts}{\scff{36}}
\newcommand{\acoA}{\scff{1A}}
\newcommand{\acdA}{\scff{2A}}
\newcommand{\acdB}{\scff{2B}}
\newcommand{\acdC}{\scff{2C}}
\newcommand{\acdD}{\scff{2D}}
\newcommand{\actA}{\scff{3A}}
\newcommand{\actB}{\scff{3B}}
\newcommand{\actC}{\scff{3C}}
%--
\newcommand{\ada}{\sdff{0}}
\newcommand{\adb}{\sdff{11}}
\newcommand{\adc}{\sdff{12}}
\newcommand{\add}{\sdff{13}}
\newcommand{\ade}{\sdff{21}}
\newcommand{\adf}{\sdff{22}}
\newcommand{\adg}{\sdff{23}}
\newcommand{\adh}{\sdff{24}}
\newcommand{\adi}{\sdff{25}}
\newcommand{\adj}{\sdff{26}}
\newcommand{\adl}{\sdff{27}}
\newcommand{\adm}{\sdff{31}}
\newcommand{\adn}{\sdff{32}}
\newcommand{\ado}{\sdff{33}}
\newcommand{\adp}{\sdff{34}}
\newcommand{\adq}{\sdff{35}}
\newcommand{\adr}{\sdff{36}}
\newcommand{\ads}{\sdff{37}}
\newcommand{\adt}{\sdff{38}}
\newcommand{\adu}{\sdff{39}}
\newcommand{\adw}{\sdff{310}}
\newcommand{\adv}{\sdff{311}}
\newcommand{\ady}{\sdff{312}}
\newcommand{\adz}{\sdff{313}}
%--
\newcommand{\admt}{\frac{\tman}{\sman}}
\newcommand{\admu}{\frac{\uman}{\sman}}
\newcommand{\frm}{\frac{3}{8}}
\newcommand{\frn}{\frac{5}{8}}
\newcommand{\fro}{\frac{15}{8}}
\newcommand{\frp}{\frac{3}{16}}
\newcommand{\frq}{\frac{5}{16}}
\newcommand{\frr}{\frac{1}{16}}
\newcommand{\frs}{\frac{7}{2}}
\newcommand{\frt}{\frac{7}{16}}
\newcommand{\fru}{\frac{1}{3}}
\newcommand{\frw}{\frac{2}{3}}
\newcommand{\frz}{\frac{4}{3}}
\newcommand{\fry}{\frac{13}{3}}
\newcommand{\fraa}{\frac{11}{4}}
%--
\newcommand{\bee}{\beta_{e}}
\newcommand{\beW}{\beta_{_\wb}}
%--
\newcommand{\toDdrh}{{\ds\frac{2}{{\hat{\varepsilon}}}}}
\newcommand{\bqas}{\begin{eqnarray*}}
\newcommand{\eqas}{\end{eqnarray*}}
%--
\newcommand{\mhcub}{M^3_{_H}}
%--
\newcommand{\adComA}{\sdff{A}}
\newcommand{\adComB}{\sdff{B}}
\newcommand{\adComC}{\sdff{C}}
\newcommand{\adComD}{\sdff{D}}
\newcommand{\adComE}{\sdff{E}}
\newcommand{\adComF}{\sdff{F}}
\newcommand{\adComG}{\sdff{G}}
\newcommand{\adComH}{\sdff{H}}
\newcommand{\adComI}{\sdff{I}}
\newcommand{\adComJ}{\sdff{J}}
\newcommand{\adComL}{\sdff{L}}
\newcommand{\adComM}{\sdff{M}}
\newcommand{\adComN}{\sdff{N}}
\newcommand{\adComO}{\sdff{O}}
\newcommand{\adComP}{\sdff{P}}
\newcommand{\adComQ}{\sdff{Q}}
\newcommand{\adComR}{\sdff{R}}
\newcommand{\adComS}{\sdff{S}}
\newcommand{\adComT}{\sdff{T}}
\newcommand{\adComU}{\sdff{U}}
%--
\newcommand{\adComAc}{\sdff{A}^c}
\newcommand{\adComBc}{\sdff{B}^c}
\newcommand{\adComCc}{\sdff{C}^c}
\newcommand{\adComDc}{\sdff{D}^c}
\newcommand{\adComEc}{\sdff{E}^c}
\newcommand{\adComFc}{\sdff{F}^c}
\newcommand{\adComGc}{\sdff{G}^c}
\newcommand{\adComHc}{\sdff{H}^c}
\newcommand{\adComIc}{\sdff{I}^c}
\newcommand{\adComJc}{\sdff{J}^c}
\newcommand{\adComLc}{\sdff{L}^c}
\newcommand{\adComMc}{\sdff{M}^c}
\newcommand{\adComNc}{\sdff{N}^c}
\newcommand{\adComOc}{\sdff{O}^c}
\newcommand{\adComPc}{\sdff{P}^c}
\newcommand{\adComQc}{\sdff{Q}^c}
\newcommand{\adComRc}{\sdff{R}^c}
\newcommand{\adComSc}{\sdff{S}^c}
\newcommand{\adComTc}{\sdff{T}^c}
\newcommand{\adComUc}{\sdff{U}^c}
%--
\newcommand{\adComAf}{\sdff{A}^f}
\newcommand{\adComBf}{\sdff{B}^f}
\newcommand{\adComCf}{\sdff{F}^f}
\newcommand{\adComDf}{\sdff{D}^f}
\newcommand{\adComEf}{\sdff{E}^f}
\newcommand{\adComFf}{\sdff{F}^f}
\newcommand{\adComGf}{\sdff{G}^f}
\newcommand{\adComHf}{\sdff{H}^f}
\newcommand{\adComIf}{\sdff{I}^f}
\newcommand{\adComJf}{\sdff{J}^f}
\newcommand{\adComLf}{\sdff{L}^f}
\newcommand{\adComMf}{\sdff{M}^f}
\newcommand{\adComNf}{\sdff{N}^f}
\newcommand{\adComOf}{\sdff{O}^f}
\newcommand{\adComPf}{\sdff{P}^f}
\newcommand{\adComQf}{\sdff{Q}^f}
\newcommand{\adComRf}{\sdff{R}^f}
\newcommand{\adComSf}{\sdff{S}^f}
\newcommand{\adComTf}{\sdff{T}^f}
\newcommand{\adComUf}{\sdff{U}^f}
%--
\newcommand{\adComBfc}{\sdff{B}^{fc}} 
\newcommand{\adComCfco}{\sdff{C}^{fc1}}
\newcommand{\adComCfcd}{\sdff{C}^{fc2}} 
\newcommand{\adComCfct}{\sdff{C}^{fc3}} 
\newcommand{\adComDfc}{\sdff{D}^{fc}}
\newcommand{\adComEfc}{\sdff{E}^{fc}}
\newcommand{\adComFfc}{\sdff{F}^{fc}}
\newcommand{\adComGfc}{\sdff{G}^{fc}}
\newcommand{\adComHfc}{\sdff{H}^{fc}}
\newcommand{\adComLfc}{\sdff{L}^{fc}}
%--
\newcommand{\afba}[1]{A^{#1}_{_{\rm FB}}}
\newcommand{\alra}[1]{A^{#1}_{_{\rm LR}}}
%--
\newcommand{\adComAt}{\sdff{A}^t}
\newcommand{\adComBt}{\sdff{B}^t}
\newcommand{\adComCt}{\sdff{T}^t}
\newcommand{\adComDt}{\sdff{D}^t}
\newcommand{\adComEt}{\sdff{E}^t}
\newcommand{\adComFt}{\sdff{T}^t}
\newcommand{\adComGt}{\sdff{G}^t}
\newcommand{\adComHt}{\sdff{H}^t}
\newcommand{\adComIt}{\sdff{I}^t}
\newcommand{\adComJt}{\sdff{J}^t}
\newcommand{\adComLt}{\sdff{L}^t}
\newcommand{\adComMt}{\sdff{M}^t}
\newcommand{\adComNt}{\sdff{N}^t}
\newcommand{\adComOt}{\sdff{O}^t}
\newcommand{\adComPt}{\sdff{P}^t}
\newcommand{\adComQt}{\sdff{Q}^t}
\newcommand{\adComRt}{\sdff{R}^t}
\newcommand{\adComSt}{\sdff{S}^t}
\newcommand{\adComTt}{\sdff{T}^t}
\newcommand{\adComUt}{\sdff{U}^t}
%--
\newcommand{\adComAtt}{\sdff{A}^{\tau}}
\newcommand{\adComBtt}{\sdff{B}^{\tau}}
\newcommand{\adComCtt}{\sdff{T}^{\tau}}
\newcommand{\adComDtt}{\sdff{D}^{\tau}}
\newcommand{\adComEtt}{\sdff{E}^{\tau}}
\newcommand{\adComFtt}{\sdff{T}^{\tau}}
\newcommand{\adComGtt}{\sdff{G}^{\tau}}
\newcommand{\adComHtt}{\sdff{H}^{\tau}}
\newcommand{\adComItt}{\sdff{I}^{\tau}}
\newcommand{\adComJtt}{\sdff{J}^{\tau}}
\newcommand{\adComLtt}{\sdff{L}^{\tau}}
\newcommand{\adComMtt}{\sdff{M}^{\tau}}
\newcommand{\adComNtt}{\sdff{N}^{\tau}}
\newcommand{\adComOtt}{\sdff{O}^{\tau}}
\newcommand{\adComPtt}{\sdff{P}^{\tau}}
\newcommand{\adComQtt}{\sdff{Q}^{\tau}}
\newcommand{\adComRtt}{\sdff{R}^{\tau}}
\newcommand{\adComStt}{\sdff{S}^{\tau}}
\newcommand{\adComTtt}{\sdff{T}^{\tau}}
\newcommand{\adComUtt}{\sdff{U}^{\tau}}
%--
\newcommand{\etavz}[1]{\eta^{\zb #1}_{_V}}
%--
\newcommand{\phanst}{$\hphantom{\sigma^{s+t}\ }$}
\newcommand{\phanat}{$\hphantom{A_{FB}^{s+t}\ }$}
\newcommand{\phanss}{$\hphantom{\sigma^{s}\ }$}
\newcommand{\phanas}{$\hphantom{A_{FB}^{s}\ }$} 
\newcommand{\pbb}{\,\mbox{\bf pb}}
\newcommand{\pe}{\,\%\:}
\newcommand{\pc}{\,\%}
%--
\newcommand{\temiv}{10^{-4}}
\newcommand{\temv}{10^{-5}}
\newcommand{\temvi}{10^{-6}}
\newcommand{\di}[1]{d_{#1}}
%--
\newcommand{\delip}[1]{\Delta_+\lpar{#1}\rpar}
\newcommand{\propbb}[5]{{{#1}\over {\lpar #2^2 + #3 - \ib\varepsilon\rpar
\lpar\lpar #4\rpar^2 + #5 -\ib\varepsilon\rpar}}}
\newcommand{\cfft}[5]{C_{#1}\lpar #2;#3,#4,#5\rpar}    
\newcommand{\ppl}[1]{p_{+{#1}}}
\newcommand{\pmi}[1]{p_{-{#1}}}
%--
\newcommand{\bpox}{\beta^2_{\xi}}
\newcommand{\bffdiff}[5]{B_{\rm d}\lpar #1;#2,#3;#4,#5\rpar}             
\newcommand{\cffdiff}[7]{C_{\rm d}\lpar #1;#2,#3,#4;#5,#6,#7\rpar}    
\newcommand{\affdiff}[2]{A_{\rm d}\lpar #1;#2\rpar}             
\newcommand{\Dqf}{\Delta\qf}
\newcommand{\bposx}{\beta^4_{\xi}}
\newcommand{\svverti}[3]{f^{#1}_{#2}\lpar{#3}\rpar}
\newcommand{\Mods}{\mbox{$M^2_{12}$}}
\newcommand{\Mots}{\mbox{$M^2_{13}$}}
\newcommand{\Motq}{\mbox{$M^4_{13}$}}
\newcommand{\Mdts}{\mbox{$M^2_{23}$}}
\newcommand{\Mdos}{\mbox{$M^2_{21}$}}
\newcommand{\Mtds}{\mbox{$M^2_{32}$}}
\newcommand{\dffpt}[3]{D_{#1}\lpar #2,#3;}           
\newcommand{\quu}{Q_{uu}}
\newcommand{\qdd}{Q_{dd}}
\newcommand{\qud}{Q_{ud}}
\newcommand{\qdu}{Q_{du}}
%--
\newcommand{\msPj}[6]{\Lambda^{#1#2#3}_{#4#5#6}}
%--
\newcommand{\bdiff}[4]{B_{\rm d}\lpar #1,#2;#3,#4\rpar}             
\newcommand{\bdifff}[7]{B_{\rm d}\lpar #1;#2;#3;#4,#5;#6,#7\rpar}             
\newcommand{\adiff}[3]{A_{\rm d}\lpar #1;#2;#3\rpar}  
%--
\newcommand{\aw}{a_{_\wb}}
\newcommand{\az}{a_{_\zb}}
%--
\newcommand{\sct}[1]{sect.~\ref{#1}}
%--
\newcommand{\dreim}[1]{\varepsilon^{\rm M}_{#1}}
\newcommand{\drem}{\varepsilon^{\rm M}}
%--
\newcommand{\hcapV}[2]{{\hat{\cal F}}^{#2}_{_{#1}}}
%--
\newcommand{\swww}{{\scriptscriptstyle \wb\wb\wb}}
\newcommand{\szhz}{{\scriptscriptstyle \zb\hb\zb}}
\newcommand{\shzh}{{\scriptscriptstyle \hb\zb\hb}}
\newcommand{\bwith}[3]{\beta^{#3}_{#1}\lpar #2\rpar}
\newcommand{\Shhh}{{\hat\Sigma}_{_{\hb\hb}}}
\newcommand{\Sphhh}{{\hat\Sigma}'_{_{\hb\hb}}}
%--
\newcommand{\seWilc}[1]{w_{#1}}
\newcommand{\seWtilc}[2]{w_{#1}^{#2}}
%--
\newcommand{\eilc}{\gamma}
\newcommand{\eilcs}{\gamma^2}
\newcommand{\eilcc}{\gamma^3}
\newcommand{\eilcb}{{\overline{\gamma}}}
\newcommand{\eilcbs}{{\overline{\gamma}^2}}
%--
\newcommand{\Sttww}{\Sigma_{_{33;\wb\wb}}}
\newcommand{\bSttww}{{\overline\Sigma}_{_{33;\wb\wb}}}
%--
\newcommand{\Pggtg}{\Pi_{\ph\ph;3Q}}

\title{\Large\bf The Twofold Way\\
a Short Disquisition of LEP Physics
\thanks{Talk given at XIV International Workshop on High Energy Physics and 
Quantum Field Theory, QFTHEP'99, Moscow, Russia, 27 May - 2 June, 1999}
}
\author{
Giampiero Passarino\\ 
Dipartimento di Fisica Teorica, Universit\`a di Torino, Italy \\
INFN, Sezione di Torino, Italy}

\date{}
\maketitle

\begin{abstract}
  \normalsize \noindent
A condensed survey of various aspects of LEP~1 physics, $e^+e^- \to \barf f$, 
is presented, with a critical examination of the ingredients that are 
available for theoretical predictions. A prototype of comparisons for 2f 
calculations can be found in hep-ph/9902452.
\footnote{
The numerical results are based on the work of 
{\tt TOPAZ0}~\cite{kn:topaz0} and {\tt ZFITTER}~\cite{kn:zfitter} teams.
At present, the following physicists are active members of the two teams: 
{\tt TOPAZ0}:  G. Montagna, O. Nicrosini, G. Passarino and F. Piccinini;  
{\tt ZFITTER}: D. Bardin, P. Christova, M. Jack, L. Kalinovskaya, 
A. Olshevski, S. Riemann, T. Riemann.
}
\end{abstract}

\section{Prolegomena}
Consider the LEP~1 measurements of $\sigma_{\mu,\rm had}$ and $\afba{l}$
as an example to discuss the data analysis strategy:
each LEP experiment extracts a set of  
%--
\bq 
{\rm PO} \stackrel{\rm def}{=} \{\mz, \gz, R_{\fe,\flm,\flt}, 
\afba{0,\fe,\flm,\flt}\},
\eq
%--
from their measured $\sigma$ and $\afba{}$ ({\em realistic observables}).  
The four sets of PO are combined, taking correlated errors between 
the LEP experiments into account, in order to obtain a LEP-average set of PO.
The latter is then interpreted, for example within the frame-work of
the Minimal Standard Model.
The practical attitude of the experiments is to stay with a 
{\em Model-Independent} (MI) fit, i.e.  
%--
\bq
{\rm from}\quad {\rm RO}  \to {\rm PO} \oplus\, {\mbox{a SM remnant}}
\eq
%--
for each experiment, and these sets of PO are averaged.  
The extraction of Lagrangian parameters, 
%--
\bq
\mz, \mt, \mh, \als(\mzs), \quad \mbox{and} \quad \alpha(\mzs), 
\eq
%--
is based on the LEP-averaged PO.
Since PO are determined by fitting RO, one has to clarify what is 
actually taken from the SM, such as 
imaginary parts,
parts which have been moved to interference terms and photon-exchange 
terms making the MI results dependent on the SM.

An alternative way to extract SM parameters from the LEP measurements
is to average directly the SM parameters which have been
obtained by each experiment through a SM fit to its own RO.  
This alternative should also be pursued by the experiments.

The 2f emergency kit is illustrated in \fig{kit}.

\section{Exegesis of comparison}

All experimental informations are passed through {\em decoders} like
{\tt TOPAZ0} and {\tt ZFITTER}. Therefore, it is of the upmost
importance to critically upgrade and to compare the complete {\tt TOPAZ0} and 
{\tt ZFITTER} predictions~\cite{kn:PCP} with a particular emphasis on 
demanding the  following criteria:

\begin{itemize}
  
\item Comparisons, after the upgrading, should be consistently better
  than what they where in earlier studies.
\item At the peak all relative deviations among total cross-sections
  and absolute deviations among asymmetries should be below $0.1$
  per-mill.
\item At the wings, typically $\sqrt{\sman} = \mz \pm 1.8\,\GeV$, they
  should be below $0.3$ per-mill.
\end{itemize}

It is important to observe that today's comparisons are at the level of
$10^{-4}$, never been attempted before.  

\section{PO, RO and all that}

Within the context of the SM the RO are described in terms of some set
of amplitudes
%--
\bq
A_{_{\rm SM}} = A_{\ph} + A_{_{\zb}} + \mbox{non-factorizable},
\eq
%--
where the last term is due to all those contributions that do not
factorize into the Born-like amplitude.
Once the matrix element $A_{_{\rm SM}}$ is computed, squared and 
integrated, a convolution with initial- and final-state
QED and final-state QCD radiation follows:
%--
\bq
\sigma(\sman) = \int d\zvar~H_{\rm in}(\zvar,\sman)\,H_{\rm fin}(\zvar,\sman)\,
{\hat\sigma}(\zvar\sman),
\eq
%--
After de-convoluting RO of QED and QCD radiation aN set of approximations 
transform RO into PO.

\section{The inner circle}

The effective couplings $\in C$, due to the  imaginary parts of the 
diagrams. 
In the past this fact had some relevance only for realistic
distributions while for PO they were {\em conventionally} taken to be real.  
The above convention has changed lately with the introduction of 
next-to-leading (NLO) corrections: imaginary parts, although not NLO in a 
strict sense, are sizeable two-loop effects.  

The explicit formulae for the $\zb\ff\barf$ vertex are always written
starting from 
a Born-like form of a pre-factor $\times$ fermionic current, where 
the Born parameters are promoted to effective, scale-dependent parameters,
%--
\bq
\rZf\gadu{\mu}\,\lrbr \lpar \tcif + \ib\,a_{\ssL}\rpar \gdp 
- 2\,\qf\kZdf{\ff} \smans + \ib\,a_{\ssQ}\rrbr
= \gadu{\mu}\,\bigl(\Gvf + \Gaf\,\gfd\bigr),
\label{prototype}
\eq
%--
where $\gdp=1+\gfd$ and $a_{\ssQ,\ssL}$ are the SM imaginary parts.
By definition, the total and partial widths of the $\zb$ boson include
also QED and QCD corrections. The partial decay width is therefore
described by the following expression:
%--
\bq
\gff\equiv\Gamma\lpar \zb\to\ff\barf\rpar = 4\,\cf\,\Gamma_0\,\bigl(
|\Gvf|^2\,R^{\ff}_{\ssV} + |\Gaf|^2\,R^{\ff}_{\ssA}\bigr) 
+\Delta_{_{\rm EW/QCD}}\;,
\label{defgammaf}
\eq
%--
where $\cf = 1$ or $3$ for leptons or quarks $(\ff=\fl,\fq)$, and
$R^{\ff}_{\ssV}$ and $R^{\ff}_{\ssA}$ describe the final state QED and
QCD corrections and take into account the fermion mass $\mf$.  
The last term,
%--
\bq \Delta_{_{\rm EW/QCD}}= 
 \Gamma^{(2)}_{_{\rm EW/QCD}} - 
 \frac{\als}{\pi}\,\Gamma^{(1)}_{_{\rm EW}}\;, 
\eq 
%--
accounts for the non-factorizable corrections.  
The standard partial width, $\Gamma_0$, is
%--
\bq
\Gamma_0 = {{\gf\mzc}\over {24\srt\,\pi}} = 82.945(7)\,\MeV.
\eq
%--
The peak hadronic and leptonic cross-sections are defined by
%--
\bq
\sigma^0_{\had} = 12\pi\,\frac{\gel\gh}{\mzs\gzs} \qquad
\sigma^0_{\ell} = 12\pi\,\frac{\gel\gl}{\mzs\gzs} \;,
\eq
%--
where $\gz$ is the total decay width of the $\zb$ boson, i.e, the sum
of all partial decay widths.
The effective electroweak mixing angles ({\em effective sinuses}) are
always defined by
%--
\bq
4\,|\qf|\seffsf{\ff} = 1-\frac{\Reb\;\Gvf}{\Reb\;\Gaf} =
1-\frac{\gvf}{\gaf}
\;,
\label{defeffsin}
\eq
%--
where we define
%--
\bq
\gvf = \Reb\;\Gvf, \qquad \gaf = \Reb\;\Gaf.
\eq
%--
The forward-backward asymmetry $\afba{}$ is defined via 
%--
\bq
\afba{}=\frac{\sigma_{_{\rm{F}}}-\sigma_{_{\rm{B}}}}
             {\sigma_{_{\rm{F}}}+\sigma_{_{\rm{B}}}}\;,
\qquad
\sigma_{_{\rm{T}}}=\sigma_{_{\rm{F}}}+\sigma_{_{\rm{B}}}\;,
\eq
%--
where $\sigma_{_{\rm{F}}}$ and $\sigma_{_{\rm{B}}}$ are the cross
sections for forward and backward scattering, respectively.  
Before analyzing the forward-backward asymmetries we have to describe the
inclusion of imaginary parts. $\afba{}$ is calculated as
%--
\bq
\afba{} = \frac{3}{4}\frac{\sigma_{_{\rm VA}}}{\sigma_{_{\rm T}}}\;,
\eq
%--
where
%---
\bqa
\sigma_{_{\rm VA}} &=& \frac{\gf\mzs}{\sqrt{2}}\,\sqrt{\rhoe\rhof}\,
\qe\qf\Reb\Bigl[\alpha^*(\mzs)\Gve\Gaf\chi(\sman)
\Bigr]  
\nl{}&&
+\frac{\gf^2\mzq}{8\,\pi}\,\rhoe\rhof 
\Reb\Bigl[\Gve\lpar\Gae\rpar^*\Bigr] 
\Reb\Bigl[\Gvf\lpar\Gaf\rpar^*\Bigr] 
\sman\,|\chi(\sman)|^2.
\nl
\label{sva}
\eqa
%--
This result is valid in the realization where $\rhof$ is a real
quantity, i.e., the imaginary parts are not re-summed in $\rhof$.  In
this case
%--
\bqa
\Gvf &=& \Reb\lpar \Gvf\rpar + \ib\,\Imb \lpar \Gvf\rpar =
\gvf + \ib\,\Imb \lpar \Gvf\rpar, 
\nl
\Gaf &=& \tcif + \ib\,\Imb \lpar \Gaf\rpar.
\eqa
%--
Otherwise $\Gaf = \tcif$ is a real quantity but $\rhof$ is complex
valued and \eqn{sva} has to be changed accordingly, i.e., we introduce
%--
\bq
\gvf = \sqrt{\rhof}\,\vc{\ff}\;, 
\qquad 
\gaf = \sqrt{\rhof}\,\tcif\;,
\eq
%--
with
%--
\bq
\vc{\ff}=\tcif-2\qf\seffsf{\ff}\;.
\eq
%--
For the peak asymmetry, the presence of $\rho$'s is irrelevant since
they will cancel in the ratio. We have
%---
\bq
\hat\afba{{{0}}\rm f} = \frac{3}{4}\,\hat A_{\fe}\,\hat A_{\ff},
\qquad
\hat A_{\ff} =
\frac{2\,\Reb\bigl[\Gvf\lpar\Gaf\rpar^*\bigr]}
{\bigl(\big|\Gvf\big|^2 + \big|\Gaf\big|^2\bigr)}\;.
\label{defasym}
\eq
%---
The question is what to do with imaginary parts in \eqn{defasym}.  For
partial widths, as they absorb all corrections, the convention is to
use
%--
\bq
\big|\Gvaf\big|^2 =\bigl(\Reb\Gvaf\bigr)^2 + \bigl(\Imb\Gvaf\bigr)^2.
\eq
%--
On the contrary, the PO peak asymmetry $\afba{0\rm f}$ will be defined by
an analogy of equation \eqn{defasym} where {\em conventionally} imaginary
parts are not included
%---
\bq
\afba{0\rm f} =\frac{3}{4}\,{\cal A}_{\fe}\,{\cal A}_{\ff},
\qquad
{\cal A}_{\ff} =\frac{2\bigl(\gvf\gaf\bigr)}{\bigl(\gvf\bigr)^2 
+ \bigr(\gaf\bigr)^2}\;.
\label{defasymre}
\eq
%---
A definition of the PO heavy quark forward-backward asymmetry
parameter which would include mass effects is
%--
\bq
{\cal A}_{\ffb} = {{2\,\gvb\gab}\over {\frac{1}{2}\,\lpar 3 - \beta^2\rpar 
(\gvb)^2 + \beta^2\,(\gab)^2}}\,\beta,
\label{ab}
\eq
%--
where $\beta$ is the $\ffb$-quark velocity.  
The difference is very small, due to an accidental cancellation of the
mass corrections between the numerator and denominator of \eqn{ab}.
This occurs for down quarks where $(\gvb)^2 \approx (\gab)^2/2$ and
where
%--
\bqa
\afba{0\rm b} & \approx &
\frac{3}{4}\,\frac{2\,\gve\gae}{(\gve)^2+(\gae)^2}\,
\frac{2\,\gvb\gab}{(\gvb)^2 + (\gab)^2}\,\lpar 1 + \delta_{\rm mass}\rpar,  
\nl
\nl
\nl
\delta_{\rm mass}&\approx& 4\,\frac{\mqs}{\sman}{{(\gab)^2/2 - (\gvb)^2}
\over{(\gvb)^2 + (\gab)^2}}.
\eqa
%---
Therefore, our definition of the PO forward--backward asymmetry and coupling
parameter will be as in \eqn{defasymre}.

\section{Status of art}

The most important upgradings in the PO calculations 
consist of the inclusion of 
higher-order QCD corrections, 
mixed electroweak-QCD corrections~\cite{kn:CK}, 
NLO two-loop corrections of $\ord{\alpha^2\mts}$~\cite{kn:DG}.
The only case that is not covered is the one of final $\ffb$-quarks,
because it involves non-universal $\ord{\alpha^2\mts}$ vertex
corrections.

Another development in the computation of radiative corrections to the
hadronic decay of the $\zb$ is contained in two papers~\cite{kn:TP} , which 
together provide complete corrections of $\ord{\alpha\als}$ to $\Gamma(\zb\to
\fq\barq)$ with $\fq=\fu,\fd,\fs,\fc$ and $\ffb$.  

\section{Improved I/O Parameters}

Since all Renormalization Schemes (RS) use $\gf$, we refer to the recent work 
of~\cite{kn:RS} giving an improved value of $\gf= 1.16637(1)\times 10^{-5}\,
\GeV^{-2}$.
An important issue concerns the evaluation of $\alpha_{\rm QED}$ at
the mass of the $\zb$. Define
%--
\bq
\alpha(\mz) = \frac{\alpha(0)}{1-\Delta\alpha^{(5)}(\mz) 
-\Delta_{\rm{top}}(\mz)-\Delta^{\alpha\als}_{\rm{top}}(\mz)}\;, 
\eq
%--
where one has 
\bq
\Delta\alpha^{(5)}(\mz) = \Delta\alpha_{\rm{lept}} +
\Delta\alpha^{(5)}_{\rm{had}}.  
\eq
%--
The input parameter is now $\Delta\alpha^{(5)}_{\rm{had}}$, 
as it is the contribution with the largest uncertainty, while 
the calculation of the top contributions to $\Delta\alpha$ is left for 
the code.  
Next, include for $\Delta\alpha_{\rm{lept}}$ the $\ord{\alpha^3}$
terms of~\cite{kn:S} 
and use as default $\Delta\alpha^{(5)}_{\rm{had}} = 0.0280398$, taken 
from~\cite{kn:EJ}. 
Therefore $1/\alpha^{(5)}(\mz) = 128.877$, to which one must
add the $\ft\bart$ contribution and the $\ord{\alpha\als}$ correction
induced by the $\ft\bart$ loop with gluon exchange~\cite{kn:K}. 

\section{MI Calculus}

To summarize the MI ansatz, one starts with the SM, which introduces
complex-valued couplings, calculated to some order in perturbation
theory; next we define $\gvf,\gaf$ as the real parts of the effective
couplings and 
$\Gamma_{\ff}$ as the physical partial width absorbing
all radiative corrections including the imaginary parts of the
couplings and fermion mass effects.  
Furthermore,
%--
\bq
R_{\fq}=\frac{\Gamma_{\fq}}{\gh}\;,
\qquad
R_{\fl}=\frac{\gh}{\Gamma_{\fl}}\;,
\eq
%--
for quarks and leptons, respectively.  

The experimental collaborations report PO for the following sets:
%--
\bq
\bigl(R_{\ff},\afba{0,\ff}\bigr), 
\qquad 
\bigl(\gvf,\gaf\bigr), 
\qquad
\bigl(\seffsf{\ff},\rho_{\ff}\bigr).
\eq
%--
In order to extract $\gvf,\gaf$ from $\Gamma_{\ff}$ one has to get the
SM-remnant from \eqn{defgammaf}, all else is trivial.  
However, the parameter transformation cannot be completely MI, due to 
the residual SM dependence appearing inside \eqn{defgammaf}.  
In conclusion, the flow of the calculation requested by the
experimental Collaborations is:

\ben
  
\item pick the Lagrangian parameters $\mt,\mh$ etc. for the explicit
  calculation of the residual SM-dependent part;
\item perform the SM initialization of everything, such as imaginary
  parts etc.  giving, among other things, the complement
  ${\overline{\rm SM}}$;
\item select $\gvf,\gaf$;
\item perform a SM-like calculation of $\Gamma_f$, but using arbitrary
  values for $\gvf, \gaf$, and only the rest, namely
%--
\bq
R^{\ff}_{\ssV}, \quad R^{\ff}_{\ssA}\;,  
\qquad
\Delta_{_{\rm EW/QCD}}\;,  
\qquad
\Imb\;\Gvf, \quad \Imb\;\Gaf\;,
\eq
%--
from the SM.

\een

An example of the parameter transformations is the following: starting
from $\mz, \gz, R_{\fe,\flm,\flt}$ and $\afba{0,\fe,\flm,\flt}$ we
first obtain
%--
\bq
\gel = \mz\gz\,\biggl[{{\sigma^{0}_{\had}}\over{12\,\pi R_{\fe}}}\biggr]^{1/2},
\qquad
\gh  = \mz\gz\,\biggl[{{\sigma^{0}_{\had}R_{\fe}}\over {12\,\pi}}\biggr]^{1/2}.
\eq
%---
With 
%--
\bq
{\cal A}_{\fe} = \frac{2}{\sqrt{3}}\sqrt{\afba{0,\fe}}\;, 
\qquad 
\mbox{and} \quad \gamma = \frac{\gf\mzc}{6\,\srt\,\pi}\;,
\eq
%--
we subtract QED radiation,
%--
\bq
\gel^0 = \frac{\gel}{\ds{1+\frac{3}{4}\,\frac{\alpha(\mzs)}{\pi}}}\;,
\eq
%--
and get
%---
\bqa
\seffsf{\fe} &=& \frac{1}{4}\,
\lpar 1 + {{\sqrt{1-{\cal A}^2_{\fe}}-1}\over{{\cal A}_{\fe}}}\rpar,  
\nl
\rho_{\fe} &=& \frac{\gel^0}{\gamma}\,
\biggl[\lpar\frac{1}{2}-2\,\seffsf{\fe}\rpar^2 
+\frac{1}{4} + \lpar\Imb\;\Gve\rpar^2 + \lpar\Imb\;\Gae\rpar^2 \biggr]^{-1}.
\qquad
\eqa
%---
With
%--
\bq
{\cal A}_{\ff} = \frac{4}{3}\,{{\afba{0,\ff}}\over{{\cal A}_{\fe}}}\;,
\eq
%--
we further obtain
%---
\bqa
\seffsf{\ff} &=& \frac{1}{4\,\big|\qf\big|}\,
\lpar 1 + {{\sqrt{1 - {\cal{A}}^2_{\ff}}-1}\over{{\cal{A}}_{\ff}}}\rpar,  
\nl
\rho_{\ff} &=& \frac{\Gamma_f^0}{\gamma}\,
\biggl[\lpar\frac{1}{2}-2|\qf|\seffsf{\ff}\rpar^2 
+\frac{1}{4} + \lpar\Imb\;\Gvf\rpar^2 + \lpar\Imb\;\Gaf\rpar^2 \biggr]^{-1},
\qquad
\eqa
%---
where $\ff = \flm,\flt$. 
For quarks one should remember to subtract first
non-factorizable terms and then to distinguish between
$R^{\ff}_{\ssV}$ and $R^{\ff}_{\ssA}$.

\section{Results for PO}

Having established a common input parameter set (IPS) we now turn to
discussing the results for pseudo-observables (PO).
The full list contains more PO and is given in \tabn{tabPO}, where
we have included the relative and absolute difference between {\tt
TOPAZ0} and {\tt ZFITTER} in units of per-mill.
%--

\begin{table}[t]
\begin{center}
\begin{tabular}{|c|c|c|c|}
\hline
Observable &{\tt TOPAZ0}&{\tt ZFITTER}&$10^3\times\frac{T-Z}{T}$\\
\hline
\hline
$1/\alpha^{(5)}(\mz)$  &  128.877 & 128.877 &       \\
\hline
$1/\alpha(\mz)$        &  128.887 & 128.887 &       \\
\hline
$\mw\,$[GeV]           &  80.3731 & 80.3738 & -0.009\\
\hline
$\sigma^0_{\had}\,$[nb]&  41.4761 & 41.4777 & -0.04 \\
\hline
$\sigma^0_{\lep}\,$[nb]&  1.9995  & 1.9997  & -0.12 \\
\hline
$\Gamma_{\had}\,$[GeV] &  1.74211 & 1.74223 & -0.07 \\
\hline
$\gz\,$[GeV]           &  2.49549 & 2.49573 & -0.10 \\
\hline
$\gn\,$[MeV]           &  167.207 & 167.234 & -0.16 \\
$\gel\,$[MeV]          &   83.983 &  83.995 & -0.14 \\
$\gmu\,$[MeV]          &   83.983 &  83.995 & -0.14 \\
$\gt\,$[MeV]           &   83.793 &  83.805 & -0.14 \\
$\gu\,$[MeV]           &  300.129 & 300.154 & -0.08 \\
$\gd\,$[MeV]           &  382.961 & 382.996 & -0.09 \\
$\gc\,$[MeV]           &  300.069 & 300.092 & -0.08 \\
$\gbq\,$[MeV]          &  375.997 & 375.993 &  0.01 \\
$\gi\,$[GeV]           &  0.50162 & 0.50170 & -0.16 \\
\hline
$R_l$                  &  20.7435 & 20.7420  & 0.07 \\
$R^0_b$                & 0.215829 & 0.215811 & 0.08 \\
$R^0_c$                & 0.172245 & 0.172246 &-0.01 \\
\hline
$\seffsf{\rm{lept}}$   & 0.231596 & 0.231601 &-0.02 \\
$\seffsf{\ffb}$        & 0.232864 & 0.232950 &-0.37 \\
$\seffsf{\fc}$         & 0.231491 & 0.231495 &-0.02 \\
\hline
$\rhoe$                & 1.00513  & 1.00528  &-0.15 \\
$\rhoi{\ffb}$          & 0.99413  & 0.99424  &-0.11 \\
$\rhoi{\fc}$           & 1.00582  & 1.00598  &-0.16 \\
\hline
\hline
Observable & {\tt TOPAZ0} & {\tt ZFITTER}    &$10^3\times(T-Z)$\\
\hline
\hline
$\afba{0,\rm l}$       & 0.016084 & 0.016074 & 0.01 \\
$\afba{0,\ffb}$        & 0.102594 & 0.102617 &-0.02 \\
$\afba{0,\fc}$         & 0.073324 & 0.073300 & 0.02 \\
\hline
${\cal A}_{\fe}$       & 0.146440 & 0.146396 & 0.04 \\
${\cal A}_{\ffb}$      & 0.934654 & 0.934607 & 0.05 \\
${\cal A}_{\fc}$       & 0.667609 & 0.667595 & 0.01\\
\hline
\end{tabular}
\end{center}
\vspace*{3mm}
\caption[]{
Complete table of PO, from {\tt TOPAZ0} and {\tt ZFITTER}.
\label{tabPO}}
%\vspace*{-3cm}
\end{table}

\section{Theretical Uncertainties for PO}

Here we discuss the theoretical uncertainties associated with PO.  In
\tbns{tabTH1}{tabTH2}. We give the central value, the minus error and
the plus error as predicted by {\tt TOPAZ0} and compare with the
current total experimental error if available. The procedure is
straightforward: both codes have a preferred calculational setup and
options to be varied, options having to do with the remaining
theoretical uncertainties and the corresponding implementation of
higher order terms.  To give an example, we have now LO and NLO
two-loop EW corrections but we are still missing the NNLO ones and
this allows for variations in the final recipe for $\rho_f$, etc.

{\tt TOPAZ0} has been run over all the remaining (after implementation
of NLO) options and all the results for PO have been collected.  We
will use
\begin{itemize}

\item[--] {\em central} for PO evaluated at the preferred setup;
  
\item[--] minus error for $\mbox{PO}_{\it central} - \min_{\rm
    opt}\,\mbox{PO}$;
  
\item[--] plus error for $\max_{\rm opt}\,\mbox{PO} - \mbox{PO}_{\it
    central}$.

\end{itemize}
%--
\begin{table}[h]
\begin{center}
\begin{tabular}{|c|c|c|c|c|}
\hline
Observable       & central & minus error & plus error & total exp. error \\
\hline
                       &          &         &            & \\
$1/\alpha^{(5)}(\mz)$  &  128.877 & -       &  -         & \\
                       &          &         &            & \\
$1/\alpha(\mz)$        &  128.887 & -       &  -         & \\
                       &          &         &            & \\
\hline
                       &          &         &            & \\
$\mw\,$[GeV]           &  80.3731 & 5.8 MeV &  0.3 MeV   & 90 MeV \\
$\sigma^0_{\had}\,$[nb]&  41.4761 & 1.0 pb  &  1.6 pb    & 58 pb \\
\hline                                   
\hline                                   
                       &          &         &            &  \\
$\gn\,$[MeV]           &  167.207 & 0.017   &  0.001     &\\
$\gel\,$[MeV]          &   83.983 & 0.010   &  0.0005    & 0.10$^*$ \\
$\gmu\,$[MeV]          &   83.983 & 0.010   &  0.0005    &\\
$\gt\,$[MeV]           &   83.793 & 0.010   &  0.0005    &\\
$\gu\,$[MeV]           &  300.129 & 0.047   &  0.013     &\\
$\gd\,$[MeV]           &  382.961 & 0.054   &  0.010     &\\
$\gc\,$[MeV]           &  300.069 & 0.047   &  0.013     &\\
$\gbq\,$[MeV]          &  375.997 & 0.208   &  0.077     &\\
$\Gamma_{\rm{had}}\,$[GeV]&1.74211& 0.26 MeV&  0.11 MeV  &  2.3 MeV$^*$\\
$\gi\,$[GeV]           &  0.50162 & 0.05 MeV& 0.002 MeV  &  1.8 MeV$^*$\\
$\gz\,$[GeV]           &  2.49549 & 0.34 MeV&  0.11 MeV  &  2.4 MeV    \\
\hline
\end{tabular}
\end{center}
\vspace*{3mm}
\caption[]{
Theoretical uncertainties for PO from {\tt TOPAZ0}. $*)$ assumes lepton
universality.
\label{tabTH1}}
\end{table}

%--
\begin{table}[t]
\begin{center}
\begin{tabular}{|c|c|c|c|c|}
\hline
Observable  & central & minus error & plus error & total exp. error \\
\hline
                       &          &          &           &          \\
$R_l$                  &  20.7435 & 0.0020   & 0.0013    & 0.026    \\
$R^0_b$                & 0.215829 & 0.000100 & 0.000031  & 0.00074  \\
$R^0_c$                & 0.172245 & 0.000005 & 0.000024  & 0.0044   \\
\hline
                       &          &          &           &          \\
$\seffsf{\rm{lept}}$   & 0.231596 & 0.000035 & 0.000033  & 0.00018  \\
%                         0.0010(0.00029$^\dagger)$       \\
$\seffsf{\ffb}$        & 0.232864 & 0.000002 & 0.000048  &          \\
$\seffsf{\fc}$         & 0.231491 & 0.000029 & 0.000033  &          \\
\hline
                       &          &          &           &          \\
$\afba{0,\rm l}$       & 0.016084 & 0.000057 & 0.000060  & 0.00096  \\
$\afba{0,\ffb}$        & 0.102594 & 0.000184 & 0.000195  & 0.0021   \\
$\afba{0,\fc}$         & 0.073324 & 0.000142 & 0.000149  & 0.0044   \\
\hline
                       &          &          &           &          \\
${\cal A}_{\fe}$       & 0.146440 & 0.000259 & 0.000275  & 0.0051   \\
${\cal A}_{\ffb}$      & 0.934654 & 0.000032 & 0.000005  & 0.035    \\
${\cal A}_{\fc}$       & 0.667609 & 0.000114 & 0.000103  & 0.040    \\
\hline
                       &          &          &           &          \\
$\rhoe$                & 1.00513  & 0.00010  & 0.000005  & 0.0012   \\
$\rhoi{\ffb}$          & 0.99413  & 0.00048  & 0.000001  &          \\
$\rhoi{\fc}$           & 1.00582  & 0.00010  & 0.000005  &          \\
\hline
\end{tabular}
\end{center}
\vspace*{3mm}
\caption[]{
Theoretical uncertainties for PO from {\tt TOPAZ0}. 
%$\dagger$ is the error 
%for $\seffsf{\rm{lept}}$ from SLD $\alr$ measurement.
}
\label{tabTH2}
\end{table}

\section{Realistic Observables}

The RO are computed in the context of the SM, however one of the goals 
will be to pin down the definition of PO;
the calculation of RO in terms of the defined PO for the
purpose of MI fits, showing that for PO with values as calculated
in the SM, the RO are {\em by construction} identical to the full
SM RO calculation.
The last point requires expressing $\rho$'s and effective mixing
angles in terms of PO, assuming the validity of the SM.  
One should remember that gauge invariance (GI) at the $\zb$
pole (on-shell GI) is entirely another story from GI
at any arbitrary scale (off-shell GI).  

Some of the re-summations that are allowed at the pole and that heavily
influence the definition of effective $\zb$ couplings are not
trivially extendible to the off-shell case. 
Therefore, the expression for RO = RO(PO), at arbitrary $\sman$, 
requires a careful examination and should be better understood as 
%--
\bq
{\rm RO} = {\rm RO}\lpar {\rm PO}, {\overline{\rm SM}}\rpar, 
\eq
%--
that is, for example:
%--
\bqa
\sigma_{\rm MI} = \sigma_{\rm SM}\lpar R_{\fl},\afba{0,l},\dots\right. 
&\to& \left. \gvf,\gaf \right.
\nl
&\to& \left. \rho_f,\seffsf{\ff}; \mbox{residual SM}\rpar.
\nl
\label{defmiapp}
\eqa
%--
As long as the procedure does not violate GI and the PO
are given SM values, there is nothing wrong with the calculations.  It
is clear that in this case the SM RO coincide with the MI RO.

\section{Final-State Radiation}

One should realize that $\smanp$ ($\smanp$ being
the centre-of-mass energy of the $\fep\fem$ system after initial state 
radiation) is not equivalent to the invariant
mass of the final-state $\ff\barf$ system, $\Mlones(\ff\barf)$, due to
final state QED and QCD radiation.  
Furthermore, in the presence of a
$\smanp$-cut the correction for final state QED radiation is simply
%--
\bq
R^{\rm FS}_{\rm{QED}} = 
  \frac{3}{4}\,\qfs\,\frac{\alpha(\sman)}{\pi}\;,
\eq
%--
while for a $\Mlones$-cut the correction is more complicated, see~\cite{kn:MNP}.

For full angular acceptance one derives the following corrections:
%---
\bqas 
\sigma\lpar\sman\rpar &=&  
\frac{\alpha}{4\pi}\qfs\sigma^0\lpar{\sman}\rpar
\biggl\{-2\xvars 
+4\biggl[
\lpar\zvar+\frac{\zvars}{2}+2\ln\xvar\rpar\ln\frac{\sman}{\mfs}
\\{}&&
+\zvar\lpar 1 +\frac{\zvar}{2}\rpar\ln\zvar 
+2\ztwo-2\li{2}{\xvar}
-2\ln\xvar + \frac{5}{4} 
- 3\zvar - \frac{\zvars}{4}   
\biggr]
\biggr\}.
\label{imc1}
\nn
\eqas
%--
Here we have introduced
%--
\bq
\xvar=1-\zvar,\qquad
\zvar=\Mlones\lpar\ff\barf\rpar/\sman. 
\eq
%--
For an $\smanp$-cut both QED and QCD final-state radiation are
included through an inclusive correction factor.  
For $\Mlones$-cut
the result remains perfectly defined for leptons, however for hadronic
final states there is a problem.  This has to do with QCD final-state
corrections. Indeed we face the following situation:

\ben
  
\item for $\fep\fem\to\ff\barf \gamma$ the exact correction factor is
  known at $\ord{\alpha}$ even in the presence of a
  $\Mlones(\ff\barf)$ cut;
\item the complete set of final-state QCD corrections are known up to
  $\ord{\als^3}$~\cite{kn:CET} 
only for the fully inclusive setup, i.e., no cut on the $\ff\barf$
 invariant mass;
\item the mixed two-loop QED/QCD final-state corrections are also
  known only for a fully inclusive setup~\cite{kn:KAT} 

\item at $\ord{\als}$ QCD final-state corrections in presence of a
  $\Mlones$-cut follow from the analogous QED calculation of\cite{kn:ABL}.

\een

\section{Convoluted Realistic Observables}

We have the following options:

\begin{itemize}  

\item[CA3] Complete RO, with QED initial state radiation implemented
  through an additive $\ord{\alpha^3}$ radiator~\cite{kn:PV} 
\item[CF3] Complete RO, with QED initial state radiation implemented
  through a factorized $\ord{\alpha^3}$ radiator~\cite{kn:YFS} 
\end{itemize}

The following two equations define cross-sections and forward-backward 
asymmetry {\em convoluted} with ISR:
%--
\bq
\sigma_{_{\rm T}}\lpar\sman\rpar = \int^1_{\zvari{_0}}\,d\zvar 
\Fluxf{\zvar}{\sman}{\hat\sigma}_{_{\rm T}}\bigl(\zvar\sman\bigr),
\label{fluxtotal}
\eq
%--
where $\zvari{_0}=\smani{0}/\sman$ and 
%--
\bq
\afb\lpar\sman\rpar=\frac{\pi\alpha^2\qes\qfs}{\sigma_{\rm tot}}\,
\int^1_{\zvari{_0}}\,d\zvar
\frac{1}{\lpar 1+\zvar\rpar^2}\,H_{_{\rm FB}}\lpar\zvar;\sman\rpar\,
{\hat\sigma}_{_{\rm FB}}\bigl(\zvar\sman\bigr), 
\eq
%--
Note that the so-called {\em radiator} (or {\em flux function}),
$\Fluxf{\zvar}{\sman}$, is known up to terms of order $\alpha^3$ while
$H_{_{\rm FB}}$ is only known up to terms of order $\alpha^2$.
The kernel cross-sections ${\hat\sigma}_{_{\rm T,FB}}$ should be
understood as 
the improved Born approximation (IBA), 
corrected with all electroweak and possibly all 
FSR (QED $\,\otimes\,$QCD) corrections,
where all coupling constants and effective vector and axial weak 
couplings are {\em running}, i.e., they depend on $\smanp=\zvar\sman$. 

\section{Uncertainty on QED Convolution}

We define the following quantities:
%--
\bq
\delta^{\rm dec}\lpar O\rpar = \frac{O}{O^{\rm SD}} - 1, \qquad
\Delta^{\rm dec}\lpar O\rpar = O - O^{\rm SD}. 
\eq
%--
For convenience of the reader we have reproduced in \tabn{tab8} the
results for the CA3 and CF3 mode.
%--
From \tabn{tab8} we derive the absolute differences, for
$\afba{\flm}$, and the relative ones, for cross-sections, between
additive and factorized versions of the QED radiators. They are shown
in \tabn{tab9}.

\section{Initial-Final QED Interference}

If initial-final QED (ISR-FSR) interference (IFI) is included in a
calculation we have a conceptual problem with the meaning of the
$\smanp$-cut.  
In this case the definition of the variable $\smanp$ is
unnatural since one does no longer know the origin of the radiative
photon (ISR or FSR).  
Only a cut on the invariant mass of the final-state $\ff\barf$ system 
would make sense.
There is another option: to select events with little initial-state
radiation one can use a cut on the acollinearity angle $\theta_{\rm
  acol}$, between the outgoing fermion and anti-fermion.  
A cut on $\theta_{\rm acol}$ is roughly equivalent to a cut on the 
invariant mass of the $\ff\barf$ system, indeed one may write
%--
\bq
\frac{\smanp}{\sman}\approx\zvari{\rm eff} =  
{{1 - \sin(\theta_{\rm acol}/2)}\over{1 + \sin(\theta_{\rm acol}/2)}}\;,
\label{xeff}
\eq
%--
therefore a cut of $\theta_{\rm acol} < 10^{\circ}$ is roughly
corresponding to the request that $\smanp/\sman > 0.84$.
The inclusion of initial-final QED interference in {\tt TOPAZ0} and in
{\tt ZFITTER} is done at $\ord{\alpha}$, i.e., the $\ord{\alpha}$
interference term is added linearly to the cross-sections without
entering the convolution with ISR and without cross-talk to FSR.  
For loose cuts the induced uncertainty is rather small and the effect of
the interference itself is minute.  
On the contrary, when we select a tight acollinearity cut the resulting 
limit on the energy of the emitted photon becomes more stringent and the
effect of interference grows.  
The corresponding theoretical uncertainty, due to missing higher-order 
corrections, is therefore expected to be larger.

I/F INT is currently treated dubiously by the experiments:
The MC used to extrapolate for efficiency and possibly acceptance do
not contain I/F interference (eg, KORALZ in multi-photon mode).  
Thus extrapolated and quoted results somehow miss I/F interference.
There is a claim of some serious discrepancy between 
{\tt KK}~\cite{kn:KKMC} / {\tt TOPAZ0} / {\tt ZFITTER} in $\sigma_{\rm int}, \afba{\rm int}$ 
for muons 
(see http://home.cern.ch/jadach)
It has to do with {\em exponentiation} of {\tt ISR} $\,\otimes\,$ 
{\tt FSR}.

\begin{table}[ht]
\begin{center}
\begin{tabular}{|c||c|c|c|c|c|}
\hline
& \multicolumn{5}{c|}{LEP\,1 energy in GeV} \\
\cline{2-6}
& $\mz - 3$ & $\mz - 1.8$ & $\mz$ & $\mz + 1.8$ & $\mz + 3$  \\
\hline 
\hline
$\delta^{\rm dec}(\sigma^{\rm CA3}_{\flm})$ {\tt T}  &
    -23.976  &   -27.616  &   -26.257  &     5.356  &    30.665  \\
$\delta^{\rm dec}(\sigma^{\rm CA3}_{\flm})$ {\tt Z}  &
    -24.007  &   -27.629  &   -26.261  &     5.341  &    30.631  \\
$\delta^{\rm dec}(\sigma^{\rm CF3}_{\flm})$ {\tt T}  &
    -23.973  &   -27.611  &   -26.253  &     5.364  &    30.671  \\
$\delta^{\rm dec}(\sigma^{\rm CF3}_{\flm})$ {\tt Z}  &
    -24.000  &   -27.624  &   -26.256  &     5.348  &    30.637  \\
\hline \hline
$\delta^{\rm dec}(\sigma^{\rm CA3}_{\rm had})$ {\tt T}  &
    -25.867  &   -28.501  &   -26.537  &     4.952  &    30.564  \\
$\delta^{\rm dec}(\sigma^{\rm CA3}_{\rm had})$ {\tt Z} &
    -25.873  &   -28.503  &   -26.538  &     4.945  &    30.550  \\
$\delta^{\rm dec}(\sigma^{\rm CF3}_{\rm had})$ {\tt T}  &
    -25.863  &   -28.497  &   -26.533  &     4.959  &    30.572  \\
$\delta^{\rm dec}(\sigma^{\rm CF3}_{\rm had})$ {\tt Z} &
    -25.869  &   -28.499  &   -26.533  &     4.953  &    30.558  \\
\hline \hline
$\Delta^{\rm dec}(\afba{\mu\rm CA3})$ {\tt T}  &
     -2.190  &    -1.967  &    -1.804  &    -6.292  &   -11.486  \\
$\Delta^{\rm dec}(\afba{\mu\rm CA3})$ {\tt Z} &
     -2.211  &    -1.975  &    -1.803  &    -6.287  &   -11.480  \\
$\Delta^{\rm dec}(\afba{\mu\rm CF3})$ {\tt T}  &
     -2.189  &    -1.966  &    -1.804  &    -6.292  &   -11.487  \\
$\Delta^{\rm dec}(\afba{\mu\rm CF3})$ {\tt Z} &
     -2.215  &    -1.977  &    -1.803  &    -6.286  &   -11.479  \\
\hline
\end{tabular}
\end{center}
\caption[]{RO: the effect in $\%$ of initial state QED radiation for CA3 and 
CF3 modes.}
\label{tab8}
\end{table}

%--
\begin{table}[ht]
\begin{center}
\begin{tabular}{|c||c|c|c|c|c|}
\hline
& \multicolumn{5}{c|}{LEP\,1 energy in GeV} \\
\cline{2-6}
& $\mz - 3$ & $\mz - 1$ & $\mz$ & $\mz + 1$ & $\mz + 3$  \\
\hline \hline
\multicolumn{6}{|c|}{$10^{4} \times\,$(fact/add-1)} \\
\hline \hline
$\sigma_{\flm}$
&  0.44 & 0.63 & 0.61 & 0.72 & 0.49 \\
&  0.88 & 0.63 & 0.68 & 0.72 & 0.49 \\
\hline
$\sigma_{\rm had}$  
&  0.58 & 0.58 & 0.64 & 0.73 & 0.59 \\
&  0.61 & 0.62 & 0.67 & 0.76 & 0.62 \\
\hline \hline
\multicolumn{6}{|c|}{fact-add [pb]} \\
\hline \hline
$\sigma_{\flm}$
&  0.01 &   0.03  &   0.09  &   0.05  &   0.02 \\
&  0.02 &   0.03  &   0.10  &   0.05  &   0.02 \\
\hline
$\sigma_{\rm had}$  
&  0.26 &   0.56  &   1.95  &   1.04  &   0.48 \\
&  0.27 &   0.60  &   2.04  &   1.08  &   0.51 \\
\hline \hline
\multicolumn{6}{|c|}{$10^{5} \times\,$(fact-add)} \\
\hline \hline
$\afba{\flm}$  
&  1.00 &  1.00 & 0.00 & 0.00 & -1.00 \\
& -4.00 & -2.00 & 0.00 & 1.00 &  1.00 \\
\hline 
\end{tabular}
\end{center}
\caption[]{Absolute and relative differences in {\tt TOPAZ0} and
in {\tt ZFITTER} for additive and factorized radiators.}
\label{tab9}
\end{table}

\section{Note added: upgrading of ISPP}

Here we discuss some recent work connected with Initial State Pair Production
(ISPP). We consider two different approaches:
the first is based on JMS~\cite{kn:JMS}, an ansatz for re-summation of both 
soft photons and soft lepton pairs. For hadron pairs it is corrected 
numerically with an uncertainty of $30\%$. 

All versions of {\tt TOPAZ0} $< 4.4$, use an effective, naive, implementation 
of ISPP ({\bf ONP = Y}) based on KKKS~\cite{kn:KKKS} interfaced with soft 
photon re-summation. The agreement with {\tt JMS} is perfect around the 
resonance but quickly deteriorates on the high energy side. 

In the latest version of {\tt TOPAZ0}, the variable {\bf ONP} has the 
additional value of  {\bf ONP = I}, where

\begin{itemize} 

\item KF approach~\cite{kn:KF} for virtual pairs and for soft and 
exponentiated pairs is used; it is also applicable to hadron pairs where we
use KKKS results for $\ord{\alpha^2}$ and write them in terms of moments. 
Then we match it to KF and generalize KF to soft and exponentiated hadrons 
pairs.

\item Next, we use generalized KF for virtual and soft pairs and cut to the 
same $\smanp$ cut of IS QED radiation. Once an $\smanp$ cut is introduced in 
one place then it should be used everywhere, even for photons + pairs. 

\item Finally, we use the soft approximation only up to some cut $\Delta$ that 
is compatible with $E \gg \Delta \gg 2\,m$. Above the cut we use 
KKKS-formalism but not added linearly to the cross-section; instead we use
a KKKS correction factor in convolution with IS QED radiation. The radiator 
used here is a leading-log (LL) radiator, evaluated at the second order or,
optionally, at (third order). 

\end{itemize}

Now, some comparison\footnote{{\tt JMS} numbers are a courtesy of B.~Pietrzyk 
and of G.~Quast}, the cross-sections at the MI-point  
$\mz= 91.1882\,$GeV,$\gz = 2.4952\,$GeV, $\sigma^0_{\had}= 41.560\,$nb
and $R_l = 20.728, \afba{{{0}}\rm l} = 0.017319$,
for ISPP\footnote{ Here ISPP means {\bf NON-SINGLET} Initial State Pair
Production. {\bf SINGLET} pairs are, in principle, subtracted together with 
two-photon contributions} on and off; ISPP effect was taken as the difference.
%--
\begin{table}[hp]\centering
\begin{tabular}{|c|c|c|c|}
\hline
$\sqrt{s}$[GeV] & {\tt JMS} & {\tt TOPAZ0}(ONP=Y) & {\tt TOPAZ0}(ONP=I)  \\
\hline
88.464  &        5.2297 & 5.2187 &  -        \\
        &        5.2170 & 5.2059 &  5.2063   \\
        &       -0.0127 &-0.0128 &  -0.0124   \\
\hline
89.455  &        10.121 &10.105  & -         \\
        &        10.094 &10.079  & 10.079    \\
        &        -0.027 &-0.026  & -0.026    \\
\hline
90.212  &        18.238 &18.227  & -         \\
        &        18.187 &18.178  & 18.179    \\
        &        -0.051 &-0.049  & -0.048    \\
\hline
91.207  &        30.529 &30.549  & -         \\
        &        30.449 &30.471  & 30.470    \\
        &        -0.080 &-0.078  & -0.079    \\
\hline
91.238  &        30.589 &30.609  & -         \\
        &        30.510 &30.531  & 30.530    \\
        &        -0.079 &-0.078  & -0.079    \\ 
\hline  
91.952  &        25.173 &25.166  & -         \\
        &        25.121 &25.117  & 25.112    \\
        &        -0.052 &-0.049  & -0.054    \\
\hline
92.952  &        14.503 &14.484  & -         \\
        &        14.488 &14.476  & 14.466    \\
        &        -0.015 &-0.008  & -0.018    \\ 
\hline
93.701  &        10.067 &10.051  & -         \\
        &        10.064 &10.059  & 10.046    \\
        &       -0.003  &0.008   &-0.005     \\
\hline
\hline
\end{tabular}
\vspace*{3mm}
\caption[]{Comparison of {\tt TOPAZ0} with {\tt JMS}-approach for the
inclusion of ISPP.}
\label{tabISPP}
\end{table}
\normalsize
%--
We report $\sigma_{\rm had}\,$[nb] in \tabn{tabISPP}:
first entry is without pairs, second entry is with pairs and
third entry is (with - without).

%--
\begin{table}[hp]\centering
\begin{tabular}{|c|c|c|c|c|}
\hline
$\sqrt{s}$[GeV] &  $\smanp/s = 0.01$ &  $\smanp/s = 0.1$ &  $\smanp/s = 0.5$ 
&  $\smanp/s = 0.9$\\
\hline
$\mz-3\,$[GeV]   &  4.45106 &  4.44444  &  4.43769  &  4.38335  \\
                 &  4.44069 &  4.43391  &  4.42705  &  4.37227  \\
                 & -2.33    & -2.37     & -2.40     & -2.52     \\
\hline
$\mz-1.8\,$[GeV] &  9.60037 &  9.59393  &  9.58742  &  9.52026  \\
                 &  9.57589 &  9.56929  &  9.56267  &  9.49497  \\
                 & -2.55    & -2.57     & -2.58     & -2.66     \\
\hline
$\mz\,$[GeV]     & 30.43633 & 30.43005  & 30.42316  & 30.31927  \\
                 & 30.35752 & 30.35109  & 30.34409  & 30.23944  \\
                 &-2.59     &-2.59      &-2.60      &-2.63      \\
\hline
$\mz+1.8\,$[GeV] & 14.18153 & 14.17541  & 14.16813  & 13.98373  \\
                 & 14.16470 & 14.15846  & 14.15117  & 13.96556  \\
                 &-1.19     &-1.20      &-1.20      &-1.30      \\
\hline
$\mz+3\,$[GeV]   &  8.19803 &  8.19206  &  8.18498  &  7.87534  \\
                 &  8.19800 &  8.19193  &  8.18460  &  7.87272  \\
                 & -0.004   & -0.02     & -0.05     & -0.33     \\
\hline
\hline
\end{tabular}
\vspace*{3mm}
\caption[]{SM results with the inclusion of ISPP.}
\label{tabSMR}
\end{table}
\normalsize
%--
\noindent
Next, some SM numbers, i.e. the hadronic cross-section for $\mz= 91.1867\,$GeV,
$\mt= 173.8\,$GeV, $\mh= 100\,$GeV and $\als(\mz)= 0.119$. In \tabn{tabSMR}
first entry is {\tt ONP = N}, second entry is {\tt ONP = I} and third entry is
with/without - 1 in per-mill.

Similar results have been obtained in~\cite{kn:ARBY}.

For an extension to LEP~2 energies~\cite{kn:RADCOR} let's start from a 
simple case, $e^+e^-$ PP-corrections to $e^+e^- \to \barb b$. 
To avoid confusion we will introduce the following
definition:
{\bf Multi-Peripheral} or {\bf MP} diagrams;
{\bf Initial State Singlet}, or {\bf ISS} diagrams;
{\bf Initial State Non-Singlet}, or {\bf ISNS} diagrams;
{\bf Final State}, or {\bf FS} diagrams.

Note that we include both $\ph$ and $\zb$ exchange, so that one could still
distinguish between ISNS$_{\ph}$, ISNS$_{\zb}$ and interference.
On top of real pair production one has to include virtual $e^+e^-$ pairs. 

So far, {\tt TOPAZ0} (and {\tt ZFITTER} too) only include ISNS$_{\ph}$ plus
virtual, with ISS optional. The approximation is well justified around the
$\zb$-peak but now we have to move to higher energies, where new thresholds 
open. Furthermore, one has to add a proper definition of {\em soft} pairs
and of {\em hard pairs}.
We can define
Soft(Hard) Invariant Mass pairs, or {\bf SIM(HIM)} pairs, according to
some pair mass cut.
Of course, one could still use a soft-hard separation based on other 
variables but, in any case, also very hard SIM pairs are 2f signal.

We need an {\em operative}, universally accepted, separation of various 
contributions.
Consider again the process $e^+e^- \to \barb b e^+ e^-$; the total 
cross-section contributes to three different processes:
a) genuine 4f events; b) $e^+e^-$ PP-correction to $e^+e^- \to \barb b$;
c) $\barb b$ PP-correction to Bhabha scattering.

A {\em naive} separation, easy-to-implement, would be the following. Let
the process be specified by $e^+e^- \to \barb b(Q^2) + e^+e^-(q^2)$. 
We can integrate and reduce the cross-section to
a two-fold integral with the following boundaries:
%--
\bqa
4\,m^2_b &<& Q^2 < (\sqrt{s}-2\,m_e)^2, \nl
4\,m^2_e &<& q^2 < (\sqrt{s}-\sqrt{Q^2})^2.
\eqa
%--
Next, we introduce two cuts $z_p,z_s$, i.e. primary and secondary cuts
and define

\begin{itemize}

\item  $e^+e^-$ PP-correction to $e^+e^- \to \barb b$:
%--
\bq
z_p\,s < Q^2 < (\sqrt{s}-2\,m_e)^2, \qquad
4\,m^2_e < q^2 < \min\{z_s\,s\,,\,(\sqrt{s}-\sqrt{Q^2})^2\}.
\eq
%--
\item $\barb b$ PP-correction to Bhabha scattering:
%--
\bq
z_p\,s < q^2 < (\sqrt{s}-2\,m_b)^2, \qquad
4\,m^2_b < Q^2 < \min\{z_s\,s\,,\,(\sqrt{s}-\sqrt{q^2})^2\}.
\eq
%--
\item while the remaining portion of the phase space is background.

\end{itemize}

At the same time, we must address what has to go in the calculation and what 
is already subtracted from the data. We have to define the exact content of 
PP-corrections to be inserted in the upgradings of the semi-analytical codes.
Then, one can apply a {\em theory} correction, actually a subtraction obtained
from MC, to go to signal definitions as employed by {\tt TOPAZ0/ZFITTER}.
This is what is done now, the only problem being that the present signal
definition in {\tt TOPAZ0/ZFITTER} is not good enough for LEP~2. Thus
signal definition must be improved and we have to find an agreement on the
optimal signal/background separation; it should be realistic enough to be
used in the experimental analysis but simple enough to allow its implementation
in the codes. 

Typical example? Using invariant-mass cuts one finds relatively large
corrections (around $5\%$ or more), most of which ares due to the
MP process. So, if the situation remains unaltered, one has to apply a 
rather large subtraction, which is rather unnatural. Of course, we can take 
the MP into account, but only if some {\em reasonable} cut is applied to
$M(\barb b)$.
At $200\,$GeV the double resonant $e^+e^- \to \zb\zb \to e^+e^- \barb b$ 
cross-section, i.e. the NC02 MC, is about $10^{-2}\,$pb, while the total 
multi-peripheral, that is not accessible with a NC02 (two diagrams) MC, is 
about $1\,$pb compared with a $\sigma_{\rm had}$ of $80-90\,$pb. 

However, this huge increase in cross-section is caused by hadronic two-photon 
collision processes, where the $\barb b$ system is created by the
two photons radiated of the $e^+e^-$. Thus the invariant mass of the 
$\barb b$ system will be very low, peaking toward $2\,m_b$. The bulk of the 
cross section will be around $10\,$GeV invariant $\barb b$ invariant mass.

Finally we come to the most complicate configurations, $e^+e^- \to \barq q
{\bar Q} Q$ or even worst, $e^+e^- \to \barq q \barq q$, e.g. $uuuu$ 
etc. configurations. 
We have to generalize our previous separation of the 2f signal from the 4f 
background. There are two options:
we require that {\em at least} one pair has an invariant mass greater
than $z_p\,s$. But only the $\barq q$ pairs or, should we also include
the $\barq Q$ pairs? And what about the $q Q, \barq Q$ pairs?
Or, we require that {\em at least} one pair has an invariant mass greater
than $z_p\,s$ while all remaining configurations have invariant mass less
than $z_s\,s$.

Now we come to the really difficult part of the problem. Since arbitrarily
low SIM pairs are allowed, in both cases, a parton-level calculation cannot
be accurate enough. 
Under the simplified assumption that one pair (the $Q^2$ one) is HIM enough
and that the remaining one (the $q^2$ one) is SIM enough we can write the
cross-section as
%--
\bqa
{{d\sigma_f}\over {dq^2dQ^2}} &=& \lpar\frac{\alpha}{\pi}\rpar^2\,\sigma_f(Q^2)
{{R_{\rm had}(q^2)}\over {3\,s q^2}}\,f\lpar s,Q^2,q^2\rpar\,  \nl
f\lpar s,Q^2,q^2\rpar &=& {{s+(Q^2+q^2)}\over {s-Q^2-q^2}}\,\ln
{{s-Q^2-q^2+\sqrt{\lambda}}\over {s-Q^2-q^2-\sqrt{\lambda}}} - 2\,
\sqrt{\lambda}\,  \nl
\lambda &=& \lambda\lpar s,Q^2,q^2\rpar.
\eqa
%--
Clearly we face two problems, one conceptual and another practical. The
first problem will be referred to as the {\em double-counting 
problem}.
At the parton level, we have $e^+e^- \to \barq q {\bar Q} Q$ and,
for instance we want to sum over $q$ in order to define the $Q$-line shape.
However, when we also sum over $Q$ in order to have the full hadronic line 
shape, proper care must be taken in order to avoid double-counting. At the
parton level this can be done but one should remember that we actually
compute $e^+e^- \to {\bar Q} Q + {\rm hadrons}$, through $R_{\rm had}(s)$,
and the problem is, by far, more severe.
Therefore, we have to agree on what should happen once $R_{\rm had}(s)$ is 
called for low $s$.

At least in principle, we can introduce three different
cuts, i.e. $s'/s$ for initial state QED radiation and $z_p,z_s$ for pair
production. From a theoretical point of view, there is no problem here.
We should reach an agreement: one common cut ($s'/s = 
z_{\rm min}$), two cuts ($s'/s \not= z_{\rm min}$ but also $s'/s = z_p$ and
$z_s$) or three cuts ($s'/s \not= z_p \not= z_s$).

Tentative conclusions are as follows:
the whole 4f must be included to compute the 2f cross-sections;
Or, the whole 4f is to be divided into two components, {\em signal}
and {\em background}. For our purposes their definition is peculiar,
{\em signal} is what you have implemented into the semi-analytical 
codes.
{\em Background} is what one subtracts by using a MC.
We go from one extreme solution to the other:

{\em Background} = $\emptyset$ if everything is included in the 
semi-analytical calculation. Multi-peripheral is an example of something 
difficult to implement into the semi-analytical approach if low-invariant 
mass regions are required.

{\em Signal} = ISNS, i.e. everything else (large effects) is subtracted
by MC. However, using different MC programs would bring to subtractions
that differ by some per-cent, which then would have to be regarded as a 
{\em theoretical} systematic uncertainty. 

One could decide to count as {\em background} 4f processes 
from {\em true} $e^+e^-$ annihilation (e.g. $\zb\zb, \wb\wb$)
and 4f processes via e-gamma collisions (e.g. $(e)e\zb,(e)\nu\wb$). 
This distinction is based on selecting sub-sets of diagrams to be 
used for the 2f signal definition. However, we should not define 
the {\em signal} according to the kind diagrams it originates from.
We need a definition based on measurable quantities, so we should not
exclude a set of diagrams, we should exclude the kinematical region where 
that particular set dominates.

\vskip 1cm
\noindent{\Large\bf Acknowledgements}

\vskip 0.5cm
\noindent
I would like to thank Edward Boos and Misha Dubinin for the invitation and 
for the very pleasant stay at the QFTHEP '99 Workshop.
I am particularly thankful to Dmitri Bardin, Frank Filthaut, Martin 
Gr\"unewald, Joachim Mnich and G\"unter Quast for many fruitful discussions.

\clearpage

\vspace*{-2mm}
\bqas
  \begin{picture}(300,300)(0,0)
  \GCirc(100,100){100}{1}
  \ArrowLine(-50,-50)(30,30)
  \ArrowLine(30,170)(-50,270)
  \ArrowLine(280,270)(170,170)
  \ArrowLine(170,30)(280,-50)
  \GCirc(15,190){5}{0}
  \Photon(15,190)(120,270){3}{7}
  \Photon(15,190)(120,250){3}{7}
  \GCirc(15,15){5}{0}
  \Photon(15,15)(110,-90){3}{7}
  \Photon(15,15)(110,-110){3}{7}
  \Text(100,100)[cb]{\huge\bf EW $\,\oplus\,$ QCD}
  \Text(140,260)[cb]{\Large\bf ISR}
  \Text(130,-100)[cb]{\Large\bf ISR}
  \GCirc(-25,-25){5}{0}
  \Photon(-25,-25)(60,-120){3}{7}
  \ArrowLine(80,-100)(60,-120)
  \ArrowLine(60,-120)(80,-140)
  \Text(100,-130)[cb]{\Large\bf ISPP}
  \Photon(-25,240)(60,240){3}{7}
  \Photon(160,240)(245,240){3}{7}
  \Text(30,250)[cb]{\Large\bf ISR $\,\otimes\,$FSR}
  \GCirc(210,205){5}{0}
  \Photon(210,205)(270,200){3}{7}
  \Photon(210,205)(270,180){3}{7}
  \Text(290,190)[cb]{\Large\bf FSR}
  \GCirc(200,10){5}{0}
  \Gluon(200,10)(270,20){3}{7}
  \Gluon(200,10)(270,-10){3}{7}
  \Text(310,5)[cb]{\Large\bf QCD FSR}
  \Text(280,100)[cb]{\huge\bf $\leadsto\,\,\seffsf{\ff},\,\gvaf\,\dots$}
  \Text(320,-20)[cb]{\Large\bf $\smanp$-cut}
  \Text(320,-40)[cb]{\Large\bf $M^2(\barf f)$-cut}
  \Text(320,-60)[cb]{\Large\bf $\theta_{\rm acol}$}
  \Text(320,-80)[cb]{\Large\bf $E_{\rm th}$}
  \Text(320,-100)[cb]{\Large\bf $\theta_{\rm acc}$}
  \Text(-20,80)[cb]{\Huge\bf $\succ$}
  \Text(-80,84)[cb]{\large\bf de-convolution}
  \end{picture}
\eqas
\vspace{5cm}
\begin{figure}[h]
\caption[]{The process $e^+e^- \to \barf f$.}
\label{kit}
\end{figure}

\clearpage

%------------------------------------------------------------------------

\end{document}